\documentclass[aps,prb,twocolumn,showpacs,amsmath,amssymb]{revtex4-1}

\usepackage{graphicx}
\usepackage{dcolumn}
\usepackage{bm}
\usepackage[colorlinks=true]{hyperref}

\newcommand{\uv}{{\mathbf u}}
\newcommand{\rv}{{\mathbf r}}
\newcommand{\kv}{{\mathbf k}}
\newcommand{\kp}{{\mathbf k'}}
\newcommand{\qv}{{\mathbf q}}
\newcommand{\delk}{{\mathbf{\Delta k}}}
\newcommand{\hep}{{H_{\rm e-ph}}}
\newcommand{\LL}{{\mathcal L}}
\newcommand{\WW}{{\mathcal W}}
\newcommand{\bra}[1]{{\langle{#1}|}}
\newcommand{\ket}[1]{{|{#1}\rangle}}
\newcommand{\ltf}{\lambda_{\rm TF}}
\newcommand{\Tm}{T_{\rm m}}
\newcommand{\kB}{k_{\rm B}}

\begin{document}

\title{Damping of mechanical vibrations by free electrons in metallic nanoresonators}

\author{Ze'ev Lindenfeld and Ron Lifshitz}
\email[Corresponding author:\ ]{ronlif@tau.ac.il}
\affiliation{Raymond and Beverly Sackler School of Physics and Astronomy,
Tel Aviv University, Tel Aviv 69978, Israel}

\date{November 2, 2012}

\begin{abstract}
  We investigate the effect of free electrons on the quality factor ($Q$) of a metallic nanomechanical resonator in the form of a thin elastic beam. The flexural and longitudinal modes of the beam are modeled using thin beam elasticity theory, and simple perturbation theory is used to calculate the rate at which an externally excited vibration mode decays due to its interaction with free electrons. We find that electron-phonon interaction significantly affects the $Q$ of longitudinal modes, and may also be of significance to the damping of flexural modes in otherwise high-$Q$ beams. The finite geometry of the beam is manifested in two important ways. Its finite length breaks translation invariance along the beam and introduces an imperfect momentum conservation law in place of the exact law. Its finite width imposes a quantization of the electronic states that introduces a temperature scale for which there exists a crossover from a high-temperature macroscopic regime, where electron-phonon damping behaves as if the electrons were in the bulk, to a low-temperature mesoscopic regime, where damping is dominated by just a few dissipation channels and exhibits sharp non-monotonic changes as parameters are varied. This suggests a novel scheme for probing the electronic spectrum of a nanoscale device by measuring the $Q$ of its mechanical vibrations.

\end{abstract}

\maketitle

\section{Introduction}\label{sec:introduction}

The design and fabrication of high-$Q$ mechanical resonators is an ongoing effort that has intensified with the advent of microelectromechanical systems (MEMS) and even more with the recent progression toward nanoelectromechanical systems (NEMS).\cite{RoukesPlenty,Cleland03,ekinci05} One requires low-loss mechanical resonators for a host of nanotechnological applications, such as low phase-noise oscillators;\cite{Kenig12} highly sensitive mass,\cite{ekinci04,*Yang06,*Hanay12,ilic04,Jensen08,lassagne08} spin,\cite{rugar} and charge detectors;\cite{cleland98} and ultra-sensitive thermometers\cite{roukes99} and displacement sensors;\cite{cleland,*knobel,ekinci02,*truitt07} as well as for basic research in the mesoscopic physics of phonons,\cite{schwab00} and the general study of the behavior of mechanical degrees of freedom at the interface between the classical and the quantum worlds.\cite{schwab05,lahaye04,*naik06,Oconnell10,katz,*katz08}  It is therefore of great importance to understand the dominant damping mechanisms in small mechanical resonators.

A variety of different mechanisms---such as internal friction due to bulk or surface defects,\cite{Mihailovich95,olkhovets,*carr2,*evoy2,liu,mohanty,zolfagharkhani,seoanez,remus,chu07,unterreithmeier} phonon-mediated damping,\cite{lifshitzTED,*lifshitzPhonon,houston,sudipto,kiselev} and clamping losses \cite{cross,photiadis1,geller3,*geller1,*geller2,schmid08,wilson-rae,*cole}---may contribute to the dissipation of energy in mechanical resonators, and thus impose limits on their quality factors. The dissipated energy is transferred from a particular mode of the resonator, which is driven externally, to energy reservoirs formed by all other degrees of freedom of the system. Here, we focus our attention on \emph{electron-phonon damping}, arising from energy transfer between the driven mode of the resonator and free electrons. This dissipation mechanism is avoided altogether by fabricating resonators from dielectric materials, but for different practical reasons one often prefers to fabricate MEMS and NEMS resonators from metals, such as platinum,\cite{husain} gold,\cite{buks02,venkatesan10} and aluminum.\cite{davis,li,*hoehne,teufel} Free electrons are also present in metallic carbon-nanotube resonators\cite{peng06,Eriksson08} and in resonating nanoparticles.\cite{min,pelton,zijlstra} All these different resonators exhibit a wide range of quality factors, from as low as about $10$ and up to  around $10^5$, yet one still lacks a full understanding of their damping mechanisms.

It is well-known from at least as early as the 1950s that electron-phonon scattering is a dominant source of attenuation of longitudinal sound waves in bulk metals at low temperatures,\cite{Bommel54,kittel,pippard,blount,ziman,kokkedee} and indications exist that it may play a significant role in the damping of longitudinal vibrations in freely suspended bi-pyramid gold nanoparticles.\cite{pelton} We note that the effect of electron-phonon scattering on electronic transport through suspended nanomechanical beams,\cite{weig04} carbon nanotubes,\cite{leroy04,leturcq09,*mariani,Steele09,*huttel09,*Laird12} fullerenes,\cite{park} atomic wires,\cite{paulsson,viljas,vega} and molecular junctions\cite{pecchia,kushmerick,wang,galperin,Tal08} is well documented and intensively studied. There is also evidence for the effect of electron-phonon scattering on heat transport in nanostructures.\cite{fon,barman} Motivated by all of these considerations, it is our aim here to estimate the contribution of electron-phonon interaction to the damping of vibrational modes in small metallic resonators, while focusing on the effects of their finite dimensions. 

We describe the interaction between electrons and phonons by means of a simple screened electrostatic potential. We assume that initially both electrons and phonons are at thermal equilibrium at the same temperature, except for a single mode, which is externally excited by the addition of just a single phonon to its thermal population. This allows us to assume that the electrons remain almost thermally distributed at all times, even though they do not actually relax back to equilibrium. The decay rate of the excited mode is calculated pertubatively,  using Fermi's Golden Rule, as the difference between the rates at which phonons enter and leave the excited mode through their interaction with free electrons. This requires us to assume that the electron and phonon energies are precisely known, or in other words that the \emph{a priori} lifetimes of both the electrons and the phonons are much longer than all other relevant time scales. 

For the phonons this means that all other damping mechanisms must be much weaker than electron-phonon damping---although we show later that additional damping mechanisms do not significantly alter the results of our calculations. For the electrons, on the other hand, this implies that we are working in the high-frequency, or unrelaxed \emph{adiabatic limit}, with $\omega_q\tau_{e}>1$, where $\omega_q$ is the vibration frequency and $\tau_{e}$ is the mean lifetime of the electron due to its scattering with other electrons, thermal phonons, defects, etc. In certain situations, as discussed in detail in section 5.12 of the book by Ziman,\cite{ziman} it is sufficient to satisfy the spatial version of this requirement, namely that $\Lambda_{e}q>1$---where $\Lambda_{e}$ is the electron mean free path and $q$ is the wavenumber of the excited mode---which is easier to satisfy because the ratio of the phonon group velocity to the electron Fermi velocity is typically very small. Intuitively speaking, it is as if the moving electron explores the elastic wave much faster than it would have, if it were standing in place and waiting for the wave to go by. In bulk metals it is difficult to satisfy the adiabatic condition, and one is inevitably required to address the relaxation of the electronic distribution by employing the Boltzmann equation or other approaches.\cite{pippard,blount,Khan87} However, in clean nanometer scale devices oscillating at very-high frequencies, and operating at sufficiently low temperatures, there is a greater chance of reaching the adiabatic limit. We therefore assume that this is the case, and alert the reader to the fact that our results may be less applicable at high temperatures.

We describe the vibrational modes of the nanomechanical resonator using continuum elasticity theory, which is often employed for treating nanomechanical systems, even in the case of carbon nanotubes,\cite{kahn,suzuura,martino}, and also in the quantum regime.\cite{blencowe99,santamore01,*santamore02,lindenfeld11} The small size of such nanostructures may raise the question of the validity of a continuum elastic approach. However, explicit comparisons with atomistic calculations and experimental results have shown that continuum elasticity is valid down to dimensions of a few nanometers,\cite{Broughton97,murray2,combe,ramirez} and may indeed be used even for carbon nanotubes, as long as one uses appropriate effective parameters.\cite{yoon,chico} 

Finally, we investigate beams with typical dimensions that are much larger than the bulk Fermi wavelength. In such systems it is usually assumed that the effect of the boundaries on the free electrons can be ignored and that the electrons are essentially unconfined. However, as we demonstrate here, the transverse dimensions of the beam set a temperature scale---between a few kelvins to more than a hundred kelvins for the beam geometries considered here---below which the confinement of the electrons can no longer be ignored, and electron-phonon damping is expected to behave qualitatively different. 

The paper is organized as follows. In section~\ref{sec:Phonon and electron states} we describe the quantized vibrational modes of a nanomechanical beam. The interaction between these modes and free electrons is addressed in section~\ref{sec:Interaction between the electrons and the phonons}, and the resulting expressions for the dimensionless damping $Q^{-1}$ are derived in section~\ref{sec:Decay rates of the longitudinal and flexural modes}. The reader who is not interested in the technical details of the calculations is invited to skip directly to section~\ref{sec:Results}, where we discuss and explain the results of the calculation and their physical implications. We conclude with a summary in section~\ref{sec:Conclusion}.

\section{Quantized vibrational modes of a nanomechanical beam}\label{sec:Phonon and electron states}

We consider a mechanical resonator in the form of a thin elastic beam with a rectangular cross section, although we expect the essence of our calculation to be independent of the specific geometry of the resonator. In what follows we describe the classical longitudinal and flexural normal modes of vibration, obtained within standard theories of thin elastic beams,\cite{graff,landl} and quantize these modes as a collection of simple harmonic oscillators. We do not consider the twist modes of the thin beam because the displacement fields of these modes have zero divergence and therefore do not couple significantly to free electrons, as explained in the next section.

We take the length $l$ of the beam to lie along the $x$ axis from $x=-l/2$ to $x=l/2$, and its transverse dimensions along the $y$ axis from $y=0$ to $y=a$, and along the $z$ axis from $z=0$ to $z=b$. In equilibrium, the beam is unstrained, unstressed, and at temperature $T$ everywhere. Departure of the beam from equilibrium is described by a displacement field $u_i$ ($i=x,y,z$). The displacement field $u_i$, as well as the strain and stress tensors $u_{ij}$ and $\sigma_{ij}$, are all functions of position and time. Yet, we assume that the temperature of the beam remains uniform and constant during the vibration, thus ignoring thermoelastic effects.\cite{lifshitzTED} We take the surfaces of the beam to be stress free, which implies that all but the $\sigma_{xx}$ component of the stress tensor vanish on the surface. Because the beam is thin, this approximately holds in its interior as well. Hooke's law for the thin beam then takes a rather simple form:
\begin{subequations}\label{hooke}
\begin{eqnarray}
u_{xx}&=&{1\over E}\sigma_{xx}, \\
u_{yy}&=&u_{zz}=-{\sigma\over E}\sigma_{xx}, \label{hooke-sigma}\\
u_{xy}&=&u_{yz}=u_{zx}=0,
\end{eqnarray}
\end{subequations}
where $E$ is Young's modulus, and $\sigma$ is Poisson's ratio.

\subsection{Quantized longitudinal modes}\label{subsubsec:Longitudinal modes}

To describe longitudinal vibrations in the thin beam we make an additional simplifying assumption of neglecting any contribution to the dynamics that arises from having a nonzero Poisson ratio. We therefore take $u_y=u_z=0$, take $u_x$ to be independent of $y$ and $z$, and ignore any deviation of the cross section of the beam from its rectangular shape. Under these assumptions, longitudinal vibrations are governed by a standard dispersionless wave equation. Thus, a mode of wavenumber $q$ vibrates at a frequency $\omega_{q} = c_{l} q$, where $c_{l} = \sqrt{E/\rho}$ is the speed of sound of longitudinal waves in the bulk, and $\rho$ is the mass density of the beam. Taking $\sigma\neq 0$ in Eq.~\eqref{hooke-sigma} would lead to the so-called Love equation\cite{graff} and to dispersive longitudinal modes, which we do not consider here.

In the  limit $l\rightarrow\infty$ of an infinitely long thin beam it is convenient to describe the longitudinal modes as travelling waves leading to a phonon field operator of the form
\begin{equation}\label{eq:inf_long_modes}
u_x = \sum_{q}\sqrt{\frac{\hbar}{2M\omega_{q}}}e^{iqx}\left(b_{q}+b^{\dag}_{-q}\right),
\end{equation}
where $b^\dag_q$ and $b_q$ are bosonic creation and annihilation operators, $M=\rho V$ is the mass of the beam, $V=abl$ is its volume, and we use a large volume normalization of the phononic wave functions as $M$ and $V$ tend to infinity along with $l$. In this limit, the sum $\sum_q$ is to be interpreted as an integral $l\int dq/2\pi$.

For beams of finite length we consider either doubly-clamped boundary conditions, with $u_x=0$ at both ends of the beam, which is a common experimental geometry, or stress-free boundary conditions, with $du_x/dx=0$ at both ends, which may be suitable for describing freely suspended resonators in solution.\cite{min,zijlstra,pelton} The phonon field operator is then given by
\begin{equation}\label{eq:finite_long_modes}
u_x = \sum_{n=1}^\infty\sqrt{\frac{\hbar}{M\omega_{q_{n}}}}\left(A_{n}\sin{q_{n}x}+B_{n}\cos{q_{n}x}\right)
\left(b_{q_{n}}+b^{\dag}_{q_{n}}\right),
\end{equation}
where $q_{n} = n\pi/l$, and for doubly-clamped modes
\begin{equation}
[A_{n},B_{n}] =
\begin{cases}
[0,1], & {\rm odd\ } n,\\
[1,0], & {\rm even\ } n,
\end{cases}
\end{equation}
while for stress-free modes
\begin{equation}
[A_{n},B_{n}] =
\begin{cases}
[1,0], & {\rm odd\ } n,\\
[0,1], & {\rm even\ } n.
\end{cases}
\end{equation}

\subsection{Quantized flexural modes}\label{subsubsec:Flexural modes}

We describe flexural vibrations in the thin beam by the transverse motion $w(x,t)$ of the neutral axis of the beam in the $z$ direction. We make the usual Euler-Bernoulli assumption that the transverse dimensions of the beam, $a$ and $b$, are sufficiently small compared with the length $l$ of the beam and the radius of curvature $R$ of the bending that any plane cross section, initially perpendicular to the axis of the beam, remains perpendicular to the neutral axis during bending. We further assume that the rectangular shape of the cross section remains unaltered during the bending motion. Such an approximation is justified for small deflections since the error it introduces is only on the order of the transverse beam dimension divided by the radius of curvature of the bending. Since we assume a non-deformable cross-section, there is in fact a neutral surface running through the length of the beam, at $z=b/2$, which suffers no extension or contraction during its bending.  One can then show\cite{landl} that the longitudinal strain component $u_{xx}$, a distance $\delta z=z-b/2$ away from the neutral surface, is equal to $\delta z/R$. By replacing the curvature of the beam $1/R$ with $-\partial^2 w/\partial x^2$ we express the non-zero components of the strain field in the beam as
\begin{subequations}\label{eq:flex-strain}
\begin{eqnarray}
    u_{xx} &= &-\left(z-\frac{b}{2}\right) \frac{\partial^2 w}{\partial x^2}, \\
    u_{yy} &= &u_{zz} = \sigma \left(z-\frac{b}{2}\right) \frac{\partial^2 w}{\partial x^2}.
\end{eqnarray}
\end{subequations}
Using this strain field it is possible to write down the Lagrangian density of the beam and derive its equation of motion. This equation contains two kinetic terms, one that is associated with the transverse motion of the beam and one that is associated with the rotation of the cross-section. The latter is smaller by a factor of $b/R$, and is therefore usually neglected, leading to the Euler-Bernoulli equation of motion
\begin{equation}\label{eq:E-B equation}
\rho A \frac{\partial^2 w}{\partial t^2}
+ \frac{\partial^2}{\partial x^2} \left(EI \frac{\partial^2 w}{\partial x^2}\right) = 0,
\end{equation}
where $A=ab$  is the area of the cross-section, and $I=ab^{3}/12$ is the moment of inertia of the cross section. The resulting flexural modes possess a quadratic dispersion, $\omega_{q}=Gq^{2}$, where $G = \sqrt{EI/\rho A} = b\sqrt{E/12\rho}$.

In the limit of an infinitely long beam, as in Eq.~\eqref{eq:inf_long_modes}, it is again convenient to describe the transverse modes as travelling waves with a quantized phonon field operator of the form
\begin{equation}\label{eq:inf_flex_modes}
w = \sum_{q}\sqrt{\frac{\hbar}{2M\omega_{q}}} e^{iqx} \left(b_{q}+b^{\dag}_{-q}\right).
\end{equation}
For beams of finite length we consider doubly clamped boundary conditions, taking $w=dw/dx=0$ at both ends. The phonon field operator is then given by
\begin{multline}\label{eq:finite_flex_modes}
w = \sum_{n=1}^\infty\sqrt{\frac{\hbar}{2M\omega_{q_{n}}}}
\left(A_{n}\sin{q_{n}x} + B_{n}\cos{q_{n}x}\right.\\ + C_{n}\sinh{q_{n}x} + D_{n}\cosh{q_{n}x})
\left(b_{q_{n}}+b^{\dag}_{q_{n}}\right),
\end{multline}
where the wavenumbers $q_{n}$ are solutions of the transcendental equation $\cos q_n l \cosh q_n l = 1$, with $q_n l$ tending to odd-multiples of $\pi/2$ as $n$ increases. The numerical coefficients $A_{n},B_{n},C_{n},D_{n}$ are determined through the boundary conditions and the normalization of the phonon wave functions, where by symmetry $A_{n}=C_{n}=0$ for odd $n$, and $B_{n}=D_{n}=0$ for even $n$.

\section{Electron-phonon interaction}\label{sec:Interaction between the electrons and the phonons}

We assume a standard screened static interaction potential \cite{fetter}
\begin{equation}\label{eq:e-ph potential}
v\left(\rv -\mathbf{r'}\right)=\frac{-4\pi Ze^{2}}{q_{\rm TF}^{2}}\delta\left(\rv-\mathbf{r'}\right),
\end{equation}
between the negative charge $-e$ of the electron density $\sum_{\sigma}\psi^{\dag}_{\sigma}(\rv)\psi_{\sigma}(\rv)$ and the positive charge $Ze$ of the disturbance in the density $n_0 \delta v(\rv')$ of the ionic background, induced by the vibration. Here $q^{\phantom{2}}_{\rm TF}$ is the Thomas-Fermi wavenumber, $e$ is the magnitude of the electron charge, $\sigma$ is the electron spin (not to be confused with Poisson's ratio), $Z$ is the number of valence electrons per atom in the material, $n_{0}$ is the atomic density, and $\delta v(\rv)$ is the local volume change induced by the vibration. The electron-phonon interaction Hamiltonian, derived from this potential, is
\begin{equation}\label{eq:e-ph general}
\hep =-g\int\limits_{V}\sum_\sigma\psi_\sigma^{\dag}(\rv)\psi_\sigma(\rv)\delta v(\rv) d^{3}\rv,
\end{equation}
where
\begin{equation}
g=\frac{4\pi Zn_{0}e^2}{q_{\rm TF}^{2}}.
\end{equation}

For small deformations, to first order in the displacement field, the local change in volume $\delta v(\rv) = \nabla\!\cdot\!\uv$. For twist modes $\nabla\!\cdot\!\uv=0$  giving $\delta v(\rv)=0$, which is why we have ignored them here; for longitudinal modes
$\nabla\cdot\uv = \partial u_x/\partial x$, where $u_x$ is given by either Eq.~\eqref{eq:inf_long_modes} or Eq.~\eqref{eq:finite_long_modes}; and for flexural modes we find from Eqs.~\eqref{eq:flex-strain} that
\begin{equation}\label{eq:div u flexural}
  \mathbf{\nabla}\!\cdot\!\mathbf{u}=(2\sigma-1)\left(z-\frac{b}{2}\right)\frac{d^{2}w}{dx^{2}},
\end{equation}
where $w$ is given by either Eq.~\eqref{eq:inf_flex_modes} or Eq.~\eqref{eq:finite_flex_modes}.

We take two approaches for describing the electron field $\psi(\rv)$, where from here onwards we suppress the spin index. As discussed in section \ref{sec:introduction}, treating the electrons as bulk-like is a common approximation for structures of the size considered in this work. Thus we can use a simple free electron field of the form
\begin{equation}\label{eq:electron_field_inf}
\psi(\rv)=\sum_{\kv}\frac{1}{\sqrt{V}}e^{i\kv\cdot\rv}c_{\kv},
\end{equation}
where $c_{\kv}$ is a fermionic annihilation operator. Within this approximation it is permissable to treat the transverse dimensions of the beam as being infinite from the point of view of the electrons. Accordingly, the sum $\sum_{\kv}$ is replaced by a 3-dimensional integral $V \int d\kv/(2\pi)^3$. This leads to a simplification of the calculations and the resulting expressions below.

Alternatively, we do not neglect the finite transverse dimensions (but the length of the beam is still considered to be infinite as far as the electrons are considered), and view the electrons as being geometrically confined. We then obtain the ``particle in a box'' field operators
\begin{equation}
\psi(x,y,z) = \sum_{\kv}\frac{2}{\sqrt{V}} e^{ik_{x}x} \sin{k_{y}y} \sin{k_{z}z}\ c_{\kv},\label{electron_field_fin}\\
\end{equation}
with $k_{y} = n_{y}\pi/a$, $k_{z} = n_{z}\pi/b$, and $n_y, n_z = 1,2,3,\ldots$ In both cases the energy of the free electrons is given by
\begin{equation}\label{eq:electron energy}
\varepsilon_{\kv} = \frac{\hbar^{2}}{2m}\left(k_{x}^{2}+k_{y}^{2}+k_{z}^{2}\right) = \frac{\hbar^2 k^2}{2m},
\end{equation}
but when the lateral confinement of the electrons is not neglected it takes the explicit form of parabolic bands
\begin{equation}\label{eq:electronic bands}
\varepsilon_{\kv} = \varepsilon^\textrm{min}_{n_y,n_z} + \frac{\hbar^{2} k_{x}^{2}}{2m}, \textrm{with\ }
\varepsilon^\textrm{min}_{n_y,n_z} = \frac{\hbar^{2}\pi^2}{2m} \left(\frac{n_y^2}{a^2} + \frac{n_z^2}{b^2}\right).
\end{equation}

\subsection{Interaction of free electrons with longitudinal phonons}\label{subsec:Longitudinal phonons}

To obtain the interaction Hamiltonian~\eqref{eq:e-ph general} for longitudinal phonons with unconfined free electrons in an infinite beam we use the derivative of Eq.~\eqref{eq:inf_long_modes}  for the quantized elastic displacement field, and the field operator~\eqref{eq:electron_field_inf} for unconfined electrons. This yields
\begin{widetext}
\begin{eqnarray}
\hep &=&-i\frac{g}{V} 
\sum_{\kv,\kp,q} \sqrt{\frac{\hbar}{2M \omega_q}} q \int\limits^{\infty}_{-\infty} dy
e^{i\left(k_{y}-k^{'}_{y}\right)y}
\int\limits^{\infty}_{-\infty} dz e^{i\left(k_{z}-k^{'}_{z}\right)z}
\int\limits^{\infty}_{-\infty} dx e^{i\left(k_{x}-k^{'}_{x}+q\right)x}
c^{\dag}_{\kp}c_{\kv}\left(b_{q}+b^{\dag}_{-q}\right)\nonumber\\
&=&-ig \sum_{\kv,q} \sqrt{\frac{\hbar}{2M \omega_q}} q c^{\dag}_{\kv+\qv}c_{\kv}\left(b_{q}+b_{-q}^{\dag}\right),
\label{eq:e-ph longitudinal infinite free electrons}
\end{eqnarray}
\end{widetext}
where $\qv=q\hat{x}$, and where we have performed the sum over $\kp$ to eliminate the three delta functions that appear as exponential integrals in the first line of Eq.~\eqref{eq:e-ph longitudinal infinite free electrons}. We note that the integrations in the $y$ and $z$ directions involve only the unconfined electron wavenumbers. Thus, the width and thickness of the beam can effectively be taken to be infinite, which leads to momentum conservation in the $y$ and $z$ directions.

When the confinement of the electrons is not neglected, the electronic field \eqref{eq:electron_field_inf} is replaced with the one given by Eq.~\eqref{electron_field_fin}. When longitudinal phonons in an infinite beam are considered, the replacement of the continuous electron spectrum~\eqref{eq:electron energy} with the set of parabolic bands given by Eq.~\eqref{eq:electronic bands} yields the same formal expression for the interaction Hamiltonian as above in Eq.~\eqref{eq:e-ph longitudinal infinite free electrons}. The only difference is that the summations over $k_{y}$ and $k_{z}$ become discrete.

To obtain the interaction Hamiltonian of longitudinal modes with unconfined electrons in a beam of finite length we use the derivative of the elastic displacement field given by Eq.~\eqref{eq:finite_long_modes}. This then yields
\begin{widetext}
\begin{eqnarray}
\hep &=&-\frac{g}{V} \sum_{\mathbf{k},\mathbf{k'},q_{n}} \sqrt{\frac{\hbar q_n}{M c_{l}}}
\int\limits^{\infty}_{-\infty} dy e^{i\left(k_{y}-k^{'}_{y}\right)y}
\int\limits^{\infty}_{-\infty} dz e^{i\left(k_{z}-k^{'}_{z}\right)z}
\int\limits^{\frac{l}{2}}_{-\frac{l}{2}} dx e^{i\left(k_{x}-k^{'}_{x}\right)x}\left(A_{n}\cos{q_{n}x}-B_{n}\sin{q_{n}x}\right)
c^{\dag}_{\kp}c_{\kv}\left(b_{q_{n}}+b^{\dag}_{q_{n}}\right)\nonumber\\
&=&-\frac{g}{2} \sum_{\kv,k'_{x},q_{n}} \sqrt{\frac{\hbar q_n}{M c_{l}}}
\left[\left(A_{n}-iB_{n}\right)\mathrm{sinc}\frac{\left(q_{n}+\Delta k_x\right) l}{2}
+\left(A_{n}+iB_{n}\right)\mathrm{sinc}\frac{\left(q_{n}-\Delta k_x\right) l}{2}\right]
c^{\dag}_{\kv+\Delta\kv}c_{\kv}\left(b_{q_{n}}+b^{\dag}_{q_{n}}\right),\label{eq:e-ph longitudinal finite free electrons}
\end{eqnarray}
\end{widetext}
where $\Delta\kv=\Delta k_x \hat{x}$, with $\Delta k_x = k'_x-k_x$, and where the $\mathrm{sinc}(\alpha) = \sin(\alpha)/\alpha$ functions replace the momentum conservation delta function that exists for the infinite beam in Eq.~\eqref{eq:e-ph longitudinal infinite free electrons}.

We note that by taking the limit of $l\rightarrow\infty$ in the first line of Eq.~\eqref{eq:e-ph longitudinal finite free electrons} we obtain the interaction Hamiltonian for an infinitely long beam as written in the basis of standing waves instead of the basis of traveling waves, used in Eq.~\eqref{eq:e-ph longitudinal infinite free electrons}. Using this Hamiltonian it is possible to calculate the decay rate of a standing wave mode in an infinitely long beam, which is identical to the one obtained from the Hamiltonian in Eq.~\eqref{eq:e-ph longitudinal infinite free electrons} for a traveling wave. Alternatively, by converting the creation and annihilation  operators of the standing waves in Eq.~\eqref{eq:e-ph longitudinal finite free electrons} into creation and annihilation operators of traveling waves, it is possible to show that the Hamiltonian in Eq.~\eqref{eq:e-ph longitudinal finite free electrons} is identical to that of Eq.~\eqref{eq:e-ph longitudinal infinite free electrons} in the limit of an infinite beam and continuous phonon wavenumbers.

\subsection{Interaction of free electrons with flexural phonons}\label{subsec:Flexural phonons}

For flexural vibrations interacting with laterally confined electrons in an infinite beam, we substitute the divergence~\eqref{eq:div u flexural} of  the quantized elastic displacement field \eqref{eq:inf_flex_modes} together with the field operator~\eqref{electron_field_fin} for laterally confined electrons into the general expression~\eqref{eq:e-ph general} of the electron-phonon interaction Hamiltonian. We then obtain
\begin{widetext}
\begin{eqnarray}
\hep &=&\frac{4g (2\sigma-1)}{V} \sum_{\mathbf{k},\mathbf{k'},q} \sqrt{\frac{\hbar q^2}{2MG}}
\int\limits^{a}_{0} dy \sin{k_{y}y}\sin{k'_{y}y} \int\limits^{b}_{0} dz \left(z-\frac{b}{2}\right)\sin{k_{z}z}\sin{k'_{z}z}
\int\limits^{\infty}_{-\infty} dx e^{i\left(k_{x}-k'_{x}+q\right)x} c^{\dag}_{\kp}c_{\kv} \left(b_{q}+b^{\dag}_{-q}\right)\nonumber\\
&&=\frac{8g(1-2\sigma)b}{\pi^{2}} \sum_{\kv,k'_{z},q} \sqrt{\frac{\hbar q^2}{2MG}} \frac{n_{z}n'_{z}}{\left(n^{2}_{z}-n'^{2}_{z}\right)^{2}}c^{\dag}_{\kv+\Delta\kv}c_{\kv} \left(b_{q}+b^{\dag}_{-q}\right),
\label{eq:e-ph flexural infinite confined electrons}
\end{eqnarray}
where now $\Delta\kv=q\hat{x} + \Delta k_z \hat{z}$, with $\Delta k_z = k'_z-k_z$, and where the last sum is restricted to pairs of integers $n_{z}$ and $n'_{z}$ that are of different parity. Note that owing to the transverse nature of the vibration the electron momentum changes both in the $x$ and in the $z$ directions.

The interaction Hamiltonian between the flexural modes and unconfined free electrons in an infinitely long beam is
\begin{eqnarray}\label{eq: hamiltonian flexural bulk electrons inf beam 1}
\hep &=&\frac{g(1-2\sigma)}{V} \sum_{\kv,\kp,q} \sqrt{\frac{\hbar q^2}{2MG}} 
\int\limits^{\infty}_{-\infty} dx e^{i\left(k_{x}-k'_{x}+q\right)x}
\int\limits^{\infty}_{-\infty} dy e^{i\left(k_{y}-k'_{y}\right)y}
\int\limits^{b}_{0} dz \left(z-\frac{b}{2}\right)e^{i\left(k_{z}-k'_{z}\right)z}
c^{\dag}_{\kp}c_{\kv} \left(b_{q}+b_{-q}^{\dag}\right)\nonumber\\
&=& i\frac{g(1-2\sigma)}{b} \sum_{\kv,k'_{z},q} \sqrt{\frac{\hbar q^2}{2M G}} 
\frac{2\sin{\frac{\Delta k_{z}b}{2}} -\Delta k_{z}b \cos{\frac{\Delta k_{z}b}{2}} }{\Delta k_{z}^{2}} e^{-i\frac{\Delta k_{z}b}{2}}
c^{\dag}_{\kv+\Delta\kv}c_{\kv}\left(b_{q}+b_{-q}^{\dag}\right),
\end{eqnarray}
\end{widetext}
where as above $\Delta\kv=q\hat{x} + \Delta k_z \hat{z}$, with $\Delta k_z = k'_z-k_z$. The linear dependence of $\delta v$ on the distance from the neutral axis of the beam, which is a result of the Euler-Bernoulli thin-beam approximation, precludes the replacement of the finite limits of integration in the $z$ direction with integration over an infinite range, even though we consider unconfined electronic states, since it leads to an unphysical divergence of the decay rate. A more realistic model for the flexural modes may saturate this effect and lead to finite results even in the limit of large $b$.

The interaction Hamiltonian for flexural vibrations in a finite beam is obtained by replacing the quantized displacement field of the infinite beam~\eqref{eq:inf_flex_modes} with the one given by Eq.~\eqref{eq:finite_flex_modes}, yielding
\begin{widetext}
\begin{equation}\label{eq:e-ph flexural finite confined electrons}
\hep =\frac{8g(1-2\sigma)b}{\pi^{2}} \sqrt{\frac{\hbar}{2MG}} \sum_{\kv,k'_{x},k'_{z},q_{n}} q_{n} \frac{n_{z}n'_{z}}{\left(n^{2}_{z}-n'^{2}_{z}\right)^{2}} \LL_n\!\left( k_{x}-k'_{x}\right)c^{\dag}_{\kv+\Delta\kv}c_{\kv} \left(b_{q_{n}}+b^{\dag}_{q_{n}}\right),
\end{equation}
where here $\Delta\kv=\Delta k_x \hat{x} + \Delta k_z \hat{z}$, and
\begin{equation}\label{eq:Lkx_k'x_q}
\LL_n\left(\kappa\right)=\frac{1}{l}\int\limits^{\frac{l}{2}}_{-\frac{l}{2}}e^{i\kappa x}\left(A_{n}\sin{q_{n}x}+B_{n}\cos{q_{n}x}
-C_{n}\sinh{q_{n}x}-D_{n}\cosh{q_{n}x}\right)dx
\end{equation}
\end{widetext}
replaces the momentum conservation delta function that exists in the infinite beam~\eqref{eq:e-ph flexural infinite confined electrons}.

\section{Inverse quality factor}\label{sec:Decay rates of the longitudinal and flexural modes}

The damping of mechanical vibrations is estimated by assuming that both electrons and phonons are in thermal equilibrium except for a single vibration mode, which is externally excited by adding just a single phonon to its thermal population. The rate at which the excited mode decays is evaluated as the difference between the rates at which phonons leave it and enter it due to their scattering with electrons. These rates are calculated with Fermi's golden-rule using the interaction Hamiltonians $\hep$ derived above. The decay rate of an externally excited vibration mode of wavenumber $q$ is thus given by
\begin{multline}\label{eq:general fermi}
\Gamma_{q}=\frac{2\pi}{\hbar}\sum_{\kv,\kp,\sigma} \left|\bra{f_-} \hep \ket{i}\right|^{2}
\delta\left(\varepsilon_{\kv}-\varepsilon_{\kp}+\hbar\omega_{q}\right)\\
-\frac{2\pi}{\hbar}\sum_{\kv,\kp,\sigma} \left|\bra{f_+} \hep \ket{|i}\right|^{2}
\delta\left(\varepsilon_{\kv}-\varepsilon_{\kp}-\hbar\omega_{q}\right),
\end{multline}
where $\ket{f_-}$ is the state with one less phonon and $\ket{f_+}$ is the state with one more phonon, namely
\begin{subequations}\label{eq:kets}
\begin{eqnarray}
\ket{i} &=& \ket{n_{\kv},n_{\kp},N_{q}+1}\label{eq:i},\\
\ket{f_-} &=& \ket{n_{\kv}-1,n_{\kp}+1,N_{q}}\label{eq:f1},\\
\ket{f_+} &=& \ket{n_{\kv}-1,n_{\kp}+1,N_{q}+2}\label{eq:f2},
\end{eqnarray}
\end{subequations}
where, without any loss of generality, for infinitely long beams we consider modes with a positive wavenumber $q>0$. 
The dimensionless damping, or inverse quality factor, of a mode with wavenumber $q$ is then defined as $Q^{-1}_q=\Gamma_{q}/\omega_{q}$. In Eqs.~\eqref{eq:kets} $n_{\kv}$ stands for the Fermi-Dirac distribution and $N_{q}$ for the Bose-Einstein distribution. We note that we do not take into account the change in the electronic chemical potential as the temperature is increased above zero, but rather assume that it remains equal to the Fermi energy of the beam. The latter is calculated for each specific thickness and width of the beam when the confinement of the electrons is taken into account.

This procedure presupposes that the electron and phonon energies that appear in the delta functions in Eq.~\eqref{eq:general fermi} are exact, or equivalently that the \emph{a priori} lifetimes of electrons and phonons are infinite. For this to be valid we require (1)~that all other vibration damping mechanisms be much weaker than electron-phonon damping; and (2)~that scattering rates of electrons---with other electrons, thermal phonons, defects, or surface imperfections---be much slower than the frequency of the mechanical vibration. As we demonstrate in section~\ref{subsubsec:Broadened phonon states}, removal of the first requirement does not significantly affect the contribution of electron-phonon scattering to the overall damping, as long as the different damping mechanisms are assumed to be independent of each other. The second requirement implies that our calculation is valid in the \emph{adiabatic limit}, $\omega_q\tau_e > 1$, where $\tau_e$ is the mean lifetime of the electron. This requirement will be better satisfied for cleaner and smaller---hence, higher-frequency---devices at sufficiently low temperatures. 

In what follows we give the derivation of the exact expressions for the inverse quality factors, using Eq.~\eqref{eq:general fermi} along with the interaction Hamiltonians of section~\ref{sec:Interaction between the electrons and the phonons}. The reader who is not interested in these rather technical derivations is welcome to skip to the next section where we discuss the physical consequences of these expressions.

\subsection{Damping of longitudinal vibrations by unconfined electrons in an infinite beam}\label{subsec:Decay rates of the longitudinal phonons}

Using the Hamiltonian of Eq.\ \eqref{eq:e-ph longitudinal infinite free electrons} and the general expression~\eqref{eq:general fermi}  for the decay rate we obtain
\begin{align}\label{eq:decay_long_inf_free initial}
Q_{q}^{-1}=&\frac{\pi g^{2}}{Mc_{l}^{2}}\nonumber\\
\times&\sum_{\kv,\sigma}\left[n_{\kv}\left(1-n_{\kv+\qv}\right)
\left(N_{q}+1\right)\delta\left(\varepsilon_{\kv}-\varepsilon_{\kv+\qv}+\hbar\omega_{q}\right)\right.\nonumber\\
&-\left.n_{\kv}\left(1-n_{\kv-\qv}\right)
\left(N_{q}+2\right)\delta\left(\varepsilon_{\kv}-\varepsilon_{\kv-\qv}-\hbar\omega_{q}\right)\right],
\end{align}
where $\qv =q\hat{\mathbf{x}}$. We convert the sum over wave vectors in Eq.~\eqref{eq:decay_long_inf_free initial} into three integrals and change the energy conservation delta functions into momentum delta functions
\begin{equation}\label{eq:delta1 long inf free}
\delta\left(\varepsilon_{\kv}-\varepsilon_{\kv\pm\qv}\pm\hbar\omega_{q}\right)=\frac{m}{\hbar^{2}q}\delta
\left(k_{x}-\left(\frac{mc_{l}}{\hbar}\mp\frac{q}{2}\right)\right).
\end{equation}
After performing the integration over $k_{x}, k_{y}$ and $k_{z}$ and summing over the spin index we obtain
\begin{align}
Q_{q}^{-1}&=\frac{g^{2}m^{2}}{2\pi\hbar^{4}\rho c_{l}^{2}\beta q}\left[\left(N_{q}+1\right)^{2}
\left(\beta\hbar\omega_{q}-\ln{\frac{1+e^{\gamma_{-}}}{1+e^{\gamma_{+}}}}\right)\right. \nonumber\\
&-\left.\left(N_{q}+2\right)N_{q}\left(\beta\hbar\omega_{q}-\ln{\frac{1+e^{\gamma_{+}}}{1+e^{\gamma_{-}}}}\right)\right],
\label{eq:decay_long_inf_free almost final}
\end{align}
where
\begin{equation}
\gamma_{\pm}= \frac{\beta\hbar^{2}}{2m}\left[\left(\frac{c_{l}m}{\hbar}\pm\frac{q}{2}\right)^{2}-k_{\mathrm{F}}^2\right].
\end{equation}

Both logarithmic terms in Eq.~\eqref{eq:decay_long_inf_free almost final} are negligible as long as $c_{l}m/\hbar$ and $q/2$ are small compared to $k_{\mathrm{F}}$ (which is the case for all reasonable values of $q$). For such vibrational wavenumbers $\gamma_{+}\simeq\gamma_{-}$ and the logarithms vanish. Combining the two terms in the square brackets in Eq.~\eqref{eq:decay_long_inf_free almost final} then yields
\begin{equation}\label{eq:decay_long_inf_free final main}
Q^{-1}_{q}=\frac{g^{2}m^{2}}{2\pi\hbar^{3}\rho c_{l}},
\end{equation}
which is the same as the result for the damping of bulk longitudinal acoustic waves in the adiabatic limit (Ziman,\cite{ziman} Eq.~8.10.9, and Kokkedee\cite{kokkedee}). The damping in Eq.~\eqref{eq:decay_long_inf_free final main} is independent of the mode wavenumber, the temperature, and the geometry of the beam. In fact, $Q$ depends only on the material parameters of the beam, and typically varies between $10$ and $10^{3}$ for different metals.

\subsection{Damping of longitudinal vibrations by unconfined electrons in a finite beam}

We substitute the Hamiltonian of Eq.~\eqref{eq:e-ph longitudinal finite free electrons} into Eq.~\eqref{eq:general fermi} and again convert the sums over the electronic wave vectors into integrals. After summing over the spin index we obtain
\begin{eqnarray}
Q_{q_{n}}^{-1}&=&\frac{g^{2}n}{16\pi^{2}\rho c_{l}^{2}q_{n}}\int \mathbf{dk} dk'_{x}n_{\mathbf{k}}\left(1-n_{\kv+\delk}\right)
\WW_{n}\left(k_{x}-k'_{x}\right)\nonumber\\
& &\times\left[\left(N_{q_{n}}+1\right)\delta\left(\epsilon_{\kv}-\epsilon_{\kv+\delk}+\hbar\omega_{q_{n}}\right)
\right.\nonumber\\
& & \left.-\left(N_{q_{n}}+2\right)\delta\left(\epsilon_{\kv}-\epsilon_{\kv+\delk}-\hbar\omega_{q_{n}}\right)\right],
\label{eq:decay_long_finite_free electrons initial}
\end{eqnarray}
where $\delk=\left(k'_{x}-k_{x}\right)\hat{\mathbf{x}}$, and
\begin{multline}\label{eq:W clamped}
\WW_n\left(\kappa\right) = A_{n}\left(\mathrm{sinc}\frac{\left(\kappa-q_{n}\right) l}{2}
+\mathrm{sinc}\frac{\left(\kappa+q_{n}\right) l}{2}\right)^{2}\\
+B_{n}\left(\mathrm{sinc}\frac{\left(\kappa-q_{n}\right) l}{2}
-\mathrm{sinc}\frac{\left(\kappa+q_{n}\right) l}{2}\right)^{2}.
\end{multline}

We perform the integration over $k_{y}$ and $k_{z}$ and change the energy delta functions into a delta function of the variable $k'_{x}$
\begin{subequations}
\begin{multline}\label{eq:delta energy to kx'}
\delta\left(\epsilon_{\kv}-\epsilon_{\kv+\delk}+\hbar\omega_{q_{n}}\right)\\
=\frac{m}{\hbar^{2}k'^{0}_{x}}\left[\delta\left(k'_{x}-k'^{0}_{x}\right)
+\delta\left(k'_{x}+k'^{0}_{x}\right)\right],
\end{multline}
and a delta function of the variable $k_{x}$
\begin{multline}\label{eq:delta energy to kx}
\delta\left(\epsilon_{\kv}-\epsilon_{\kv+\delk}-\hbar\omega_{q_{n}}\right)\\
=\frac{m}{\hbar^{2}k^{0}_{x}}\left[\delta\left(k_{x}-k^{0}_{x}\right)
+\delta\left(k_{x}+k^{0}_{x}\right)\right],
\end{multline}
\end{subequations}
where $\left(k'^{0}_{x}\right)^{2} = k^{2}_{x} + 2mc_{l}q_{n}/\hbar$, and $\left(k^{0}_{x}\right)^{2} = k'^{2}_{x} + 2mc_{l}q_{n}/\hbar$.
We integrate over $k'_{x}$ for the first delta function and over $k_{x}$ for the second delta function and obtain
\begin{widetext}
\begin{eqnarray}
Q_{q_{n}}^{-1} &=&\frac{ng^{2}m^{2}}{8\pi\hbar^{4}\rho c_{l}^{2}\beta q_{n}}\nonumber\\
&\times&\left[\left(N_{q_{n}}+1\right)^{2}\int\limits^{\infty}_{-\infty} \frac{dk_{x}}{k'^{0}_{x}}
\left(\beta\hbar\omega_{q_{n}}+\ln{\frac{1+e^{\frac{\beta\hbar^{2}}{2m}\left(k^{2}_{x}-k^{2}_{\mathrm{F}}\right)}}
{1+e^{\frac{\beta\hbar^{2}}{2m}\left(k^{2}_{x}+\frac{2mc_{l}q_{n}}{\hbar}-k^{2}_{\mathrm{F}}\right)}}}\right)
\left[\WW_{n}\left(k_{x}-k'^{0}_{x}\right) + \WW_{n}\left(k_{x}+k'^{0}_{x}\right)\right]
\right.\nonumber\\
&-&\left.  N_{q_{n}}\left(N_{q_{n}}+2\right) \int\limits^{\infty}_{-\infty} \frac{dk_{x}}{k'^{0}_{x}}
\left(\beta\hbar\omega_{q_{n}}+\ln{\frac{1+e^{\frac{\beta\hbar^{2}}{2m}\left(k^{2}_{x}+\frac{2mc_{l}q_{n}}{\hbar}-k^{2}_{\mathrm{F}}\right)}}
{1+e^{\frac{\beta\hbar^{2}}{2m}\left(k^{2}_{x}-k^{2}_{\mathrm{F}}\right)}}}\right)
\left[\WW_{n}\left(k'^{0}_{x}-k_{x}\right) + \WW_{n}\left(-k'^{0}_{x}-k_{x}\right)\right]\right]
,\label{eq:decay_long_finite_free electrons 2}
\end{eqnarray}
\end{widetext}
where we have dropped the prime from the integration variable.

The integrands in Eq.~\eqref{eq:decay_long_finite_free electrons 2} nearly vanish for most values of $k_{x}$ unless it is relatively close to  $\pm\left(mc_{l}/\hbar-q_{n}/2\right)$. For these values of $k_{x}$ (and as long as $q_{n}$ is small compared to $k_{\mathrm{F}}$) both $k^{2}_{x}-k^{2}_{\mathrm{F}}$ and $k^{2}_{x} + 2mc_{l}q_{n}/\hbar - k^{2}_{\mathrm{F}}$ are nearly equal to $-k^{2}_{\mathrm{F}}$ and the logarithmic functions in Eq.~\eqref{eq:decay_long_finite_free electrons 2} can be neglected. Finally, by using the fact that $\WW_{n}(\kappa)=\WW_{n}(-\kappa)$ we find that
\begin{multline}\label{eq:decay_long_finite_free electrons final main}
Q^{-1}_{q_{n}} = \frac{n g^{2} m^{2}}{8\pi \hbar^{3}\rho c_{l}} \int\limits^{\infty}_{-\infty}\frac{dk_{x}}
{k'^{0}_{x}} \left[\WW_n\left(k_{x}-k'^{0}_{x}\right)\right.\\ + \left.\WW_n\left(k_{x}+k'^{0}_{x}\right)\right].
\end{multline}
We note that the damping for the infinite beam in Eq.~\eqref{eq:decay_long_inf_free final main} and for the finite beam in Eq.~\eqref{eq:decay_long_finite_free electrons final main}, due to interaction with unconfined electrons, are both independent of temperature and of the length of the beam for a given mode.

\subsection{Damping of longitudinal vibrations by laterally confined electrons in an infinite beam}

As noted in section~\ref{sec:Interaction between the electrons and the phonons}, the Hamiltonian for the interaction between longitudinal phonons and laterally confined electrons is formally identical to the one describing the interaction between longitudinal phonons and unconfined free electrons, with the integrals over $k_{y}$ and $k_{z}$ in Eq.~\eqref{eq:e-ph longitudinal infinite free electrons} replaced by discrete sums over the allowed values of $k_{y}$ and $k_{z}$. These discrete sums yield
\begin{align}
Q^{-1}_{q} = &\frac{g^{2}m}{\hbar^{2} E A q}\sum_{k_{y},k_{z}}\nonumber\\
&\left[n_{\left(k_x^-,k_{y},k_{z}\right)}
\left(1-n_{\left(k_x^+,k_{y},k_{z}\right)}\right) \left(N_{q}+1\right)\right.\nonumber\\
&\left.-n_{\left(k_x^+,k_{y},k_{z}\right)} \left(1-n_{\left(k_x^-,k_{y},k_{z}\right)}\right) \left(N_{q}+2\right)\right],
\label{eq:decay_long_infinite_confined electrons final}
\end{align}
where
\begin{equation}\label{eq:kx-pm}
k_x^{\pm} = \frac{mc_{l}}{\hbar} \pm \frac{q}{2}.
\end{equation}

\subsection{Damping of flexural vibrations  by laterally confined electrons in an infinite beam}\label{subsec:Decay rates of the flexural phonons}

The Hamiltonian of Eq.\ \eqref{eq:e-ph flexural infinite confined electrons} which describes the interaction between flexural modes in an infinite beam with confined electrons together with the general expression for the decay rate \eqref{eq:general fermi} yield
\begin{align}
Q_{q}^{-1}=&\frac{64\pi g^{2}(2\sigma-1)^{2}}{G^{2}b^{2}M}\sum_{\kv,k'_{z},\sigma}\frac{k^{2}_{z}k'^{2}_{z}}{\left(k^{2}_{z}-k'^{2}_{z}\right)^{4}}
\nonumber\\
& \times\left[n_{\kv}\left(1-n_{\kv+\qv+\delk}\right)\left(N_{q}+1\right)\right.\nonumber\\
& \times\left.\delta\left(\varepsilon_{\kv}-\varepsilon_{\kv+\qv+\delk}+\hbar\omega_{q}\right)\right.\nonumber\\
&-\left.n_{\kv}\left(1-n_{\kv-\qv+\delk}\right)\left(N_{q}+2\right)\right.\nonumber\\
&\times\left.\delta\left(\varepsilon_{\kv}-\varepsilon_{\kv-\qv+\delk}-\hbar\omega_{q}\right)\right],\label{eq:decay_flex_infinite_confined electrons initial}
\end{align}
where here $\delk=\left(k'_{z}-k_{z}\right)\hat{\mathbf{z}}$.

We sum over the spin index and change the delta functions in Eq.~\eqref{eq:decay_flex_infinite_confined electrons initial} into delta functions of the variable $k_{x}$
\begin{equation}
\delta\left(\epsilon_{\kv}-\epsilon_{\kv+\delk\pm\qv}\pm\hbar\omega_{q}\right)=\frac{m}{\hbar^{2}q}\delta\left(k_{x}-\kappa_{\pm}\right),
\label{eq:delta energy to kx+-}
\end{equation}
where
\begin{equation}\label{eq:kappa-pm}
\kappa_{\pm} = \frac{mGq}{\hbar} \pm \frac{k^{2}_{z}-k'^{2}_{z}}{2q} \mp \frac{q}{2}.
\end{equation}
We express the sum over $k_{x}$ as an integral, and use Eqs.~\eqref{eq:delta energy to kx+-} and \eqref{eq:kappa-pm} to obtain
\begin{eqnarray}
Q^{-1}_{q}&=&\frac{768 g^{2} m (2\sigma-1)^{2}}{\pi^{4} \hbar^{2} E A q}
\sum_{k_{y},k_{z},k'_{z}}\frac{n^{2}_{z}n'^{2}_{z}}{\left(n^{2}_{z}-n'^{2}_{z}\right)^{4}}\nonumber\\
&\times&\left[n_{\left(\kappa_+,k_{y},k_{z}\right)}
\left(1-n_{\left(\kappa_+ +q,k_{y},k'_{z}\right)}\right) \left(N_{q}+1\right)\right.\nonumber\\
&-&\left.n_{\left(\kappa_-,k_{y},k_{z}\right)}
\left(1-n_{\left(\kappa_- -q,k_{y},k'_{z}\right)}\right)\left(N_{q}+2\right)\right].
\label{eq:decay_flex_infinite_confined electrons final}
\end{eqnarray}

\subsection{Damping of flexural vibrations  by laterally confined electrons in a finite beam}

The inverse quality factors of flexural modes in a finite beam are obtained in a similar way but without exact momentum conservation. Thus, the momentum delta functions are replaced with the $\LL_{n}$ functions, given by Eq.~\eqref{eq:Lkx_k'x_q}, and the integral over $k_{x}$ remains in the final expression
\begin{widetext}
\begin{eqnarray}
Q^{-1}_{q_{n}}
&=&\frac{384(2\sigma-1)^{2}g^{2} m l^2}{\pi^{5} \hbar^{2} E A}
\sum_{k_{y},k_{z},k'^{2}_{z}<k^{2}_{z}+\frac{2m\omega_{q_{n}}}{\hbar}}
\frac{n^{2}_{z}n'^{2}_{z}}{\left(n^{2}_{z}-n'^{2}_{z}\right)^{4}} \int\limits_{-\infty}^{\infty} \frac{dk_{x}}{\sqrt{k^{2}_{x}+k^{2}_{z}-k'^{2}_{z}+\frac{2m\omega_{q_{n}}}{\hbar}}}\nonumber\\
&\times&\left(\left|\LL_n\left(k_{x}-\sqrt{k^{2}_{x}+k^{2}_{z}-k'^{2}_{z}+\frac{2m\omega_{q_{n}}}{\hbar}}\right)\right|^{2}
+\left|\LL_n\left(k_{x}+\sqrt{k^{2}_{x}+k^{2}_{z}-k'^{2}_{z}+\frac{2m\omega_{q_{n}}}{\hbar}}\right)\right|^{2}\right)\nonumber\\
&\times& \left[\left(N_{q_{n}}+1\right)n\left(\frac{\hbar^{2}k^2}{2m}\right)
\left(1-n\left(\frac{\hbar^{2}k^2}{2m} + \hbar\omega_{q_{n}}\right)\right)
-\left(N_{q_{n}}+2\right)n\left(\frac{\hbar^{2}k^2}{2m} + \hbar\omega_{q_{n}}\right)
\left(1-n\left(\frac{\hbar^{2}k^2}{2m}\right)\right)\right]\nonumber\\
&+&\sum_{k_{y},k_{z},k'^{2}_{z}<k^{2}_{z}-\frac{2m\omega_{q_{n}}}{\hbar}}
\frac{n^{2}_{z}n'^{2}_{z}}{\left(n^{2}_{z}-n'^{2}_{z}\right)^{4}}\int\limits_{-\infty}^{\infty} \frac{dk_{x}}{\sqrt{k^{2}_{x}+k_{z}^{2}-k'^{2}_{z}-\frac{2m\omega_{q_{n}}}{\hbar}}}\nonumber\\
&\times&\left(\left|\LL_n\left(-k_{x} + \sqrt{k^{2}_{x}+k_{z}^{2}-k'^{2}_{z}-\frac{2m\omega_{q_{n}}}{\hbar}}\right)\right|^{2}
+\left|\LL_n\left(-k_{x}-\sqrt{k^{2}_{x}+k_{z}^{2}-k'^{2}_{z}+\frac{2m\omega_{q_{n}}}{\hbar}}\right)\right|^{2}\right)\nonumber\\
&\times&\left[\left(N_{q_{n}}+1\right)n\left(\frac{\hbar^{2}k^2}{2m} - \hbar\omega_{q_{n}}\right)
\left(1-n\left(\frac{\hbar^{2}k^2}{2m}\right)\right)
-\left(N_{q_{n}}+2\right)n\left(\frac{\hbar^{2}k^2}{2m}\right)
\left(1-n\left(\frac{\hbar^{2}k^2}{2m} - \hbar\omega_{q_{n}}\right)\right)\right].\label{eq:decay_flex_finite_confined electrons final}
\end{eqnarray}
\end{widetext}

\section{Results and Discussion}\label{sec:Results}

We use the material properties of aluminum in our numerical calculations below, because, according to our model, it is expected to have the highest dissipation among the typical materials used in the fabrication of metallic nanomechanical beams. We use the room temperature values of these properties, listed in Table~\ref{properties}, as they do not change much between room-temperature and cryogenic temperatures.\cite{simmons} We consider beams with cross-sectional dimensions that vary between 10nm and 50nm, and finite lengths that vary between 500nm and $2.0\mathrm{\mu m}$. Because our model involves many  approximations, the calculated values of $Q^{-1}$ given below should be considered only as a rough order of magnitude estimate of electron-phonon damping. Nevertheless, we believe that we describe correctly the qualitative behavior of the damping as a function of the different physical and geometrical parameters.

\renewcommand{\arraystretch}{1.2}
\begin{table}
 \begin{center}
 \begin{tabular}{|l|l|}
 \hline\hline
 $E$ \cite{simmons}   & 70 GPa\\
 $\sigma$ \cite{simmons}   & 0.35\\
 $\rho$ \cite{simmons} \hspace{1mm}   & 2.7 gr/cm$^3$\\
 $Z$ & 3\\
 $\ltf$ \cite{ashcroft}   & 0.049 nm\\ 
\hline\hline
 \end{tabular}
 \end{center}
\caption{\label{properties}
 Material properties of bulk aluminum at room temperature, used for calculating the results in section \ref{sec:Results}.}
\end{table}
\renewcommand{\arraystretch}{1.0}

\subsection{Damping of longitudinal vibrations as a function of temperature and wavenumber}\label{subsec:Longitudinal modes}

\begin{figure}
 \includegraphics[width=1.0\columnwidth]{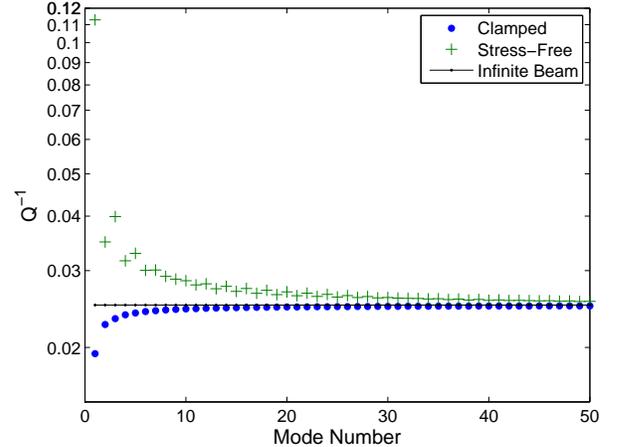}
 \caption{\label{qf finite beam long}
  (color online) Inverse quality factors, on a logarithmic scale, of the first 50 longitudinal modes interacting with unconfined electrons, in a finite beam with stress-free and clamped boundary conditions, compared with $Q^{-1}$ of an infinite beam.}
\end{figure}

The inverse quality factors of the first 50 longitudinal modes of a $1\mathrm{\mu m}$-long beam---calculated using Eq.~\eqref{eq:decay_long_finite_free electrons final main}, which assumes that the electrons are unconfined by the lateral dimensions of the beam---are shown in Fig.~\ref{qf finite beam long}, for both clamped and stress-free boundary conditions. As one should expect, for sufficiently short wavelengths the damping is not affected by the finite length of the beam. Accordingly, both sets of results converge to the value of $Q^{-1}$, calculated for an infinite beam using Eq.~\eqref{eq:decay_long_inf_free final main}, which gives an estimated $Q$ of about 40 for short-wavelength (large $q$) modes in aluminum beams. For long-wavelength (small $q$) modes the finite length of the beam has a significant effect on the damping. The inverse quality factor of the lowest \emph{clamped} mode is about 25\% smaller than $Q^{-1}$ of an infinite beam, and that of the lowest \emph{stress-free} mode is almost 4 times larger than $Q^{-1}$ of an infinite beam, with a difference larger than 10\% persisting up to about the tenth stress-free mode. 

\begin{figure}
 \includegraphics[width=1.0\columnwidth]{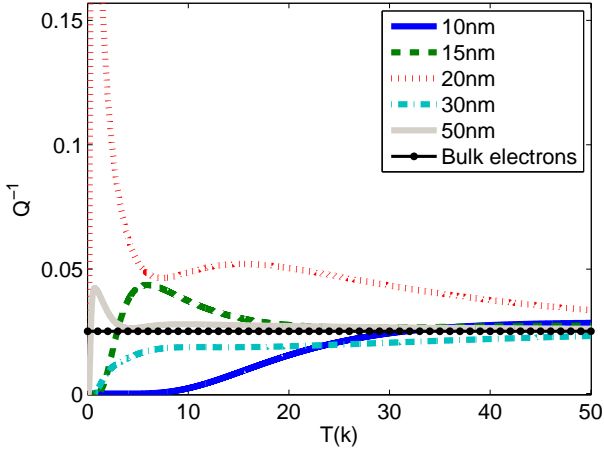}
 \caption{\label{finite_width_infinite_length_beam_longitudinal_phonons_zoomin}
  (color online) Inverse quality factor of the longitudinal mode $q=\pi/10^{3}\mathrm{nm^{-1}}$ as a function of temperature in infinitely long beams with square cross sections, where the lateral confinement of the electrons is taken into account. The value of $Q^{-1}$ for unconfined electrons is shown for comparison.}
\end{figure}

The lack of an intrinsic energy scale in the spectrum of the unconfined electrons---other than the Fermi energy which is too high to be relevant---results in electron-phonon damping that is temperature independent for both infinite and finite beams, as is apparent from Eqs.~\eqref{eq:decay_long_inf_free final main} and \eqref{eq:decay_long_finite_free electrons final main}. This changes when the lateral confinement of the electrons is no longer ignored, introducing a mesoscopic energy scale $\Delta E_{\rm m}=k_{\rm B}\Tm$.
This energy scale---in the simple case of longitudinal vibrations---is determined by the transverse dimensions $a$ and $b$ of the beam through the typical energy difference between successive electronic band minima, given by Eq.~\eqref{eq:electronic bands}. It is on the order of $\hbar^2\pi^2/2ma^2$ to $\hbar^2\pi^2/2mb^2$, to within a multiplicative numerical factor arising from the values of $n_y$ and $n_z$ near the Fermi energy. Indeed, Fig.~\ref{finite_width_infinite_length_beam_longitudinal_phonons_zoomin} shows that $Q^{-1}$, when calculated using Eq.~\eqref{eq:decay_long_infinite_confined electrons final} which takes into account the lateral confinement of the electrons, is a function of temperature up to some $\Tm\propto 1/a^2$ for beams with square cross sections. For example, in the case of aluminum, the crossover occurs at $\Tm\simeq 250\mathrm{k}$ for $a=b=10\mathrm{nm}$ (not shown in the Figure), and at $\Tm\simeq 10$k for $a=b=50\mathrm{nm}$. 

\begin{figure}
 \includegraphics[width=1.0\columnwidth]{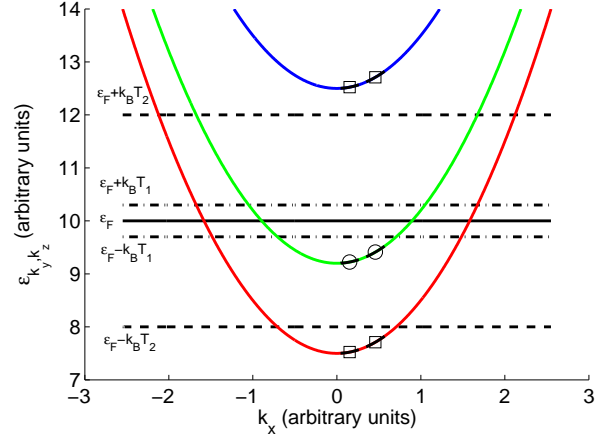}
 \caption{\label{schematic}
  (color online) Schematic plot of the electronic band structure near the Fermi energy. The scattering of a longitudinal  phonon with a positive wavenumber $q$ is considered. At a very low temperature ($T_{1}$) the damping is very weak because there is no dissipation channel within the narrow temperature window around $\varepsilon_{\mathrm{F}}$. At a somewhat higher temperature ($T_{2}$) the temperature window is wider and the dissipation channel with $k_{x}^{\pm}$, marked by open circles, becomes active.  The two nearby dissipation channels (marked by open squares) remain inactive as they are outside the temperature window. The black intervals around the markers represent schematically the electronic states that can scatter a phonon when the length of the beam is finite and momentum conservation is imperfect.}
\end{figure}

Thus, there are two qualitatively distinct regimes in the behavior of electron-phonon damping in our model of metallic nanomechanical resonators. For $T$ sufficiently greater than $\Tm$, which is obtained either by increasing the temperature or by increasing the transverse dimensions of the beam, the resonator operates in a \emph{macroscopic} or \emph{bulk-like regime}, where damping approaches its temperature-independent value, as calculated for unconfined electrons. In the macroscopic regime electrons behave as they do in 3-dimensional bulk systems and one can safely ignore their confinement by the finite geometry of the resonator. For $T$ sufficiently less than $\Tm$, which is obtained either by lowering the temperature or by decreasing the transverse dimensions of the beam, the resonator is in a \emph{mesoscopic regime}, where damping, as we explain below, depends non-monotonically on temperature and other parameters. In this regime the electrons behave as a collection of effective 1-dimensional particles, each characterized by its own parabolic band, given by Eq.~\eqref{eq:electronic bands}. 

It follows from Eq.~\eqref{eq:decay_long_infinite_confined electrons final}, due to the lack of any displacement in the $y$ and $z$ directions, that a longitudinal phonon can be created or annihilated only via intra-band electronic transitions. Moreover, due to momentum conservation in the $x$ direction along with energy conservation, a phonon of wavenumber $q$ can interact with only two specific electronic states in any given band, whose momentum in the $x$-direction is given by $k_{x}^{\pm}$, as defined in Eq.~\eqref{eq:kx-pm}. An electron in the state with $k_{x}^{-}$ can absorb a phonon and scatter into the state with $k_{x}^{+}$, whereas an electron with $k_{x}^{+}$ can emit a phonon and scatter into the state with $k_{x}^{-}$. The difference between these two processes, for each such pair of electronic states, constitutes a \emph{dissipation channel}, which may contribute to electron-phonon damping. It then follows from the form of the Fermi-Dirac factors $n(1-n)$, whose width is of the order of $\kB T$, that if the two states happen to lie on either side of the Fermi energy, the corresponding dissipation channel is active at any temperature. More generically, the channel is active only if the two states are located within a window of order $k_{\mathrm{B}}T$ near the Fermi energy. This is demonstrated schematically in Fig.~\ref{schematic}.

\begin{figure}
 \includegraphics[width=1.0\columnwidth]{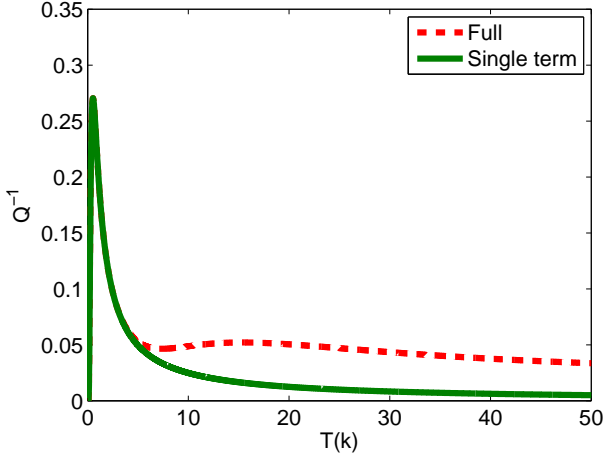}
 \caption{\label{finite_width_infinite_length_beam_longitudinal_phonons_with_single_level}
  (color online) Inverse quality factor of the longitudinal mode $q=\pi/10^{3}\mathrm{nm^{-1}}$ of an infinitely long beam with $a=b=20\mathrm{nm}$ plotted as a function of temperature, showing that up to $T\simeq5$k electron-phonon damping is dominated by a single dissipation channel as explained in the text.}
\end{figure}

In generic situations, at a sufficiently low temperature all the dissipation channels are inactive and electron-phonon damping vanishes, as can be seen in all the curves in Fig.~\ref{finite_width_infinite_length_beam_longitudinal_phonons_zoomin}. As the temperature increases, at some point the dissipation channel that is closest to the Fermi energy becomes active, and electron-phonon damping is dominated by this single dissipation channel. To confirm this, we compare in Fig.~\ref{finite_width_infinite_length_beam_longitudinal_phonons_with_single_level} the total value of $Q^{-1}$, as obtained from the full sum in Eq.~\eqref{eq:decay_long_infinite_confined electrons final}, to the contribution of just a single term in the sum, corresponding to the dissipation channel that is closest to the Fermi energy. The contribution of a given dissipation channel to the overall damping is maximal at a temperature that is roughly given by $k_{\mathrm{B}}T \simeq \left|\frac{\hbar^{2}}{2m}\left(\left(k_{x}^{\pm}\right)^{2} + k^{2}_{y} + k^{2}_{z}\right) - \varepsilon_{\mathrm{F}}\right|$. These maxima are well pronounced unless a second dissipation channel happens to be relatively close to the Fermi energy (as is the case for $a = 30\mathrm{nm}$ in Fig.~\ref{finite_width_infinite_length_beam_longitudinal_phonons_zoomin}). As the temperature is further increased the number of active dissipation channels gradually increases until eventually the behavior crosses over to the bulk regime.

\subsection{Damping of flexural vibrations in an infinite beam as a function of temperature and wavenumber}\label{subsubsec:Infinite Beam-T}

\begin{figure}
 \includegraphics[width=1.0\columnwidth]{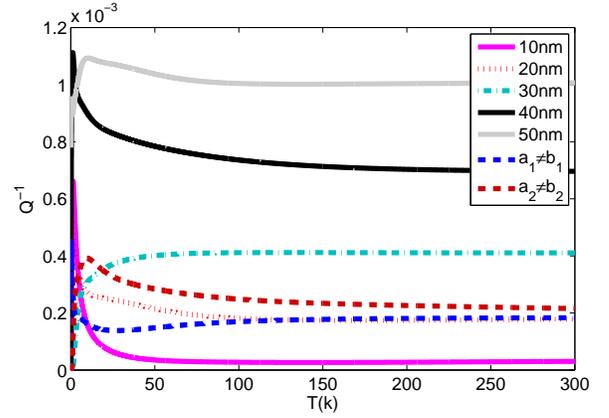}
\caption{\label{flex_ph_dis_ele_Q_q_as_fun_of_T_a=b=10_50}
  (color online) Inverse quality factor of the flexural mode $q=8\pi/(0.5\times10^{3})\mathrm{nm^{-1}}$ in an infinitely long beam with square cross sections $a=b=10,20,30,40,50\mathrm{nm}$, and rectangular cross sections $a_1 = 40\mathrm{nm}, b_1 = 20\mathrm{nm}$ and $a_2 = 10\mathrm{nm}, b_2 = 20\mathrm{nm}$, as a function of temperature. Note the independence of the asymptotic value on $a$.}
\end{figure}

\begin{figure}
 \includegraphics[width=1.0\columnwidth]{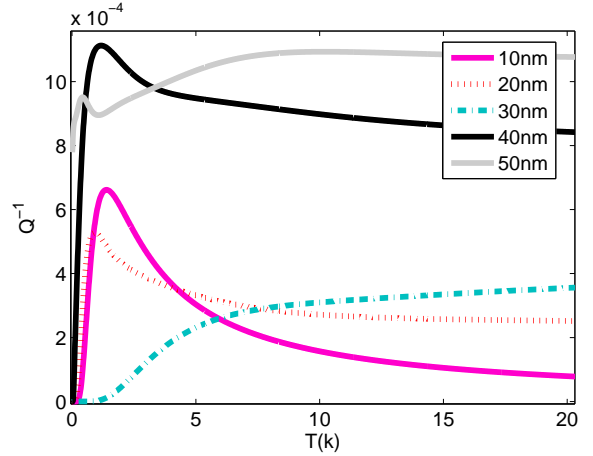}
\caption{\label{flex_ph_dis_ele_Q_q_as_fun_of_T_a=b=10_50_zoomin}
  (color online) Zoom-in on the low temperature behavior of $Q^{-1}$ of the flexural mode given in Fig.\ \ref{flex_ph_dis_ele_Q_q_as_fun_of_T_a=b=10_50}.}
\end{figure}

The inverse quality factors for flexural modes of a given wavenumber $q$ in infinitely long beams---as calculated using Eq.~\eqref{eq:decay_flex_infinite_confined electrons final}, taking into account the lateral confinement of the electrons---are plotted in Fig.~\ref{flex_ph_dis_ele_Q_q_as_fun_of_T_a=b=10_50} as a function of temperature. We again identify two qualitatively distinct regimes of behavior---a low-temperature \emph{mesoscopic regime}, where electron-phonon damping is dominated by just a single or a few dissipation channels, and a high-temperature \emph{macroscopic regime}, where electron-phonon damping approaches its temperature-independent value as expected for unconfined bulk electrons. We show below that the high-temperature limit of the expression for $Q^{-1}$ in Eq.~\eqref{eq:decay_flex_infinite_confined electrons final} is indeed temperature-independent  and independent of $a$, and that for long-wavelength modes, with $qb\ll1$, it behaves approximately as $(qb)^2$. This is indeed the behavior observed in Fig.~\ref{flex_ph_dis_ele_Q_q_as_fun_of_T_a=b=10_50}. The mesoscopic regime, which is shown in greater detail in Fig.~\ref{flex_ph_dis_ele_Q_q_as_fun_of_T_a=b=10_50_zoomin}, is again characterized by electron-phonon damping that tends to zero at very low temperatures, followed by peaks at slightly higher temperatures that may result in damping that is several times larger than the high-temperature constant asymptotic value. 

For flexural modes, it follows from Eq.~\eqref{eq:decay_flex_infinite_confined electrons final} that the emission and absorption of phonons involve inter-band electronic transitions, where the value of $n_{z}$ increases or decreases by an odd integer. Momentum and energy conservation again restrict the number of electronic states that can interact with a given phonon of wavenumber $q$, leaving two possible dissipation channels for each pair of bands $[n_{y}, n_{z}]$ and $[n_{y}, n'_{z}]$, with $n_z+n'_z$ an odd integer. The contribution of these dissipation channels to $Q^{-1}$ decreases rapidly as the difference between $n_{z}$ and $n'_{z}$ increases, due to the prefactor $n^{2}_{z}n'^{2}_{z}/\left(n^{2}_{z}-n'^{2}_{z}\right)^{4}$ in Eq.~\eqref{eq:decay_flex_infinite_confined electrons final}. Thus, the major contribution to $Q^{-1}$ comes from \emph{dominant channels} with $n'_{z}=n_{z}\pm 1$, while the remaining channels can usually be neglected.
 
As before, a dissipation channel is active when its two electronic states are within a temperature window of the order of $k_{\mathrm{B}}T$ around the Fermi energy, yet for flexural modes the situation is slightly more involved. To see this, let us examine the dissipation channels in more detail. It follows from Eq.~\eqref{eq:decay_flex_infinite_confined electrons final} that an electron in the band labeled $[n_{y}, n_{z}]$ with momentum in the $x$ direction given by $k_1 = \kappa_+$ [as defined in Eq.~\eqref{eq:kappa-pm}] and a corresponding energy $\varepsilon_{1}$ [according to Eq.~\eqref{eq:electron energy}], can absorb a phonon with a wavenumber $q$ and scatter into a state in the band $[n_{y}, n'_{z}]$ with momentum in the $x$ direction $k_2 = \kappa_+ + q$ and energy $\varepsilon_{2}=\varepsilon_{1}+\hbar\omega_{q}$. The reverse is possible via phonon emission.  Alternatively, an electron in the band labeled $[n_{y}, n_{z}]$ with momentum in the $x$ direction given by $k_3 = \kappa_-$ and energy denoted by $\varepsilon_{3}$, can emit a phonon with a wavenumber $q$ and scatter into a state in the band $[n_{y}, n'_{z}]$ with momentum in the $x$ direction $k_4 = \kappa_- - q$ and energy $\varepsilon_{4}=\varepsilon_{3} - \hbar\omega_{q}$. The reverse occurs via phonon absorption.

\begin{figure}
 \includegraphics[width=1.0\columnwidth]{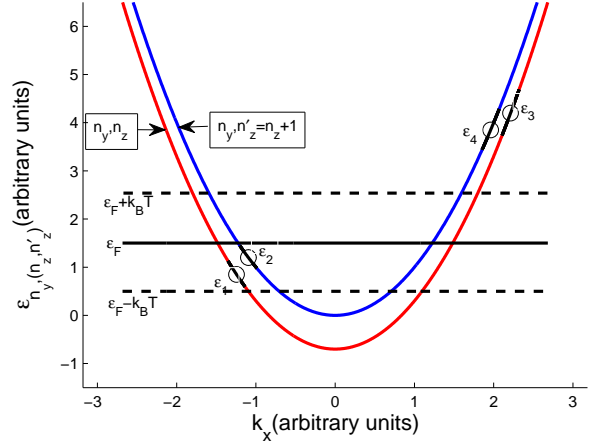}
 \caption{\label{schematic_ef}
 (color online) A schematic plot of two electronic energy bands characterized by the same $n_{y}$ and with $n'_{z}=n_{z}+1$ showing the pair of dissipation channels associated with a flexural mode with wavenumber $q>0$ as explained in the text. At the temperature $T$, indicated by a dashed line, only one of the two channels is active; at a higher temperature both channels would be active and one would lose the ability to resolve them, as demonstrated below in Fig.~\ref{flex_inf_dis_ele_a=b=10_T=10_qPiDIV0_5x103_ EF10_25_zoomin}. The black intervals around the four markers represent schematically the electronic states that can scatter a phonon in finite-length beams where there is only approximate conservation of momentum.}
\end{figure}

Such a pair of dissipation channels, associated with a given pair of bands, are depicted schematically in Fig.~\ref{schematic_ef} for dominant channels with $n'_z=n_z+1$. The energy scale $\Delta\varepsilon(n_y,n_z)$---associated with the difference in energy between the two dissipation channels---is given by
\begin{equation}\label{eq:delta-e}
\Delta\varepsilon(n_y,n_z) = \frac{\hbar G \pi^2}{b^2} \left(2n_z+1\right) \propto \frac{1}{b};
\end{equation}
whereas the average energy $\varepsilon(n_y,n_z)$ associated with both channels, in the limit of long-wavelength vibrations, is given by
\begin{equation}\label{eq:e-approx}
\varepsilon(n_y,n_z) = \lim_{qb\to 0}\varepsilon_i = \frac{\hbar^2 \pi^2}{2m} \left[\frac{n_y^2}{a^2} + \left(\frac{\pi}{qb^2}\right)^2 \left(n_z + \frac{1}{2}\right)^2 \right],
\end{equation}
where $\varepsilon_i$ is any one of the four energies $\varepsilon_1\ldots\varepsilon_4$ mentioned earlier and shown schematically in Fig.~\ref{schematic_ef}.

At high temperatures, the leading term in an expansion of Eq.~\eqref{eq:decay_flex_infinite_confined electrons final}  in powers of $\beta\hbar\omega_q\ll1$, shows that each active pair of dominant dissipation channels with $n'_{z}=n_{z}\pm1$ is responsible for a contribution to the damping that is proportional to $\beta\hbar\omega_q/abq$, which in turn is proportional to $\beta q/a$. To estimate the total damping, one simply needs to count the number of integer pairs $\left(n_{y},n_{z}\right)$ corresponding to dissipation channels whose electronic energies $\varepsilon(n_y,n_z)$ are a distance of order $\kB T$ within $\varepsilon_{\rm F}$. The density of such states, obtained directly from Eq.~\eqref{eq:e-approx}, yields a number that is proportional to $qab^2\kB T$, and therefore to a total damping proportional to $(qb)^2$ and independent of $a$ and $T$, as mentioned earlier.

\begin{figure}
 \includegraphics[width=1.0\columnwidth]{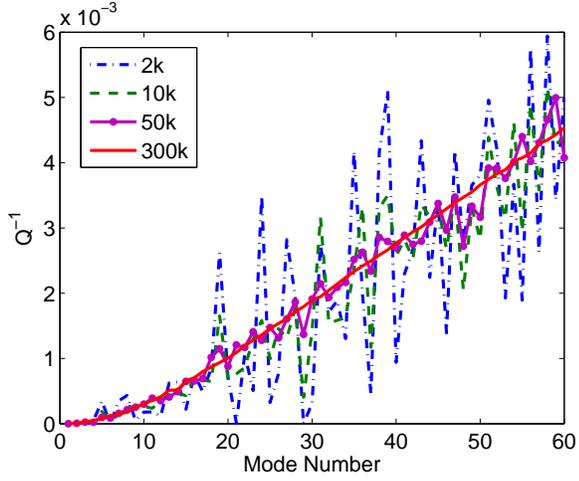}
 \caption{\label{a=b=20_T=01_1_10_100_300_500_n=1_60}
  (color online) Inverse quality factor of flexural modes with $q=\pi/(0.5\times10^{3})\mathrm{nm^{-1}}$ to $60\times\pi/(0.5\times10^{3})nm^{-1}$ (which approximately correspond to the first 60 flexural modes of a $l=0.5\mathrm{\mu m}$ beam) in an infinitely long beam with $a=b=20\mathrm{nm}$ for different temperatures.}
\end{figure}

In Fig.~\ref{a=b=20_T=01_1_10_100_300_500_n=1_60} we plot $Q^{-1}$ as a function of phonon wave\-number $q$ for several temperatures. At low temperatures, in the mesoscopic regime, $Q^{-1}$ fluctuates greatly as $q$ is increased, although the general trend is an increase in the damping. However, as the temperature crosses over to the macroscopic regime the fluctuations are suppressed and the damping converges to a single monotonically rising curve. The large fluctuations in $Q^{-1}$ in the mesoscopic regime again stem from the small number of active dissipation channels, whose positions depend very sensitively on the value of $q$. Thus, for narrow temperature windows around the Fermi energy, small changes in $q$ can easily cause an active dissipation channel to become inactive or \emph{vice versa}. On the other hand, in the macroscopic regime the number of active channels is large, and small changes in $q$ affect the damping only monotonically. Fig.~\ref{a=b=20_T=01_1_10_100_300_500_n=1_60} confirms that for small $q$ in the macroscopic regime $Q^{-1}$ is indeed proportional to $q^{2}$.

\subsection{Damping of flexural vibrations in an infinite beam as a function of $\varepsilon_{\mathrm{F}}$}\label{subsec:Flexural modes}

As an example of the delicate sensitivity of electron-phonon damping to changing system parameters in the mesoscopic regime, we consider the dependence on the Fermi energy. The latter may be varied in real experiments by means of an external gate voltage, suggesting an intriguing way to probe the electronic spectrum by measuring the $Q$ of mechanical vibrations. As for longitudinal modes, the existence of individual dissipation channels can be discerned as long as $\kB T$ is smaller than the energy scales characterizing the difference between channels. For flexural modes there are two such energy scales. The first is the energy difference~\eqref{eq:delta-e} between two dissipation channels belonging to the same pair of energy bands; and the second and larger energy scale is the difference between pairs of dissipation channels belonging to different pairs of electronic energy bands~\eqref{eq:e-approx}. 

\begin{figure}
 \includegraphics[width=1.0\columnwidth]{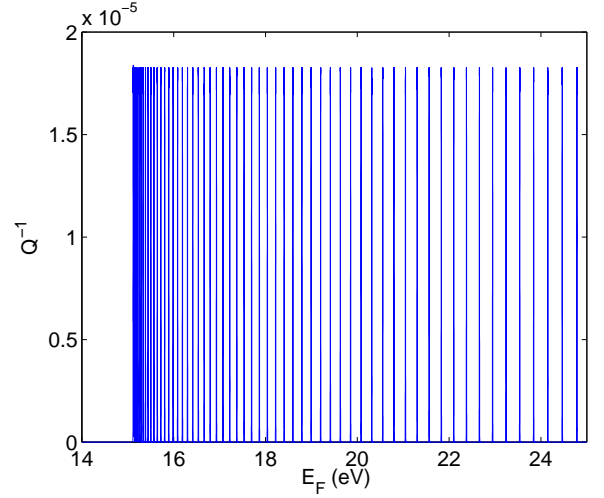}
 \caption{\label{flex_inf_dis_ele_a=b=10_T=10_qPiDIV0_5x103_ EF10_25}
  (color online) Inverse quality factor of the flexural mode $q=2\pi/10^{3}\mathrm{nm^{-1}}$ in an infinitely long beam with $a=b=10\mathrm{nm}$ at $T=10\mathrm{k}$ as a function of the Fermi energy.}
\end{figure}

Low-temperature results for $T=10$k are shown in Fig.~\ref{flex_inf_dis_ele_a=b=10_T=10_qPiDIV0_5x103_ EF10_25} for flexural vibrations in an infinite beam with a square cross-section. The damping is zero up to a threshold value of $\varepsilon_{\mathrm{F}}$, which according to Eq.~\eqref{eq:e-approx} is approximately equal to $\varepsilon(1,1) \simeq 9\pi^4\hbar^2/8mq^2b^4$. As the Fermi energy is raised above this threshold, the damping exhibits a series of narrow peaks, while remaining zero for most values of $\varepsilon_{\mathrm{F}}$. For the particular beam geometry and energy range considered in Fig.~\ref{flex_inf_dis_ele_a=b=10_T=10_qPiDIV0_5x103_ EF10_25}, the observed damping peaks are associated with pairs of dissipation channels belonging to bands $[n_{y},n_{z}=1]$ and $[n_{y},n'_{z}=2]$, with $n_{y}$ increasing by 1 from one peak to the next. The increase of $n_{y}$ by 1 between consecutive peaks results in a linear increase of the energy separation between the peaks as the Fermi energy increases, in accordance with Eq.~\eqref{eq:e-approx}.  

\begin{figure}
 \includegraphics[width=1.0\columnwidth]{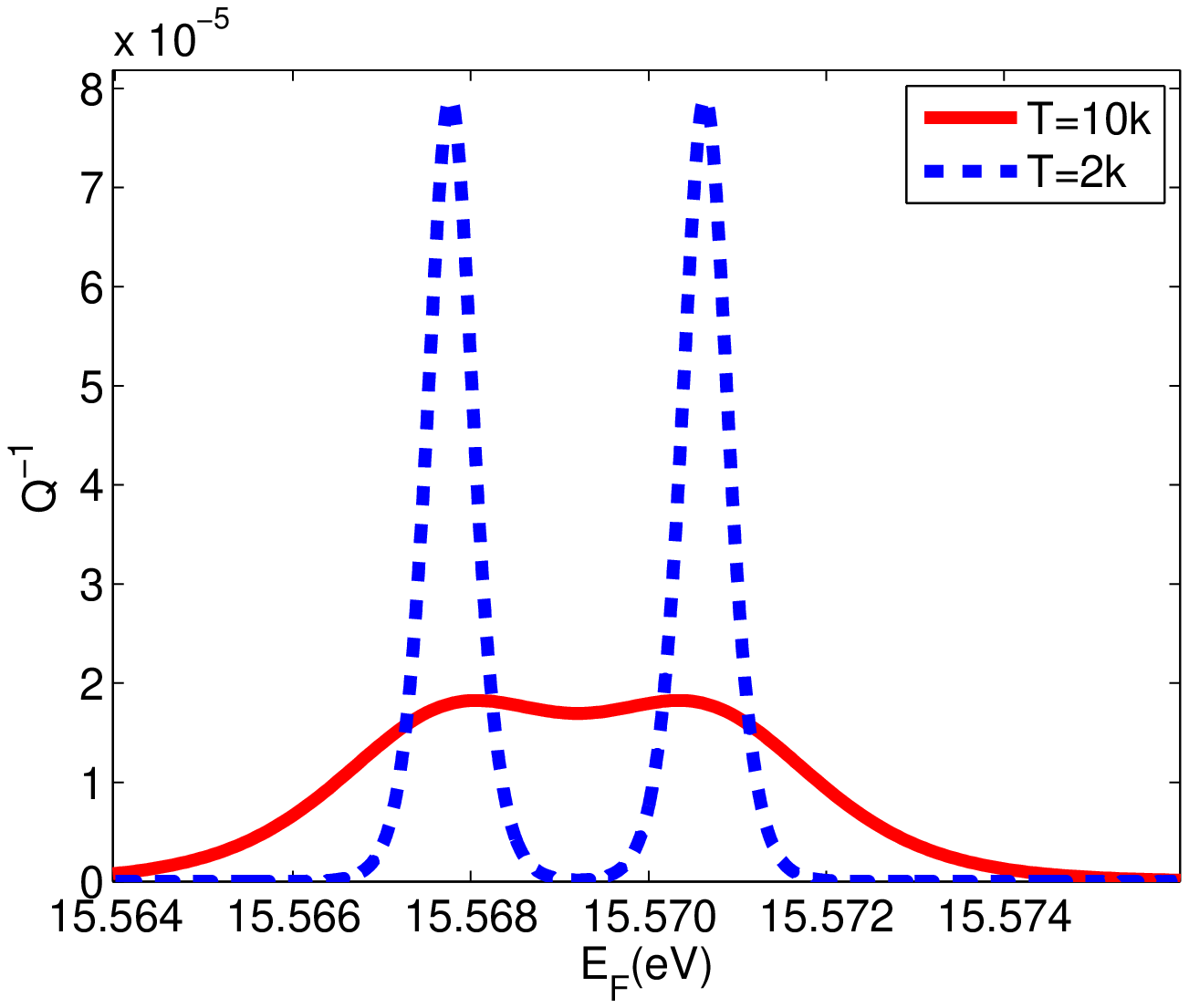}
 \caption{\label{flex_inf_dis_ele_a=b=10_T=10_qPiDIV0_5x103_ EF10_25_zoomin}
  (color online) Inverse quality factor of the flexural mode $q=2\pi/(10^{3})\mathrm{nm^{-1}}$ in an infinitely long beam with $a=b=10\mathrm{nm}$ at $T=2$k and 10k  as a function of the Fermi energy.}
\end{figure}

In Fig.~\ref{flex_inf_dis_ele_a=b=10_T=10_qPiDIV0_5x103_ EF10_25_zoomin} we zoom-in on one of the peaks and lower the temperature even further to reveal the expected inner structure, consisting of two sub-peaks. The two sub-peaks are distinct from each other if the temperature is small compared to the energy difference $\Delta\varepsilon$ between the two energy channels as is the case for $T=2\mathrm{k}$ in Fig.~\ref{flex_inf_dis_ele_a=b=10_T=10_qPiDIV0_5x103_ EF10_25_zoomin}. As the temperature increases the sub-peaks become less distinct---as can be seen in Fig.~\ref{flex_inf_dis_ele_a=b=10_T=10_qPiDIV0_5x103_ EF10_25_zoomin} for $T=10\mathrm{k}$---and eventually merge as the temperature becomes comparable to the energy difference between the adjacent channels as given by Eq.~\eqref{eq:delta-e}.  

\begin{figure}
 \includegraphics[width=1.0\columnwidth]{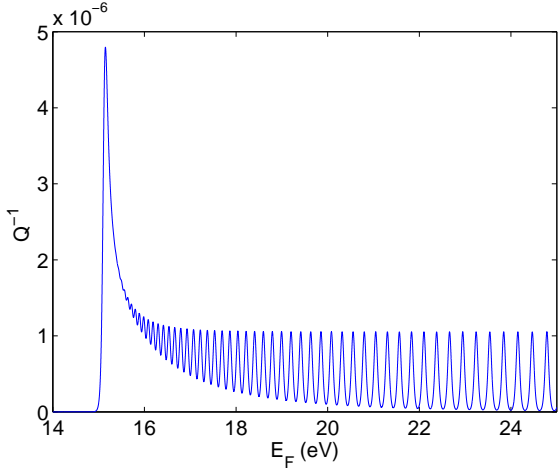}
 \caption{\label{flex_inf_a10_b10_t300_q4PiDIV0_5x103_EF14_25}
  (color online) Inverse quality factor of the flexural mode $q=2\pi/10^{3}\mathrm{nm^{-1}}$ in an infinitely long beam with $a=b=10\mathrm{nm}$ at $T=300\mathrm{k}$ as a function of the Fermi energy.}
\end{figure}

In Fig.~\ref{flex_inf_a10_b10_t300_q4PiDIV0_5x103_EF14_25} we show how $Q^{-1}$ varies with $\varepsilon_{\mathrm{F}}$ at a high temperature of $T=300\mathrm{k}$. The dense peaks near $\varepsilon_{\mathrm{F}}=14\mathrm{eV}$ in Fig. \ref{flex_inf_dis_ele_a=b=10_T=10_qPiDIV0_5x103_ EF10_25} broaden and merge into a single large asymmetric peak. As the Fermi energy increases---and along with it so does the separation between the probed pairs of dissipation channels---individual peaks start to emerge and become visible. However, the sub-peak structure within each main peak remains indiscernible.

Eqs.~\eqref{eq:delta-e} and~\eqref{eq:e-approx} continue to provide a good qualitative understanding of the calculated damping peaks for other system parameters, as long as $qb$ remains smaller than 1. For example, doubling the transverse dimensions of the beam to $a=b=20\mathrm{nm}$ lowers the threshold energy by a factor of $2^4=16$ to somewhat below $1\mathrm{eV}$, and the entire peak structure shown in Fig.~\ref{flex_inf_dis_ele_a=b=10_T=10_qPiDIV0_5x103_ EF10_25} is shifted to between about $1\mathrm{eV}$ and $2.5\mathrm{eV}$. At higher Fermi energies the structure becomes less ordered than the one shown in Fig.~\ref{flex_inf_dis_ele_a=b=10_T=10_qPiDIV0_5x103_ EF10_25}, because the energy is sufficiently large for the value of $n_z$ to start changing between peaks as well.

\subsection{Electron-phonon damping in the presence of other dissipation mechanisms}\label{subsubsec:Broadened phonon states} 

\begin{figure}
 \includegraphics[width=1.0\columnwidth]{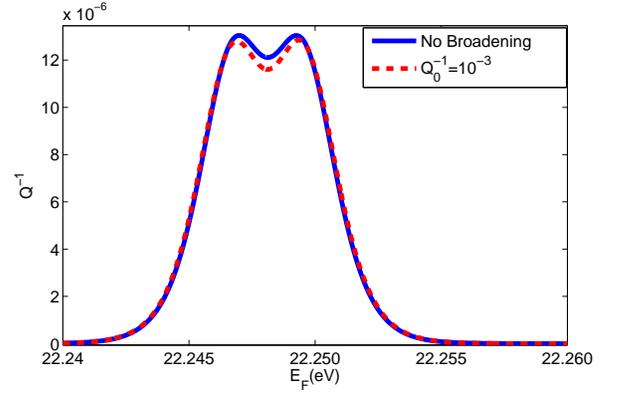}
 \caption{\label{inf_ph_dis_ele_a=b=10_T=10_Lorentizan_ph_Gamma_10_minus3_EF22_24ev_to_22_26ev}
  (color online) Variation of $Q^{-1}$ of the flexural mode $q=2\pi/10^{3}\mathrm{nm^{-1}}$ in an infinite beam with $a=b=10\mathrm{nm}$ at $T=10\mathrm{k}$ as a function of the Fermi energy. A solid (blue) line shows the results obtained if one ignores other damping mechanisms, while a dashed (red) line shows the results obtained using a broadened phonon spectral function in the form of a Lorentzian with $Q_{0}^{-1} = \Gamma/\hbar\omega_{q} = 10^{-3}$.}
\end{figure}

The calculation of the phonon decay rate in Eq.~\eqref{eq:general fermi} assumes that initially both electron and phonon lifetimes are infinite, and therefore their spectral functions are given by Dirac delta functions. This assumes that the electron mean-free path is greater than the wavelength of the vibration mode, and that all other damping mechanisms are much weaker than the damping caused by electron-phonon scattering. As a quick aside, we wish to account for the effect of other damping mechanisms by adapting our calculation to allow for phonon states with \emph{a priori} finite lifetimes, or broadened spectral functions. We adopt an approximate phenomenological method used in electron transport simulations,\cite{chang,reggiani,kim,giulio} in which collisional broadening of electrons is taken into account. We replace the sharp energy conservation delta functions in the expression for the decay rate~\eqref{eq:general fermi} with a convolution over a broadened spectral function of the phonon in the form of a Lorentzian, 
\begin{equation}\label{eq:spectral fun}
A(\varepsilon)=\frac{1}{\pi}\frac{\hbar\Gamma/2}{\varepsilon^{2}+\left(\hbar\Gamma/2\right)^{2}}.
\end{equation}
We ignore any shift in the energy of the phonon that might be caused by other damping mechanisms and continue to take the electron spectral functions to be delta functions. Thus, we replace the delta functions $\delta\left(\varepsilon_{\mathbf{k}}-\varepsilon_{\mathbf{k'}}\pm\hbar\omega_{q}\right)$ in Eq.~\eqref{eq:general fermi} with their convolution with the Lorentzian spectral function~\eqref{eq:spectral fun} of the phonons,
\begin{multline}\label{eq:replaced delta phonon}
A\left(\varepsilon_{\mathbf{k}}-\varepsilon_{\mathbf{k'}}\pm\hbar\omega_{q} \right)\\
=\int A\left(E-\hbar\omega_{q}\right)
\delta\left(\varepsilon_{\mathbf{k}}-\varepsilon_{\mathbf{k'}}\pm E\right) dE.
\end{multline}

We assume an initial broadening with $Q_{0}^{-1}=\Gamma/\hbar\omega_{q}$ that is much larger than the values calculated above for electron-phonon damping. This corresponds to situations where other sources of damping completely obscure the effect of electron-phonon damping. Nevertheless, if for such initial broadening the contribution of electron-phonon damping is not significantly altered, then it will not be affected when the initial damping is smaller and electron-phonon damping is a dominant effect. As an example, we take $Q_{0}^{-1}$ to be equal to $10^{-3}$ (which reflects a metallic beam with a relatively poor quality factor) and recalculate one of the damping peaks obtained in the previous section as function of $\varepsilon_{\rm F}$ for a flexural mode with $q=2\pi/10^{3}\mathrm{nm^{-1}}$ in an infinitely long beam with $a=b=10\mathrm{nm}$ at $T=10\mathrm{k}$. As can be seen from Fig.~\ref{flex_inf_dis_ele_a=b=10_T=10_qPiDIV0_5x103_ EF10_25}, $Q^{-1}_{q}$ of this mode is smaller by at least an order of magnitude compared to $Q_{0}^{-1}$. In Fig.~\ref{inf_ph_dis_ele_a=b=10_T=10_Lorentizan_ph_Gamma_10_minus3_EF22_24ev_to_22_26ev} we show that the effect of an initial broadening on such a peak is very small---a result which holds also for the rest of the peaks as well as for low temperatures. This indicates that even for relatively high damping by other dissipation mechanisms (compared to ``stand-alone'' electron-phonon damping), the contribution of electron-phonon interaction to the total damping is nearly unaffected.

\subsection{Damping of flexural vibrations in a finite beam}\label{subsubsec:Finite Beam}

\begin{figure}
 \includegraphics[width=1.0\columnwidth]{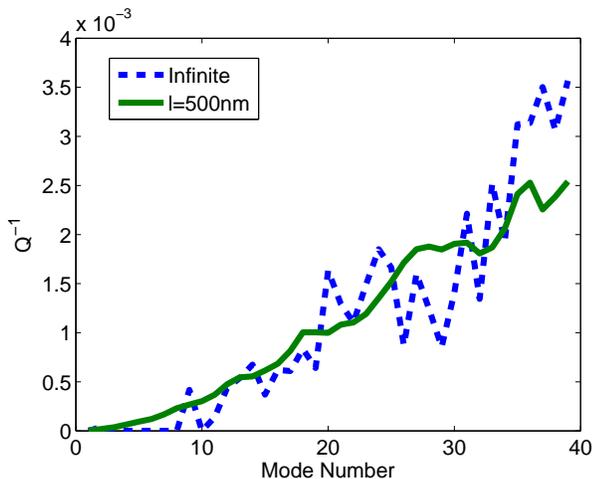}
 \caption{\label{t10_ab20_q1_39_l_500_finite_vs_infinite}
  (color online) Inverse quality factor of the lowest forty flexural modes of a finite beam with $l=0.5\mathrm{\mu m}$ and $a=b=20\mathrm{nm}$ at $T=10\mathrm{k}$. The results obtained for the same values of $q$ for an infinite beam are given for comparison.}
\end{figure}

Finally, we consider the effect of the finite length of the beam by calculating the inverse quality factors of flexural modes in a finite beam using Eq.~\eqref{eq:decay_flex_finite_confined electrons final}, which takes into account the lateral confinement of the electrons. Fig.~\ref{t10_ab20_q1_39_l_500_finite_vs_infinite} shows the low-temperature behavior of $Q^{-1}$ as a function of wavenumber $q$ in a finite beam, as compared with the values calculated earlier for an infinite beam. We can still identify the non-monotonic dependence on parameters, which is characteristic of the low-temperature mesoscopic regime, although the sharp features average out in the finite beam. The damping in the finite beam deviates significantly from the damping in the infinite beam, exhibiting smaller fluctuations away from the general increasing trend, although the relative difference between the two results decreases with increasing wavenumber $q$.

The difference between the finite beam and the infinite beam is due to the replacement of the momentum delta functions in Eq.~\eqref{eq:e-ph flexural infinite confined electrons} for an infinite beam with the $\LL_{n}$ functions appearing in Eq.~\eqref{eq:e-ph flexural finite confined electrons} for a finite beam. Thus, exact momentum conservation is replaced by an approximate or an imperfect momentum conservation. As a consequence, the discrete electronic states that form a dissipation channel in an infinite beam are replaced by small intervals of states---shown schematically in Fig.~\ref{schematic_ef}---causing the smearing or averaging effect that we observe.

\begin{figure}
 \includegraphics[width=1.0\columnwidth]{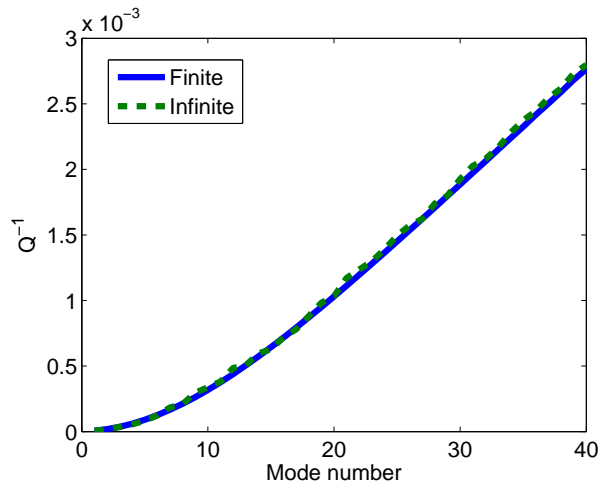}
 \caption{\label{t300_ab20_q1_40_l_500_finite_vs_infinite}
  (color online) Inverse quality factors of the lowest forty flexural modes of the same beam as in Fig.~\ref{t10_ab20_q1_39_l_500_finite_vs_infinite}, but  at $T=300\mathrm{k}$, compared again with the inverse quality factors of an infinite beam.}
\end{figure}

In order to recover the infinite beam results, the $\LL_{n}$ functions need to be sufficiently sharp compared to all other scales related to the momentum conservation in the $x$ direction. At high temperatures this is simply the case for short-wavelength modes, where $q_{n}l\gg1$. However, at low temperatures we need to require in addition that the $\LL_{n}$ functions be sufficiently narrow compared to the width of the Fermi-Dirac factors of the form $n(1-n)$. Thus, as the temperature increases and with it the width of the Fermi-Dirac factors, the difference between the damping of the finite beam and of the infinite beam decreases, as shown in Fig.~\ref{t300_ab20_q1_40_l_500_finite_vs_infinite}.

\section{Summary and Conclusions}\label{sec:Conclusion}

We have estimated the effect of free electrons in thermal equilibrium on the damping of mechanical vibrations in metallic nanoresonators in the form of thin long beams. In doing so we made a sequence of assumptions: (1)~phonon wave functions were obtained by quantizing the normal modes of vibration obtained by the simplest thin-beam elasticity theories; (2) electrons were considered as free noninteracting particles either totally unconfined, as in bulk material, or confined by the lateral surfaces of the resonators; (3) electron-phonon interaction was described by means of a short-range electrostatic potential, where electrons couple to phonons only through local volume changes, which are largest for longitudinal modes leading to $Q$'s on the order of 10, are smaller for flexural modes leading to $Q$'s on the order of $10^3$ to $10^6$, and are non-existent for twist modes; (4) the calculation itself was performed at the level of Fermi's Golden Rule, considering an electron-phonon system at thermal equilibrium with just a single additional phonon in the vibration mode of interest; (5) the \emph{a priori} lifetimes of the electrons and the phonons were assumed infinite, thereby imposing the adiabatic or high-frequency limit, $\omega_q\tau_e>1$, and assuming that all other vibration damping mechanisms are much weaker (although we showed that the latter requirement is not too stringent); and (6) to obtain actual results we used the material properties of bulk aluminum, neglecting their dependence on size and temperature. Aluminum was chosen because it is expected to exhibit a relatively high degree of electron-phonon damping among the common metals used to fabricate nanomechanical resonators. Despite all of our assumptions and approximations, we believe that our results provide an adequate qualitative understanding of the behavior of electron-phonon damping in metallic nanoresonators in the adiabatic limit.

Within these assumptions, electron-phonon interaction is restricted by energy conservation (through Fermi's golden rule), and by momentum conservation along the length of the beam when the beam is considered to be infinite. These conservation laws severely restrict the number of electronic states that can contribute to the damping of a given vibration mode, leading to the appearance of discrete dissipation channels. The energy separation between dissipation channels is determined by the dimensions of the cross-section of the beam, and for flexural modes also by the phonon wavenumber. This in turn sets a temperature scale $\Tm$ for the crossover from high-temperature macroscopic behavior, where electron-phonon damping behaves as if the electrons were in the bulk, to low-temperature mesoscopic behavior, where damping is dominated by just a few dissipation channels and exhibits sharp non-monotonic changes as parameters are varied. As a consequence, for example, at sufficiently low temperatures one can perform spectroscopy of the electronic states by measuring the quality factor of a given vibration mode, while varying an external gate potential.

In beams of finite length one lacks translation invariance, and momentum conservation along the length of the beam is lost. Nevertheless, for sufficiently long beams compared to the wavelength of the vibration, one still retains approximate, or imperfect, momentum conservation, where delta functions are replaced by finite yet sharply peaked functions, whose width is inversely proportional to the length of the beam. The infinite-length limit is recovered when the width of these functions becomes sufficiently small compared to all other relevant momentum scales along the length of the beam.  At high temperatures the only relevant scale is the wavelength of the mode, and the infinite-length limit is simply recovered when the wavelength becomes small compared to the length of the beam. At low temperatures the infinite limit is obtained only when, in addition, the width of the imperfect conservation functions becomes smaller than the widths of the Fermi-Dirac factors that appear in the expressions for the inverse quality factor. Thus, longer beams are required as temperature is decreased. 

To conclude, we have demonstrated several interesting features of the damping of vibrational modes in metallic nanoresonators. These features arise from finite size effects that lead to a quantization of the electronic spectrum as well as to imperfect momentum conservation, and are dominant in the low-temperature mesoscopic regime. At high temperatures the damping becomes similar to the one obtained for unconfined electrons. Our analysis was carried out within a simple specific geometry. Nevertheless, we believe that the different regimes we have identified are relevant to electron-phonon damping in other types and geometries of metallic nanoresonators.

\begin{acknowledgments} 
We wish to thank Mike Cross, Eli Eisenberg, Omri Gat, Uri London, and John Sader for fruitful discussions during the preparation of this work. This research was supported by the German-Israeli Foundation (GIF) through Grant No.~981-185.14/2007.  \end{acknowledgments}

\bibliography{EPD}

\begin{thebibliography}{107}%
\makeatletter
\providecommand \@ifxundefined [1]{%
 \@ifx{#1\undefined}
}%
\providecommand \@ifnum [1]{%
 \ifnum #1\expandafter \@firstoftwo
 \else \expandafter \@secondoftwo
 \fi
}%
\providecommand \@ifx [1]{%
 \ifx #1\expandafter \@firstoftwo
 \else \expandafter \@secondoftwo
 \fi
}%
\providecommand \natexlab [1]{#1}%
\providecommand \enquote  [1]{``#1''}%
\providecommand \bibnamefont  [1]{#1}%
\providecommand \bibfnamefont [1]{#1}%
\providecommand \citenamefont [1]{#1}%
\providecommand \href@noop [0]{\@secondoftwo}%
\providecommand \href [0]{\begingroup \@sanitize@url \@href}%
\providecommand \@href[1]{\@@startlink{#1}\@@href}%
\providecommand \@@href[1]{\endgroup#1\@@endlink}%
\providecommand \@sanitize@url [0]{\catcode `\\12\catcode `\$12\catcode
  `\&12\catcode `\#12\catcode `\^12\catcode `\_12\catcode `\%12\relax}%
\providecommand \@@startlink[1]{}%
\providecommand \@@endlink[0]{}%
\providecommand \url  [0]{\begingroup\@sanitize@url \@url }%
\providecommand \@url [1]{\endgroup\@href {#1}{\urlprefix }}%
\providecommand \urlprefix  [0]{URL }%
\providecommand \Eprint [0]{\href }%
\providecommand \doibase [0]{http://dx.doi.org/}%
\providecommand \selectlanguage [0]{\@gobble}%
\providecommand \bibinfo  [0]{\@secondoftwo}%
\providecommand \bibfield  [0]{\@secondoftwo}%
\providecommand \translation [1]{[#1]}%
\providecommand \BibitemOpen [0]{}%
\providecommand \bibitemStop [0]{}%
\providecommand \bibitemNoStop [0]{.\EOS\space}%
\providecommand \EOS [0]{\spacefactor3000\relax}%
\providecommand \BibitemShut  [1]{\csname bibitem#1\endcsname}%
\let\auto@bib@innerbib\@empty
\bibitem [{\citenamefont {Roukes}(2001)}]{RoukesPlenty}%
  \BibitemOpen
  \bibfield  {author} {\bibinfo {author} {\bibfnamefont {M.~L.}\ \bibnamefont
  {Roukes}},\ }\href@noop {} {\bibfield  {journal} {\bibinfo  {journal}
  {Scientific American}\ }\textbf {\bibinfo {volume} {285}},\ \bibinfo {pages}
  {42} (\bibinfo {year} {2001})}\BibitemShut {NoStop}%
\bibitem [{\citenamefont {Cleland}(2003)}]{Cleland03}%
  \BibitemOpen
  \bibfield  {author} {\bibinfo {author} {\bibfnamefont {A.}~\bibnamefont
  {Cleland}},\ }\href@noop {} {\emph {\bibinfo {title} {Foundations of
  Nanomechanics}}}\ (\bibinfo  {publisher} {Springer},\ \bibinfo {address}
  {Berlin},\ \bibinfo {year} {2003})\BibitemShut {NoStop}%
\bibitem [{\citenamefont {Ekinci}\ and\ \citenamefont
  {Roukes}(2005)}]{ekinci05}%
  \BibitemOpen
  \bibfield  {author} {\bibinfo {author} {\bibfnamefont {K.~L.}\ \bibnamefont
  {Ekinci}}\ and\ \bibinfo {author} {\bibfnamefont {M.~L.}\ \bibnamefont
  {Roukes}},\ }\href {\doibase doi: 10.1063/1.1927327} {\bibfield  {journal}
  {\bibinfo  {journal} {Rev.\ Sci.\ Instrum.}\ }\textbf {\bibinfo {volume}
  {76}},\ \bibinfo {pages} {061101} (\bibinfo {year} {2005})}\BibitemShut
  {NoStop}%
\bibitem [{\citenamefont {Kenig}\ \emph {et~al.}(2012)\citenamefont {Kenig},
  \citenamefont {Cross}, \citenamefont {Lifshitz}, \citenamefont {Karabalin},
  \citenamefont {Villanueva}, \citenamefont {Matheny},\ and\ \citenamefont
  {Roukes}}]{Kenig12}%
  \BibitemOpen
  \bibfield  {author} {\bibinfo {author} {\bibfnamefont {E.}~\bibnamefont
  {Kenig}}, \bibinfo {author} {\bibfnamefont {M.~C.}\ \bibnamefont {Cross}},
  \bibinfo {author} {\bibfnamefont {R.}~\bibnamefont {Lifshitz}}, \bibinfo
  {author} {\bibfnamefont {R.~B.}\ \bibnamefont {Karabalin}}, \bibinfo {author}
  {\bibfnamefont {L.~G.}\ \bibnamefont {Villanueva}}, \bibinfo {author}
  {\bibfnamefont {M.~H.}\ \bibnamefont {Matheny}}, \ and\ \bibinfo {author}
  {\bibfnamefont {M.~L.}\ \bibnamefont {Roukes}},\ }\href {\doibase
  10.1103/PhysRevLett.108.264102} {\bibfield  {journal} {\bibinfo  {journal}
  {Phys. Rev. Lett.}\ }\textbf {\bibinfo {volume} {108}},\ \bibinfo {pages}
  {264102} (\bibinfo {year} {2012})}\BibitemShut {NoStop}%
\bibitem [{\citenamefont {Ekinci}\ \emph {et~al.}(2004)\citenamefont {Ekinci},
  \citenamefont {Huang},\ and\ \citenamefont {Roukes}}]{ekinci04}%
  \BibitemOpen
  \bibfield  {author} {\bibinfo {author} {\bibfnamefont {K.~L.}\ \bibnamefont
  {Ekinci}}, \bibinfo {author} {\bibfnamefont {X.~M.~H.}\ \bibnamefont
  {Huang}}, \ and\ \bibinfo {author} {\bibfnamefont {M.~L.}\ \bibnamefont
  {Roukes}},\ }\href {\doibase 10.1063/1.1755417} {\bibfield  {journal}
  {\bibinfo  {journal} {App. Phys. Lett.}\ }\textbf {\bibinfo {volume} {84}},\
  \bibinfo {pages} {4469} (\bibinfo {year} {2004})}\BibitemShut {NoStop}%
\bibitem [{\citenamefont {Yang}\ \emph {et~al.}(2006)\citenamefont {Yang},
  \citenamefont {Callegari}, \citenamefont {Feng}, \citenamefont {Ekinci},\
  and\ \citenamefont {Roukes}}]{Yang06}%
  \BibitemOpen
  \bibfield  {author} {\bibinfo {author} {\bibfnamefont {Y.~T.}\ \bibnamefont
  {Yang}}, \bibinfo {author} {\bibfnamefont {C.}~\bibnamefont {Callegari}},
  \bibinfo {author} {\bibfnamefont {X.~L.}\ \bibnamefont {Feng}}, \bibinfo
  {author} {\bibfnamefont {K.~L.}\ \bibnamefont {Ekinci}}, \ and\ \bibinfo
  {author} {\bibfnamefont {M.~L.}\ \bibnamefont {Roukes}},\ }\href {\doibase
  10.1021/nl052134m} {\bibfield  {journal} {\bibinfo  {journal} {Nano.\ Lett.}\
  }\textbf {\bibinfo {volume} {6}},\ \bibinfo {pages} {583} (\bibinfo {year}
  {2006})}\BibitemShut {NoStop}%
\bibitem [{\citenamefont {Hanay}\ \emph {et~al.}(2012)\citenamefont {Hanay},
  \citenamefont {Kelber}, \citenamefont {Naik}, \citenamefont {Chi},
  \citenamefont {Hentz}, \citenamefont {Bullard}, \citenamefont {Colinet},
  \citenamefont {Duraffourg},\ and\ \citenamefont {Roukes}}]{Hanay12}%
  \BibitemOpen
  \bibfield  {author} {\bibinfo {author} {\bibfnamefont {M.~S.}\ \bibnamefont
  {Hanay}}, \bibinfo {author} {\bibfnamefont {S.}~\bibnamefont {Kelber}},
  \bibinfo {author} {\bibfnamefont {A.~K.}\ \bibnamefont {Naik}}, \bibinfo
  {author} {\bibfnamefont {D.}~\bibnamefont {Chi}}, \bibinfo {author}
  {\bibfnamefont {S.}~\bibnamefont {Hentz}}, \bibinfo {author} {\bibfnamefont
  {E.~C.}\ \bibnamefont {Bullard}}, \bibinfo {author} {\bibfnamefont
  {E.}~\bibnamefont {Colinet}}, \bibinfo {author} {\bibfnamefont
  {L.}~\bibnamefont {Duraffourg}}, \ and\ \bibinfo {author} {\bibfnamefont
  {M.~L.}\ \bibnamefont {Roukes}},\ }\href {\doibase 10.1038/nnano.2012.119}
  {\bibfield  {journal} {\bibinfo  {journal} {Nature Nanotechnology}\ }\textbf
  {\bibinfo {volume} {7}},\ \bibinfo {pages} {602} (\bibinfo {year}
  {2012})}\BibitemShut {NoStop}%
\bibitem [{\citenamefont {Ilic}\ \emph {et~al.}(2004)\citenamefont {Ilic},
  \citenamefont {Craighead}, \citenamefont {Krylov}, \citenamefont {Senaratne},
  \citenamefont {Ober},\ and\ \citenamefont {Neuzil}}]{ilic04}%
  \BibitemOpen
  \bibfield  {author} {\bibinfo {author} {\bibfnamefont {B.}~\bibnamefont
  {Ilic}}, \bibinfo {author} {\bibfnamefont {H.}~\bibnamefont {Craighead}},
  \bibinfo {author} {\bibfnamefont {S.}~\bibnamefont {Krylov}}, \bibinfo
  {author} {\bibfnamefont {W.}~\bibnamefont {Senaratne}}, \bibinfo {author}
  {\bibfnamefont {C.}~\bibnamefont {Ober}}, \ and\ \bibinfo {author}
  {\bibfnamefont {P.}~\bibnamefont {Neuzil}},\ }\href {\doibase
  10.1063/1.1650542} {\bibfield  {journal} {\bibinfo  {journal} {J.\ Appl.\
  Phys.}\ }\textbf {\bibinfo {volume} {95}},\ \bibinfo {pages} {3694} (\bibinfo
  {year} {2004})}\BibitemShut {NoStop}%
\bibitem [{\citenamefont {Jensen}\ \emph {et~al.}(2008)\citenamefont {Jensen},
  \citenamefont {Kim},\ and\ \citenamefont {Zettl}}]{Jensen08}%
  \BibitemOpen
  \bibfield  {author} {\bibinfo {author} {\bibfnamefont {K.}~\bibnamefont
  {Jensen}}, \bibinfo {author} {\bibfnamefont {K.}~\bibnamefont {Kim}}, \ and\
  \bibinfo {author} {\bibfnamefont {A.}~\bibnamefont {Zettl}},\ }\href
  {\doibase 10.1038/nnano.2008.200} {\bibfield  {journal} {\bibinfo  {journal}
  {Nat Nano}\ }\textbf {\bibinfo {volume} {3}},\ \bibinfo {pages} {533}
  (\bibinfo {year} {2008})}\BibitemShut {NoStop}%
\bibitem [{\citenamefont {Lassagne}\ \emph {et~al.}(2008)\citenamefont
  {Lassagne}, \citenamefont {Garcia-Sanchez}, \citenamefont {Aguasca},\ and\
  \citenamefont {Bachtold}}]{lassagne08}%
  \BibitemOpen
  \bibfield  {author} {\bibinfo {author} {\bibfnamefont {B.}~\bibnamefont
  {Lassagne}}, \bibinfo {author} {\bibfnamefont {D.}~\bibnamefont
  {Garcia-Sanchez}}, \bibinfo {author} {\bibfnamefont {A.}~\bibnamefont
  {Aguasca}}, \ and\ \bibinfo {author} {\bibfnamefont {A.}~\bibnamefont
  {Bachtold}},\ }\href {\doibase 10.1021/nl801982v} {\bibfield  {journal}
  {\bibinfo  {journal} {Nano Letters}\ }\textbf {\bibinfo {volume} {8}},\
  \bibinfo {pages} {3735} (\bibinfo {year} {2008})}\BibitemShut {NoStop}%
\bibitem [{\citenamefont {Rugar}\ \emph {et~al.}(2004)\citenamefont {Rugar},
  \citenamefont {Budakian}, \citenamefont {Mamin},\ and\ \citenamefont
  {Chui}}]{rugar}%
  \BibitemOpen
  \bibfield  {author} {\bibinfo {author} {\bibfnamefont {D.}~\bibnamefont
  {Rugar}}, \bibinfo {author} {\bibfnamefont {R.}~\bibnamefont {Budakian}},
  \bibinfo {author} {\bibfnamefont {H.~J.}\ \bibnamefont {Mamin}}, \ and\
  \bibinfo {author} {\bibfnamefont {B.~W.}\ \bibnamefont {Chui}},\ }\href
  {\doibase 10.1038/nature02658} {\bibfield  {journal} {\bibinfo  {journal}
  {Nature}\ }\textbf {\bibinfo {volume} {430}},\ \bibinfo {pages} {329}
  (\bibinfo {year} {2004})}\BibitemShut {NoStop}%
\bibitem [{\citenamefont {Cleland}\ and\ \citenamefont
  {Roukes}(1998)}]{cleland98}%
  \BibitemOpen
  \bibfield  {author} {\bibinfo {author} {\bibfnamefont {A.~N.}\ \bibnamefont
  {Cleland}}\ and\ \bibinfo {author} {\bibfnamefont {M.~L.}\ \bibnamefont
  {Roukes}},\ }\href@noop {} {\bibfield  {journal} {\bibinfo  {journal}
  {Nature}\ }\textbf {\bibinfo {volume} {392}},\ \bibinfo {pages} {160}
  (\bibinfo {year} {1998})}\BibitemShut {NoStop}%
\bibitem [{\citenamefont {Roukes}(1999)}]{roukes99}%
  \BibitemOpen
  \bibfield  {author} {\bibinfo {author} {\bibfnamefont {M.~L.}\ \bibnamefont
  {Roukes}},\ }\href {\doibase 10.1016/S0921-4526(98)01482-3} {\bibfield
  {journal} {\bibinfo  {journal} {Physica B}\ }\textbf {\bibinfo {volume}
  {263-264}},\ \bibinfo {pages} {1} (\bibinfo {year} {1999})}\BibitemShut
  {NoStop}%
\bibitem [{\citenamefont {Cleland}\ \emph {et~al.}(2002)\citenamefont
  {Cleland}, \citenamefont {Aldridge}, \citenamefont {Driscoll},\ and\
  \citenamefont {Gossard}}]{cleland}%
  \BibitemOpen
  \bibfield  {author} {\bibinfo {author} {\bibfnamefont {A.~N.}\ \bibnamefont
  {Cleland}}, \bibinfo {author} {\bibfnamefont {J.~S.}\ \bibnamefont
  {Aldridge}}, \bibinfo {author} {\bibfnamefont {D.~C.}\ \bibnamefont
  {Driscoll}}, \ and\ \bibinfo {author} {\bibfnamefont {A.~C.}\ \bibnamefont
  {Gossard}},\ }\href {\doibase 10.1063/1.1497436} {\bibfield  {journal}
  {\bibinfo  {journal} {Appl.\ Phys.\ Lett.}\ }\textbf {\bibinfo {volume}
  {81}},\ \bibinfo {pages} {1699} (\bibinfo {year} {2002})}\BibitemShut
  {NoStop}%
\bibitem [{\citenamefont {Knobel}\ and\ \citenamefont
  {Cleland}(2003)}]{knobel}%
  \BibitemOpen
  \bibfield  {author} {\bibinfo {author} {\bibfnamefont {R.~G.}\ \bibnamefont
  {Knobel}}\ and\ \bibinfo {author} {\bibfnamefont {A.~N.}\ \bibnamefont
  {Cleland}},\ }\href {\doibase 10.1038/nature01773} {\bibfield  {journal}
  {\bibinfo  {journal} {Nature}\ }\textbf {\bibinfo {volume} {424}},\ \bibinfo
  {pages} {291} (\bibinfo {year} {2003})}\BibitemShut {NoStop}%
\bibitem [{\citenamefont {Ekinci}\ \emph {et~al.}(2002)\citenamefont {Ekinci},
  \citenamefont {Yang}, \citenamefont {Huang},\ and\ \citenamefont
  {Roukes}}]{ekinci02}%
  \BibitemOpen
  \bibfield  {author} {\bibinfo {author} {\bibfnamefont {K.~L.}\ \bibnamefont
  {Ekinci}}, \bibinfo {author} {\bibfnamefont {Y.~T.}\ \bibnamefont {Yang}},
  \bibinfo {author} {\bibfnamefont {X.~M.~H.}\ \bibnamefont {Huang}}, \ and\
  \bibinfo {author} {\bibfnamefont {M.~L.}\ \bibnamefont {Roukes}},\ }\href
  {\doibase 10.1063/1.1507833} {\bibfield  {journal} {\bibinfo  {journal}
  {Appl.\ Phys.\ Lett.}\ }\textbf {\bibinfo {volume} {81}},\ \bibinfo {pages}
  {2253} (\bibinfo {year} {2002})}\BibitemShut {NoStop}%
\bibitem [{\citenamefont {Truitt}\ \emph {et~al.}(2007)\citenamefont {Truitt},
  \citenamefont {Hertzberg}, \citenamefont {Huang}, \citenamefont {Ekinci},\
  and\ \citenamefont {Schwab}}]{truitt07}%
  \BibitemOpen
  \bibfield  {author} {\bibinfo {author} {\bibfnamefont {P.~A.}\ \bibnamefont
  {Truitt}}, \bibinfo {author} {\bibfnamefont {J.~B.}\ \bibnamefont
  {Hertzberg}}, \bibinfo {author} {\bibfnamefont {C.~C.}\ \bibnamefont
  {Huang}}, \bibinfo {author} {\bibfnamefont {K.~L.}\ \bibnamefont {Ekinci}}, \
  and\ \bibinfo {author} {\bibfnamefont {K.~C.}\ \bibnamefont {Schwab}},\
  }\href {\doibase 10.1021/nl062278g} {\bibfield  {journal} {\bibinfo
  {journal} {Nano Letters}\ }\textbf {\bibinfo {volume} {7}},\ \bibinfo {pages}
  {120} (\bibinfo {year} {2007})}\BibitemShut {NoStop}%
\bibitem [{\citenamefont {Schwab}\ \emph {et~al.}(2000)\citenamefont {Schwab},
  \citenamefont {Henriksen}, \citenamefont {Worlock},\ and\ \citenamefont
  {Roukes}}]{schwab00}%
  \BibitemOpen
  \bibfield  {author} {\bibinfo {author} {\bibfnamefont {K.}~\bibnamefont
  {Schwab}}, \bibinfo {author} {\bibfnamefont {E.~A.}\ \bibnamefont
  {Henriksen}}, \bibinfo {author} {\bibfnamefont {J.~M.}\ \bibnamefont
  {Worlock}}, \ and\ \bibinfo {author} {\bibfnamefont {M.~L.}\ \bibnamefont
  {Roukes}},\ }\href {\doibase 10.1038/35010065} {\bibfield  {journal}
  {\bibinfo  {journal} {Nature}\ }\textbf {\bibinfo {volume} {404}},\ \bibinfo
  {pages} {974} (\bibinfo {year} {2000})}\BibitemShut {NoStop}%
\bibitem [{\citenamefont {{Schwab}}\ and\ \citenamefont
  {{Roukes}}(2005)}]{schwab05}%
  \BibitemOpen
  \bibfield  {author} {\bibinfo {author} {\bibfnamefont {K.~C.}\ \bibnamefont
  {{Schwab}}}\ and\ \bibinfo {author} {\bibfnamefont {M.~L.}\ \bibnamefont
  {{Roukes}}},\ }\href {\doibase 10.1063/1.2012461} {\bibfield  {journal}
  {\bibinfo  {journal} {Physics Today}\ }\textbf {\bibinfo {volume} {58}},\
  \bibinfo {pages} {36} (\bibinfo {year} {2005})}\BibitemShut {NoStop}%
\bibitem [{\citenamefont {LaHaye}\ \emph {et~al.}(2004)\citenamefont {LaHaye},
  \citenamefont {Buu}, \citenamefont {Camarota},\ and\ \citenamefont
  {Schwab}}]{lahaye04}%
  \BibitemOpen
  \bibfield  {author} {\bibinfo {author} {\bibfnamefont {M.~D.}\ \bibnamefont
  {LaHaye}}, \bibinfo {author} {\bibfnamefont {O.}~\bibnamefont {Buu}},
  \bibinfo {author} {\bibfnamefont {B.}~\bibnamefont {Camarota}}, \ and\
  \bibinfo {author} {\bibfnamefont {K.~C.}\ \bibnamefont {Schwab}},\ }\href
  {\doibase 10.1126/science.1094419} {\bibfield  {journal} {\bibinfo  {journal}
  {Science}\ }\textbf {\bibinfo {volume} {304}},\ \bibinfo {pages} {74}
  (\bibinfo {year} {2004})}\BibitemShut {NoStop}%
\bibitem [{\citenamefont {Naik}\ \emph {et~al.}(2006)\citenamefont {Naik},
  \citenamefont {Buu}, \citenamefont {LaHaye}, \citenamefont {Armour},
  \citenamefont {Clerk}, \citenamefont {Blencowe},\ and\ \citenamefont
  {Schwab}}]{naik06}%
  \BibitemOpen
  \bibfield  {author} {\bibinfo {author} {\bibfnamefont {A.}~\bibnamefont
  {Naik}}, \bibinfo {author} {\bibfnamefont {O.}~\bibnamefont {Buu}}, \bibinfo
  {author} {\bibfnamefont {M.}~\bibnamefont {LaHaye}}, \bibinfo {author}
  {\bibfnamefont {A.}~\bibnamefont {Armour}}, \bibinfo {author} {\bibfnamefont
  {A.}~\bibnamefont {Clerk}}, \bibinfo {author} {\bibfnamefont
  {M.}~\bibnamefont {Blencowe}}, \ and\ \bibinfo {author} {\bibfnamefont
  {K.}~\bibnamefont {Schwab}},\ }\href {\doibase 10.1038/nature05027}
  {\bibfield  {journal} {\bibinfo  {journal} {Nature}\ }\textbf {\bibinfo
  {volume} {443}},\ \bibinfo {pages} {193} (\bibinfo {year}
  {2006})}\BibitemShut {NoStop}%
\bibitem [{\citenamefont {O'Connell}\ \emph {et~al.}(2010)\citenamefont
  {O'Connell}, \citenamefont {Hofheinz}, \citenamefont {Ansmann}, \citenamefont
  {Bialczak}, \citenamefont {Lenander}, \citenamefont {Lucero}, \citenamefont
  {Neeley}, \citenamefont {Sank}, \citenamefont {Wang}, \citenamefont {Weides}
  \emph {et~al.}}]{Oconnell10}%
  \BibitemOpen
  \bibfield  {author} {\bibinfo {author} {\bibfnamefont {A.}~\bibnamefont
  {O'Connell}}, \bibinfo {author} {\bibfnamefont {M.}~\bibnamefont {Hofheinz}},
  \bibinfo {author} {\bibfnamefont {M.}~\bibnamefont {Ansmann}}, \bibinfo
  {author} {\bibfnamefont {R.}~\bibnamefont {Bialczak}}, \bibinfo {author}
  {\bibfnamefont {M.}~\bibnamefont {Lenander}}, \bibinfo {author}
  {\bibfnamefont {E.}~\bibnamefont {Lucero}}, \bibinfo {author} {\bibfnamefont
  {M.}~\bibnamefont {Neeley}}, \bibinfo {author} {\bibfnamefont
  {D.}~\bibnamefont {Sank}}, \bibinfo {author} {\bibfnamefont {H.}~\bibnamefont
  {Wang}}, \bibinfo {author} {\bibfnamefont {M.}~\bibnamefont {Weides}},  \emph
  {et~al.},\ }\href {\doibase 10.1038/nature08967} {\bibfield  {journal}
  {\bibinfo  {journal} {Nature}\ }\textbf {\bibinfo {volume} {464}},\ \bibinfo
  {pages} {697} (\bibinfo {year} {2010})}\BibitemShut {NoStop}%
\bibitem [{\citenamefont {Katz}\ \emph {et~al.}(2007)\citenamefont {Katz},
  \citenamefont {Retzker}, \citenamefont {Straub},\ and\ \citenamefont
  {Lifshitz}}]{katz}%
  \BibitemOpen
  \bibfield  {author} {\bibinfo {author} {\bibfnamefont {I.}~\bibnamefont
  {Katz}}, \bibinfo {author} {\bibfnamefont {A.}~\bibnamefont {Retzker}},
  \bibinfo {author} {\bibfnamefont {R.}~\bibnamefont {Straub}}, \ and\ \bibinfo
  {author} {\bibfnamefont {R.}~\bibnamefont {Lifshitz}},\ }\href {\doibase
  10.1103/PhysRevLett.99.040404} {\bibfield  {journal} {\bibinfo  {journal}
  {Phys. Rev. Lett.}\ }\textbf {\bibinfo {volume} {99}},\ \bibinfo {pages}
  {040404} (\bibinfo {year} {2007})}\BibitemShut {NoStop}%
\bibitem [{\citenamefont {Katz}\ \emph {et~al.}(2008)\citenamefont {Katz},
  \citenamefont {Lifshitz}, \citenamefont {Retzker},\ and\ \citenamefont
  {Straub}}]{katz08}%
  \BibitemOpen
  \bibfield  {author} {\bibinfo {author} {\bibfnamefont {I.}~\bibnamefont
  {Katz}}, \bibinfo {author} {\bibfnamefont {R.}~\bibnamefont {Lifshitz}},
  \bibinfo {author} {\bibfnamefont {A.}~\bibnamefont {Retzker}}, \ and\
  \bibinfo {author} {\bibfnamefont {R.}~\bibnamefont {Straub}},\ }\href
  {\doibase 10.1088/1367-2630/10/12/125023} {\bibfield  {journal} {\bibinfo
  {journal} {New J.\ Phys.}\ }\textbf {\bibinfo {volume} {10}},\ \bibinfo
  {pages} {125023} (\bibinfo {year} {2008})}\BibitemShut {NoStop}%
\bibitem [{\citenamefont {Mihailovich}\ and\ \citenamefont
  {MacDonald}(1995)}]{Mihailovich95}%
  \BibitemOpen
  \bibfield  {author} {\bibinfo {author} {\bibfnamefont {R.~E.}\ \bibnamefont
  {Mihailovich}}\ and\ \bibinfo {author} {\bibfnamefont {N.~C.}\ \bibnamefont
  {MacDonald}},\ }\href {\doibase 10.1016/0924-4247(95)01080-7} {\bibfield
  {journal} {\bibinfo  {journal} {Sensors and Actuators A: Physical}\ }\textbf
  {\bibinfo {volume} {50}},\ \bibinfo {pages} {199 } (\bibinfo {year}
  {1995})}\BibitemShut {NoStop}%
\bibitem [{\citenamefont {Olkhovets}\ \emph {et~al.}(2000)\citenamefont
  {Olkhovets}, \citenamefont {Evoy}, \citenamefont {Carr}, \citenamefont
  {Parpia},\ and\ \citenamefont {Craighead}}]{olkhovets}%
  \BibitemOpen
  \bibfield  {author} {\bibinfo {author} {\bibfnamefont {A.}~\bibnamefont
  {Olkhovets}}, \bibinfo {author} {\bibfnamefont {S.}~\bibnamefont {Evoy}},
  \bibinfo {author} {\bibfnamefont {D.~W.}\ \bibnamefont {Carr}}, \bibinfo
  {author} {\bibfnamefont {J.~M.}\ \bibnamefont {Parpia}}, \ and\ \bibinfo
  {author} {\bibfnamefont {H.~G.}\ \bibnamefont {Craighead}},\ }\href {\doibase
  10.1116/1.1313571} {\bibfield  {journal} {\bibinfo  {journal} {J.\ Vac.\
  Sci.\ Technol.}\ }\textbf {\bibinfo {volume} {18}},\ \bibinfo {pages} {3549 }
  (\bibinfo {year} {2000})}\BibitemShut {NoStop}%
\bibitem [{\citenamefont {Carr}\ \emph {et~al.}(1999)\citenamefont {Carr},
  \citenamefont {Evoy}, \citenamefont {Sekaric}, \citenamefont {Craighead},\
  and\ \citenamefont {Parpia}}]{carr2}%
  \BibitemOpen
  \bibfield  {author} {\bibinfo {author} {\bibfnamefont {D.~W.}\ \bibnamefont
  {Carr}}, \bibinfo {author} {\bibfnamefont {S.}~\bibnamefont {Evoy}}, \bibinfo
  {author} {\bibfnamefont {L.}~\bibnamefont {Sekaric}}, \bibinfo {author}
  {\bibfnamefont {H.~G.}\ \bibnamefont {Craighead}}, \ and\ \bibinfo {author}
  {\bibfnamefont {J.~M.}\ \bibnamefont {Parpia}},\ }\href {\doibase
  10.1063/1.124554} {\bibfield  {journal} {\bibinfo  {journal} {Appl. Phys.
  Lett.}\ }\textbf {\bibinfo {volume} {75}},\ \bibinfo {pages} {920} (\bibinfo
  {year} {1999})}\BibitemShut {NoStop}%
\bibitem [{\citenamefont {Evoy}\ \emph {et~al.}(2000)\citenamefont {Evoy},
  \citenamefont {Olkhovets}, \citenamefont {Sekaric}, \citenamefont {Parpia},
  \citenamefont {Craighead},\ and\ \citenamefont {Carr}}]{evoy2}%
  \BibitemOpen
  \bibfield  {author} {\bibinfo {author} {\bibfnamefont {S.}~\bibnamefont
  {Evoy}}, \bibinfo {author} {\bibfnamefont {A.}~\bibnamefont {Olkhovets}},
  \bibinfo {author} {\bibfnamefont {L.}~\bibnamefont {Sekaric}}, \bibinfo
  {author} {\bibfnamefont {J.~M.}\ \bibnamefont {Parpia}}, \bibinfo {author}
  {\bibfnamefont {H.~G.}\ \bibnamefont {Craighead}}, \ and\ \bibinfo {author}
  {\bibfnamefont {D.~W.}\ \bibnamefont {Carr}},\ }\href {\doibase
  10.1063/1.1316071} {\bibfield  {journal} {\bibinfo  {journal} {Appl. Phys.
  Lett.}\ }\textbf {\bibinfo {volume} {77}},\ \bibinfo {pages} {2397} (\bibinfo
  {year} {2000})}\BibitemShut {NoStop}%
\bibitem [{\citenamefont {Liu}\ \emph {et~al.}(2005)\citenamefont {Liu},
  \citenamefont {Vignola}, \citenamefont {Simpson}, \citenamefont {Lemon},
  \citenamefont {Houston},\ and\ \citenamefont {Photiadis}}]{liu}%
  \BibitemOpen
  \bibfield  {author} {\bibinfo {author} {\bibfnamefont {X.}~\bibnamefont
  {Liu}}, \bibinfo {author} {\bibfnamefont {J.~F.}\ \bibnamefont {Vignola}},
  \bibinfo {author} {\bibfnamefont {H.~J.}\ \bibnamefont {Simpson}}, \bibinfo
  {author} {\bibfnamefont {B.~R.}\ \bibnamefont {Lemon}}, \bibinfo {author}
  {\bibfnamefont {B.~H.}\ \bibnamefont {Houston}}, \ and\ \bibinfo {author}
  {\bibfnamefont {D.~M.}\ \bibnamefont {Photiadis}},\ }\href {\doibase
  10.1063/1.1819980} {\bibfield  {journal} {\bibinfo  {journal} {J.\ App.\
  Phys.}\ }\textbf {\bibinfo {volume} {97}},\ \bibinfo {pages} {023524}
  (\bibinfo {year} {2005})}\BibitemShut {NoStop}%
\bibitem [{\citenamefont {Mohanty}\ \emph {et~al.}(2002)\citenamefont
  {Mohanty}, \citenamefont {Harrington}, \citenamefont {Ekinci}, \citenamefont
  {Yang}, \citenamefont {Murphy},\ and\ \citenamefont {Roukes}}]{mohanty}%
  \BibitemOpen
  \bibfield  {author} {\bibinfo {author} {\bibfnamefont {P.}~\bibnamefont
  {Mohanty}}, \bibinfo {author} {\bibfnamefont {D.~A.}\ \bibnamefont
  {Harrington}}, \bibinfo {author} {\bibfnamefont {K.~L.}\ \bibnamefont
  {Ekinci}}, \bibinfo {author} {\bibfnamefont {Y.~T.}\ \bibnamefont {Yang}},
  \bibinfo {author} {\bibfnamefont {M.~J.}\ \bibnamefont {Murphy}}, \ and\
  \bibinfo {author} {\bibfnamefont {M.~L.}\ \bibnamefont {Roukes}},\ }\href
  {\doibase 10.1103/PhysRevB.66.085416} {\bibfield  {journal} {\bibinfo
  {journal} {Phys. Rev. B}\ }\textbf {\bibinfo {volume} {66}},\ \bibinfo
  {pages} {085416} (\bibinfo {year} {2002})}\BibitemShut {NoStop}%
\bibitem [{\citenamefont {Zolfagharkhani}\ \emph {et~al.}(2005)\citenamefont
  {Zolfagharkhani}, \citenamefont {Gaidarzhy}, \citenamefont {Shim},
  \citenamefont {Badzey},\ and\ \citenamefont {Mohanty}}]{zolfagharkhani}%
  \BibitemOpen
  \bibfield  {author} {\bibinfo {author} {\bibfnamefont {G.}~\bibnamefont
  {Zolfagharkhani}}, \bibinfo {author} {\bibfnamefont {A.}~\bibnamefont
  {Gaidarzhy}}, \bibinfo {author} {\bibfnamefont {S.-B.}\ \bibnamefont {Shim}},
  \bibinfo {author} {\bibfnamefont {R.~L.}\ \bibnamefont {Badzey}}, \ and\
  \bibinfo {author} {\bibfnamefont {P.}~\bibnamefont {Mohanty}},\ }\href
  {\doibase 10.1103/PhysRevB.72.224101} {\bibfield  {journal} {\bibinfo
  {journal} {Phys. Rev. B}\ }\textbf {\bibinfo {volume} {72}},\ \bibinfo
  {pages} {224101} (\bibinfo {year} {2005})}\BibitemShut {NoStop}%
\bibitem [{\citenamefont {Seo\'anez}\ \emph {et~al.}(2008)\citenamefont
  {Seo\'anez}, \citenamefont {Guinea},\ and\ \citenamefont
  {Castro~Neto}}]{seoanez}%
  \BibitemOpen
  \bibfield  {author} {\bibinfo {author} {\bibfnamefont {C.}~\bibnamefont
  {Seo\'anez}}, \bibinfo {author} {\bibfnamefont {F.}~\bibnamefont {Guinea}}, \
  and\ \bibinfo {author} {\bibfnamefont {A.~H.}\ \bibnamefont {Castro~Neto}},\
  }\href {\doibase 10.1103/PhysRevB.77.125107} {\bibfield  {journal} {\bibinfo
  {journal} {Phys. Rev. B}\ }\textbf {\bibinfo {volume} {77}},\ \bibinfo {eid}
  {125107} (\bibinfo {year} {2008})}\BibitemShut {NoStop}%
\bibitem [{\citenamefont {Remus}\ \emph {et~al.}(2009)\citenamefont {Remus},
  \citenamefont {Blencowe},\ and\ \citenamefont {Tanaka}}]{remus}%
  \BibitemOpen
  \bibfield  {author} {\bibinfo {author} {\bibfnamefont {L.~G.}\ \bibnamefont
  {Remus}}, \bibinfo {author} {\bibfnamefont {M.~P.}\ \bibnamefont {Blencowe}},
  \ and\ \bibinfo {author} {\bibfnamefont {Y.}~\bibnamefont {Tanaka}},\ }\href
  {\doibase 10.1103/PhysRevB.80.174103} {\bibfield  {journal} {\bibinfo
  {journal} {Phys. Rev. B}\ }\textbf {\bibinfo {volume} {80}},\ \bibinfo
  {pages} {174103} (\bibinfo {year} {2009})}\BibitemShut {NoStop}%
\bibitem [{\citenamefont {Chu}\ \emph {et~al.}(2007)\citenamefont {Chu},
  \citenamefont {Rudd},\ and\ \citenamefont {Blencowe}}]{chu07}%
  \BibitemOpen
  \bibfield  {author} {\bibinfo {author} {\bibfnamefont {M.}~\bibnamefont
  {Chu}}, \bibinfo {author} {\bibfnamefont {R.~E.}\ \bibnamefont {Rudd}}, \
  and\ \bibinfo {author} {\bibfnamefont {M.~P.}\ \bibnamefont {Blencowe}},\
  }\href@noop {} {\enquote {\bibinfo {title} {The role of reconstructed
  surfaces in the intrinsic dissipative dynamics of silicon nanoresonators},}\
  } (\bibinfo {year} {2007}),\ \bibinfo {note} {preprint},\ \Eprint
  {http://arxiv.org/abs/0705.0015} {arXiv:0705.0015} \BibitemShut {NoStop}%
\bibitem [{\citenamefont {Unterreithmeier}\ \emph {et~al.}(2010)\citenamefont
  {Unterreithmeier}, \citenamefont {Faust},\ and\ \citenamefont
  {Kotthaus}}]{unterreithmeier}%
  \BibitemOpen
  \bibfield  {author} {\bibinfo {author} {\bibfnamefont {Q.~P.}\ \bibnamefont
  {Unterreithmeier}}, \bibinfo {author} {\bibfnamefont {T.}~\bibnamefont
  {Faust}}, \ and\ \bibinfo {author} {\bibfnamefont {J.~P.}\ \bibnamefont
  {Kotthaus}},\ }\href {\doibase 10.1103/PhysRevLett.105.027205} {\bibfield
  {journal} {\bibinfo  {journal} {Phys. Rev. Lett.}\ }\textbf {\bibinfo
  {volume} {105}},\ \bibinfo {pages} {027205} (\bibinfo {year}
  {2010})}\BibitemShut {NoStop}%
\bibitem [{\citenamefont {Lifshitz}\ and\ \citenamefont
  {Roukes}(2000)}]{lifshitzTED}%
  \BibitemOpen
  \bibfield  {author} {\bibinfo {author} {\bibfnamefont {R.}~\bibnamefont
  {Lifshitz}}\ and\ \bibinfo {author} {\bibfnamefont {M.~L.}\ \bibnamefont
  {Roukes}},\ }\href {\doibase 10.1103/PhysRevB.61.5600} {\bibfield  {journal}
  {\bibinfo  {journal} {Phys.\ Rev.\ B}\ }\textbf {\bibinfo {volume} {61}},\
  \bibinfo {pages} {5600} (\bibinfo {year} {2000})}\BibitemShut {NoStop}%
\bibitem [{\citenamefont {Lifshitz}(2002)}]{lifshitzPhonon}%
  \BibitemOpen
  \bibfield  {author} {\bibinfo {author} {\bibfnamefont {R.}~\bibnamefont
  {Lifshitz}},\ }\href {\doibase 10.1016/S0921-4526(02)00524-0} {\bibfield
  {journal} {\bibinfo  {journal} {Physica B}\ }\textbf {\bibinfo {volume}
  {316}},\ \bibinfo {pages} {397} (\bibinfo {year} {2002})}\BibitemShut
  {NoStop}%
\bibitem [{\citenamefont {Houston}\ \emph {et~al.}(2002)\citenamefont
  {Houston}, \citenamefont {Photiadis}, \citenamefont {Marcus}, \citenamefont
  {Bucaro}, \citenamefont {Liu},\ and\ \citenamefont {Vignola}}]{houston}%
  \BibitemOpen
  \bibfield  {author} {\bibinfo {author} {\bibfnamefont {B.~H.}\ \bibnamefont
  {Houston}}, \bibinfo {author} {\bibfnamefont {D.~M.}\ \bibnamefont
  {Photiadis}}, \bibinfo {author} {\bibfnamefont {M.~H.}\ \bibnamefont
  {Marcus}}, \bibinfo {author} {\bibfnamefont {J.~A.}\ \bibnamefont {Bucaro}},
  \bibinfo {author} {\bibfnamefont {X.}~\bibnamefont {Liu}}, \ and\ \bibinfo
  {author} {\bibfnamefont {J.~F.}\ \bibnamefont {Vignola}},\ }\href {\doibase
  10.1063/1.1449534} {\bibfield  {journal} {\bibinfo  {journal} {Appl.\ Phys.\
  Lett.}\ }\textbf {\bibinfo {volume} {80}},\ \bibinfo {pages} {1300} (\bibinfo
  {year} {2002})}\BibitemShut {NoStop}%
\bibitem [{\citenamefont {De}\ and\ \citenamefont {Aluru}(2006)}]{sudipto}%
  \BibitemOpen
  \bibfield  {author} {\bibinfo {author} {\bibfnamefont {S.~K.}\ \bibnamefont
  {De}}\ and\ \bibinfo {author} {\bibfnamefont {N.~R.}\ \bibnamefont {Aluru}},\
  }\href {\doibase 10.1103/PhysRevB.74.144305} {\bibfield  {journal} {\bibinfo
  {journal} {Phys. Rev. B}\ }\textbf {\bibinfo {volume} {74}},\ \bibinfo {eid}
  {144305} (\bibinfo {year} {2006})}\BibitemShut {NoStop}%
\bibitem [{\citenamefont {Kiselev}\ and\ \citenamefont
  {Iafrate}(2008)}]{kiselev}%
  \BibitemOpen
  \bibfield  {author} {\bibinfo {author} {\bibfnamefont {A.~A.}\ \bibnamefont
  {Kiselev}}\ and\ \bibinfo {author} {\bibfnamefont {G.~J.}\ \bibnamefont
  {Iafrate}},\ }\href {\doibase 10.1103/PhysRevB.77.205436} {\bibfield
  {journal} {\bibinfo  {journal} {Phys. Rev. B}\ }\textbf {\bibinfo {volume}
  {77}},\ \bibinfo {pages} {205436} (\bibinfo {year} {2008})}\BibitemShut
  {NoStop}%
\bibitem [{\citenamefont {Cross}\ and\ \citenamefont {Lifshitz}(2001)}]{cross}%
  \BibitemOpen
  \bibfield  {author} {\bibinfo {author} {\bibfnamefont {M.~C.}\ \bibnamefont
  {Cross}}\ and\ \bibinfo {author} {\bibfnamefont {R.}~\bibnamefont
  {Lifshitz}},\ }\href {\doibase 10.1103/PhysRevB.64.085324} {\bibfield
  {journal} {\bibinfo  {journal} {Phys. Rev. B}\ }\textbf {\bibinfo {volume}
  {64}},\ \bibinfo {pages} {085324} (\bibinfo {year} {2001})}\BibitemShut
  {NoStop}%
\bibitem [{\citenamefont {Photiadis}\ and\ \citenamefont
  {Judge}(2004)}]{photiadis1}%
  \BibitemOpen
  \bibfield  {author} {\bibinfo {author} {\bibfnamefont {D.~M.}\ \bibnamefont
  {Photiadis}}\ and\ \bibinfo {author} {\bibfnamefont {J.~A.}\ \bibnamefont
  {Judge}},\ }\href {\doibase 10.1063/1.1773928} {\bibfield  {journal}
  {\bibinfo  {journal} {Appl.\ Phys.\ Lett.}\ }\textbf {\bibinfo {volume}
  {85}},\ \bibinfo {pages} {482} (\bibinfo {year} {2004})}\BibitemShut
  {NoStop}%
\bibitem [{\citenamefont {Patton}\ and\ \citenamefont
  {Geller}(2003)}]{geller3}%
  \BibitemOpen
  \bibfield  {author} {\bibinfo {author} {\bibfnamefont {K.~R.}\ \bibnamefont
  {Patton}}\ and\ \bibinfo {author} {\bibfnamefont {M.~R.}\ \bibnamefont
  {Geller}},\ }\href {\doibase 10.1103/PhysRevB.67.155418} {\bibfield
  {journal} {\bibinfo  {journal} {Phys.\ Rev.\ B}\ }\textbf {\bibinfo {volume}
  {67}},\ \bibinfo {pages} {155418} (\bibinfo {year} {2003})}\BibitemShut
  {NoStop}%
\bibitem [{\citenamefont {Chang}\ and\ \citenamefont {Geller}(2005)}]{geller1}%
  \BibitemOpen
  \bibfield  {author} {\bibinfo {author} {\bibfnamefont {C.-M.}\ \bibnamefont
  {Chang}}\ and\ \bibinfo {author} {\bibfnamefont {M.~R.}\ \bibnamefont
  {Geller}},\ }\href {\doibase 10.1103/PhysRevB.71.125304} {\bibfield
  {journal} {\bibinfo  {journal} {Phys. Rev. B}\ }\textbf {\bibinfo {volume}
  {71}},\ \bibinfo {pages} {125304} (\bibinfo {year} {2005})}\BibitemShut
  {NoStop}%
\bibitem [{\citenamefont {Geller}\ and\ \citenamefont
  {Varley}(2005)}]{geller2}%
  \BibitemOpen
  \bibfield  {author} {\bibinfo {author} {\bibfnamefont {M.~R.}\ \bibnamefont
  {Geller}}\ and\ \bibinfo {author} {\bibfnamefont {J.~B.}\ \bibnamefont
  {Varley}},\ }\href@noop {} {\enquote {\bibinfo {title} {Friction in
  nanoelectromechanical systems: Clamping loss in the {GHz} regime},}\ }
  (\bibinfo {year} {2005}),\ \bibinfo {note} {preprint},\ \Eprint
  {http://arxiv.org/abs/cond-mat/0512710.v1} {arXiv:cond-mat/0512710.v1}
  \BibitemShut {NoStop}%
\bibitem [{\citenamefont {Schmid}\ and\ \citenamefont
  {Hierold}(2008)}]{schmid08}%
  \BibitemOpen
  \bibfield  {author} {\bibinfo {author} {\bibfnamefont {S.}~\bibnamefont
  {Schmid}}\ and\ \bibinfo {author} {\bibfnamefont {C.}~\bibnamefont
  {Hierold}},\ }\href {\doibase 10.1063/1.3008032} {\bibfield  {journal}
  {\bibinfo  {journal} {J.\ Appl.\ Phys.}\ }\textbf {\bibinfo {volume} {104}},\
  \bibinfo {pages} {093516} (\bibinfo {year} {2008})}\BibitemShut {NoStop}%
\bibitem [{\citenamefont {Wilson-Rae}(2008)}]{wilson-rae}%
  \BibitemOpen
  \bibfield  {author} {\bibinfo {author} {\bibfnamefont {I.}~\bibnamefont
  {Wilson-Rae}},\ }\href {\doibase 10.1103/PhysRevB.77.245418} {\bibfield
  {journal} {\bibinfo  {journal} {Phys. Rev. B}\ }\textbf {\bibinfo {volume}
  {77}},\ \bibinfo {pages} {245418} (\bibinfo {year} {2008})}\BibitemShut
  {NoStop}%
\bibitem [{\citenamefont {Cole}\ \emph {et~al.}(2011)\citenamefont {Cole},
  \citenamefont {Wilson-Rae}, \citenamefont {Werbach}, \citenamefont {Vanner},\
  and\ \citenamefont {Aspelmeyer}}]{cole}%
  \BibitemOpen
  \bibfield  {author} {\bibinfo {author} {\bibfnamefont {G.~D.}\ \bibnamefont
  {Cole}}, \bibinfo {author} {\bibfnamefont {I.}~\bibnamefont {Wilson-Rae}},
  \bibinfo {author} {\bibfnamefont {K.}~\bibnamefont {Werbach}}, \bibinfo
  {author} {\bibfnamefont {M.~R.}\ \bibnamefont {Vanner}}, \ and\ \bibinfo
  {author} {\bibfnamefont {M.}~\bibnamefont {Aspelmeyer}},\ }\href {\doibase
  doi: 10.1038/ncomms1212} {\bibfield  {journal} {\bibinfo  {journal} {Nature
  Communications}\ }\textbf {\bibinfo {volume} {2}},\ \bibinfo {pages} {231}
  (\bibinfo {year} {2011})}\BibitemShut {NoStop}%
\bibitem [{\citenamefont {Husain}\ \emph {et~al.}(2003)\citenamefont {Husain},
  \citenamefont {Hone}, \citenamefont {Postma}, \citenamefont {Huang},
  \citenamefont {Drake}, \citenamefont {Barbic}, \citenamefont {Scherer},\ and\
  \citenamefont {Roukes}}]{husain}%
  \BibitemOpen
  \bibfield  {author} {\bibinfo {author} {\bibfnamefont {A.}~\bibnamefont
  {Husain}}, \bibinfo {author} {\bibfnamefont {J.}~\bibnamefont {Hone}},
  \bibinfo {author} {\bibfnamefont {H.~W.~C.}\ \bibnamefont {Postma}}, \bibinfo
  {author} {\bibfnamefont {X.~M.~H.}\ \bibnamefont {Huang}}, \bibinfo {author}
  {\bibfnamefont {T.}~\bibnamefont {Drake}}, \bibinfo {author} {\bibfnamefont
  {M.}~\bibnamefont {Barbic}}, \bibinfo {author} {\bibfnamefont
  {A.}~\bibnamefont {Scherer}}, \ and\ \bibinfo {author} {\bibfnamefont
  {M.~L.}\ \bibnamefont {Roukes}},\ }\href {\doibase 10.1063/1.1601311}
  {\bibfield  {journal} {\bibinfo  {journal} {Appl.\ Phys.\ Lett.}\ }\textbf
  {\bibinfo {volume} {83}},\ \bibinfo {pages} {1240} (\bibinfo {year}
  {2003})}\BibitemShut {NoStop}%
\bibitem [{\citenamefont {Buks}\ and\ \citenamefont {Roukes}(2002)}]{buks02}%
  \BibitemOpen
  \bibfield  {author} {\bibinfo {author} {\bibfnamefont {E.}~\bibnamefont
  {Buks}}\ and\ \bibinfo {author} {\bibfnamefont {M.~L.}\ \bibnamefont
  {Roukes}},\ }\href {\doibase 10.1109/JMEMS.2002.805056} {\bibfield  {journal}
  {\bibinfo  {journal} {J. Microelectromech. Syst.}\ }\textbf {\bibinfo
  {volume} {11}},\ \bibinfo {pages} {802} (\bibinfo {year} {2002})}\BibitemShut
  {NoStop}%
\bibitem [{\citenamefont {Venkatesan}\ \emph {et~al.}(2010)\citenamefont
  {Venkatesan}, \citenamefont {Lulla}, \citenamefont {Patton}, \citenamefont
  {Armour}, \citenamefont {Mellor},\ and\ \citenamefont
  {Owers-Bradley}}]{venkatesan10}%
  \BibitemOpen
  \bibfield  {author} {\bibinfo {author} {\bibfnamefont {A.}~\bibnamefont
  {Venkatesan}}, \bibinfo {author} {\bibfnamefont {K.~J.}\ \bibnamefont
  {Lulla}}, \bibinfo {author} {\bibfnamefont {M.~J.}\ \bibnamefont {Patton}},
  \bibinfo {author} {\bibfnamefont {A.~D.}\ \bibnamefont {Armour}}, \bibinfo
  {author} {\bibfnamefont {C.~J.}\ \bibnamefont {Mellor}}, \ and\ \bibinfo
  {author} {\bibfnamefont {J.~R.}\ \bibnamefont {Owers-Bradley}},\ }\href
  {\doibase 10.1103/PhysRevB.81.073410} {\bibfield  {journal} {\bibinfo
  {journal} {Phys. Rev. B}\ }\textbf {\bibinfo {volume} {81}},\ \bibinfo
  {pages} {073410} (\bibinfo {year} {2010})}\BibitemShut {NoStop}%
\bibitem [{\citenamefont {Davis}\ and\ \citenamefont {Boisen}(2005)}]{davis}%
  \BibitemOpen
  \bibfield  {author} {\bibinfo {author} {\bibfnamefont {Z.~J.}\ \bibnamefont
  {Davis}}\ and\ \bibinfo {author} {\bibfnamefont {A.}~\bibnamefont {Boisen}},\
  }\href {\doibase 10.1063/1.1984092} {\bibfield  {journal} {\bibinfo
  {journal} {Appl.\ Phys.\ Lett.}\ }\textbf {\bibinfo {volume} {87}},\ \bibinfo
  {pages} {013102} (\bibinfo {year} {2005})}\BibitemShut {NoStop}%
\bibitem [{\citenamefont {Li}\ \emph {et~al.}(2008)\citenamefont {Li},
  \citenamefont {Pashkin}, \citenamefont {Astafiev}, \citenamefont {Nakamura},
  \citenamefont {Tsai},\ and\ \citenamefont {Im}}]{li}%
  \BibitemOpen
  \bibfield  {author} {\bibinfo {author} {\bibfnamefont {T.~F.}\ \bibnamefont
  {Li}}, \bibinfo {author} {\bibfnamefont {Y.~A.}\ \bibnamefont {Pashkin}},
  \bibinfo {author} {\bibfnamefont {O.}~\bibnamefont {Astafiev}}, \bibinfo
  {author} {\bibfnamefont {Y.}~\bibnamefont {Nakamura}}, \bibinfo {author}
  {\bibfnamefont {J.~S.}\ \bibnamefont {Tsai}}, \ and\ \bibinfo {author}
  {\bibfnamefont {H.}~\bibnamefont {Im}},\ }\href {\doibase 10.1063/1.2838749}
  {\bibfield  {journal} {\bibinfo  {journal} {Appl.\ Phys.\ Lett.}\ }\textbf
  {\bibinfo {volume} {92}},\ \bibinfo {pages} {043112} (\bibinfo {year}
  {2008})}\BibitemShut {NoStop}%
\bibitem [{\citenamefont {Hoehne}\ \emph {et~al.}(2010)\citenamefont {Hoehne},
  \citenamefont {Pashkin}, \citenamefont {Astafiev}, \citenamefont {Faoro},
  \citenamefont {Ioffe}, \citenamefont {Nakamura},\ and\ \citenamefont
  {Tsai}}]{hoehne}%
  \BibitemOpen
  \bibfield  {author} {\bibinfo {author} {\bibfnamefont {F.}~\bibnamefont
  {Hoehne}}, \bibinfo {author} {\bibfnamefont {Y.~A.}\ \bibnamefont {Pashkin}},
  \bibinfo {author} {\bibfnamefont {O.}~\bibnamefont {Astafiev}}, \bibinfo
  {author} {\bibfnamefont {L.}~\bibnamefont {Faoro}}, \bibinfo {author}
  {\bibfnamefont {L.~B.}\ \bibnamefont {Ioffe}}, \bibinfo {author}
  {\bibfnamefont {Y.}~\bibnamefont {Nakamura}}, \ and\ \bibinfo {author}
  {\bibfnamefont {J.~S.}\ \bibnamefont {Tsai}},\ }\href {\doibase
  10.1103/PhysRevB.81.184112} {\bibfield  {journal} {\bibinfo  {journal} {Phys.
  Rev. B}\ }\textbf {\bibinfo {volume} {81}},\ \bibinfo {pages} {184112}
  (\bibinfo {year} {2010})}\BibitemShut {NoStop}%
\bibitem [{\citenamefont {Teufel}\ \emph {et~al.}(2008)\citenamefont {Teufel},
  \citenamefont {Regal},\ and\ \citenamefont {Lehnert}}]{teufel}%
  \BibitemOpen
  \bibfield  {author} {\bibinfo {author} {\bibfnamefont {J.~D.}\ \bibnamefont
  {Teufel}}, \bibinfo {author} {\bibfnamefont {C.~A.}\ \bibnamefont {Regal}}, \
  and\ \bibinfo {author} {\bibfnamefont {K.~W.}\ \bibnamefont {Lehnert}},\
  }\href {\doibase 10.1088/1367-2630/10/9/095002} {\bibfield  {journal}
  {\bibinfo  {journal} {New J.\ Phys.}\ }\textbf {\bibinfo {volume} {10}},\
  \bibinfo {pages} {095002} (\bibinfo {year} {2008})}\BibitemShut {NoStop}%
\bibitem [{\citenamefont {Peng}\ \emph {et~al.}(2006)\citenamefont {Peng},
  \citenamefont {Chang}, \citenamefont {Aloni}, \citenamefont {Yuzvinsky},\
  and\ \citenamefont {Zettl}}]{peng06}%
  \BibitemOpen
  \bibfield  {author} {\bibinfo {author} {\bibfnamefont {H.~B.}\ \bibnamefont
  {Peng}}, \bibinfo {author} {\bibfnamefont {C.~W.}\ \bibnamefont {Chang}},
  \bibinfo {author} {\bibfnamefont {S.}~\bibnamefont {Aloni}}, \bibinfo
  {author} {\bibfnamefont {T.~D.}\ \bibnamefont {Yuzvinsky}}, \ and\ \bibinfo
  {author} {\bibfnamefont {A.}~\bibnamefont {Zettl}},\ }\href {\doibase
  10.1103/PhysRevLett.97.087203} {\bibfield  {journal} {\bibinfo  {journal}
  {Phys. Rev. Lett.}\ }\textbf {\bibinfo {volume} {97}},\ \bibinfo {pages}
  {087203} (\bibinfo {year} {2006})}\BibitemShut {NoStop}%
\bibitem [{\citenamefont {Eriksson}\ \emph {et~al.}(2008)\citenamefont
  {Eriksson}, \citenamefont {Lee}, \citenamefont {Sourab}, \citenamefont
  {Isacsson}, \citenamefont {Kaunisto}, \citenamefont {Kinaret},\ and\
  \citenamefont {Campbell}}]{Eriksson08}%
  \BibitemOpen
  \bibfield  {author} {\bibinfo {author} {\bibfnamefont {A.}~\bibnamefont
  {Eriksson}}, \bibinfo {author} {\bibfnamefont {S.}~\bibnamefont {Lee}},
  \bibinfo {author} {\bibfnamefont {A.~A.}\ \bibnamefont {Sourab}}, \bibinfo
  {author} {\bibfnamefont {A.}~\bibnamefont {Isacsson}}, \bibinfo {author}
  {\bibfnamefont {R.}~\bibnamefont {Kaunisto}}, \bibinfo {author}
  {\bibfnamefont {J.~M.}\ \bibnamefont {Kinaret}}, \ and\ \bibinfo {author}
  {\bibfnamefont {E.~E.~B.}\ \bibnamefont {Campbell}},\ }\href {\doibase
  10.1021/nl080345w} {\bibfield  {journal} {\bibinfo  {journal} {Nano Letters}\
  }\textbf {\bibinfo {volume} {8}},\ \bibinfo {pages} {1224} (\bibinfo {year}
  {2008})}\BibitemShut {NoStop}%
\bibitem [{\citenamefont {Hu}\ \emph {et~al.}(2003)\citenamefont {Hu},
  \citenamefont {Wang}, \citenamefont {Hartland}, \citenamefont {Mulvaney},
  \citenamefont {Juste},\ and\ \citenamefont {Sader}}]{min}%
  \BibitemOpen
  \bibfield  {author} {\bibinfo {author} {\bibfnamefont {M.}~\bibnamefont
  {Hu}}, \bibinfo {author} {\bibfnamefont {X.}~\bibnamefont {Wang}}, \bibinfo
  {author} {\bibfnamefont {G.~V.}\ \bibnamefont {Hartland}}, \bibinfo {author}
  {\bibfnamefont {P.}~\bibnamefont {Mulvaney}}, \bibinfo {author}
  {\bibfnamefont {J.~P.}\ \bibnamefont {Juste}}, \ and\ \bibinfo {author}
  {\bibfnamefont {J.~E.}\ \bibnamefont {Sader}},\ }\href {\doibase
  10.1021/ja037443y} {\bibfield  {journal} {\bibinfo  {journal} {J.\ Am.\
  Chem.\ Soc.}\ }\textbf {\bibinfo {volume} {125}},\ \bibinfo {pages} {14925}
  (\bibinfo {year} {2003})}\BibitemShut {NoStop}%
\bibitem [{\citenamefont {Pelton}\ \emph {et~al.}(2009)\citenamefont {Pelton},
  \citenamefont {Sader}, \citenamefont {Burgin}, \citenamefont {Liu},
  \citenamefont {Guyot-Sionnest},\ and\ \citenamefont {Gosztola}}]{pelton}%
  \BibitemOpen
  \bibfield  {author} {\bibinfo {author} {\bibfnamefont {M.}~\bibnamefont
  {Pelton}}, \bibinfo {author} {\bibfnamefont {J.~E.}\ \bibnamefont {Sader}},
  \bibinfo {author} {\bibfnamefont {J.}~\bibnamefont {Burgin}}, \bibinfo
  {author} {\bibfnamefont {M.}~\bibnamefont {Liu}}, \bibinfo {author}
  {\bibfnamefont {P.}~\bibnamefont {Guyot-Sionnest}}, \ and\ \bibinfo {author}
  {\bibfnamefont {D.}~\bibnamefont {Gosztola}},\ }\href {\doibase
  10.1038/nnano.2009.192} {\bibfield  {journal} {\bibinfo  {journal} {Nature
  Nanotechnology}\ }\textbf {\bibinfo {volume} {4}},\ \bibinfo {pages} {492}
  (\bibinfo {year} {2009})}\BibitemShut {NoStop}%
\bibitem [{\citenamefont {Zijlstra}\ \emph {et~al.}(2008)\citenamefont
  {Zijlstra}, \citenamefont {Tchebotareva}, \citenamefont {Chon}, \citenamefont
  {Gu},\ and\ \citenamefont {Orrit}}]{zijlstra}%
  \BibitemOpen
  \bibfield  {author} {\bibinfo {author} {\bibfnamefont {P.}~\bibnamefont
  {Zijlstra}}, \bibinfo {author} {\bibfnamefont {A.~L.}\ \bibnamefont
  {Tchebotareva}}, \bibinfo {author} {\bibfnamefont {J.~W.~M.}\ \bibnamefont
  {Chon}}, \bibinfo {author} {\bibfnamefont {M.}~\bibnamefont {Gu}}, \ and\
  \bibinfo {author} {\bibfnamefont {M.}~\bibnamefont {Orrit}},\ }\href
  {\doibase 10.1021/nl802480q} {\bibfield  {journal} {\bibinfo  {journal} {Nano
  Lett.}\ }\textbf {\bibinfo {volume} {8}},\ \bibinfo {pages} {3493} (\bibinfo
  {year} {2008})}\BibitemShut {NoStop}%
\bibitem [{\citenamefont {B\"{o}mmel}(1954)}]{Bommel54}%
  \BibitemOpen
  \bibfield  {author} {\bibinfo {author} {\bibfnamefont {H.~E.}\ \bibnamefont
  {B\"{o}mmel}},\ }\href {\doibase 10.1103/PhysRev.96.220} {\bibfield
  {journal} {\bibinfo  {journal} {Phys. Rev.}\ }\textbf {\bibinfo {volume}
  {96}},\ \bibinfo {pages} {220} (\bibinfo {year} {1954})}\BibitemShut
  {NoStop}%
\bibitem [{\citenamefont {Kittel}(1955)}]{kittel}%
  \BibitemOpen
  \bibfield  {author} {\bibinfo {author} {\bibfnamefont {C.}~\bibnamefont
  {Kittel}},\ }\href {\doibase 10.1016/0001-6160(55)90070-5} {\bibfield
  {journal} {\bibinfo  {journal} {Acta Metallurgica}\ }\textbf {\bibinfo
  {volume} {3}},\ \bibinfo {pages} {295 } (\bibinfo {year} {1955})}\BibitemShut
  {NoStop}%
\bibitem [{\citenamefont {Pippard}(1955)}]{pippard}%
  \BibitemOpen
  \bibfield  {author} {\bibinfo {author} {\bibfnamefont {A.~B.}\ \bibnamefont
  {Pippard}},\ }\href {\doibase 10.1080/14786441008521122} {\bibfield
  {journal} {\bibinfo  {journal} {Phil.\ Mag.}\ }\textbf {\bibinfo {volume}
  {46}},\ \bibinfo {pages} {1104} (\bibinfo {year} {1955})}\BibitemShut
  {NoStop}%
\bibitem [{\citenamefont {Blount}(1959)}]{blount}%
  \BibitemOpen
  \bibfield  {author} {\bibinfo {author} {\bibfnamefont {E.~I.}\ \bibnamefont
  {Blount}},\ }\href {\doibase 10.1103/PhysRev.114.418} {\bibfield  {journal}
  {\bibinfo  {journal} {Phys.\ Rev.}\ }\textbf {\bibinfo {volume} {114}},\
  \bibinfo {pages} {418} (\bibinfo {year} {1959})}\BibitemShut {NoStop}%
\bibitem [{\citenamefont {Ziman}(1960)}]{ziman}%
  \BibitemOpen
  \bibfield  {author} {\bibinfo {author} {\bibfnamefont {J.~M.}\ \bibnamefont
  {Ziman}},\ }\href@noop {} {\emph {\bibinfo {title} {Elecrtrons and
  Phonons}}}\ (\bibinfo  {publisher} {Oxford University Press},\ \bibinfo
  {address} {Oxford},\ \bibinfo {year} {1960})\BibitemShut {NoStop}%
\bibitem [{\citenamefont {Kokkedee}(1962)}]{kokkedee}%
  \BibitemOpen
  \bibfield  {author} {\bibinfo {author} {\bibfnamefont {J.~J.~J.}\
  \bibnamefont {Kokkedee}},\ }\href {\doibase 10.1016/0031-8914(62)90077-0}
  {\bibfield  {journal} {\bibinfo  {journal} {Physica}\ }\textbf {\bibinfo
  {volume} {28}},\ \bibinfo {pages} {893} (\bibinfo {year} {1962})}\BibitemShut
  {NoStop}%
\bibitem [{\citenamefont {Weig}\ \emph {et~al.}(2004)\citenamefont {Weig},
  \citenamefont {Blick}, \citenamefont {Brandes}, \citenamefont {Kirschbaum},
  \citenamefont {Wegscheider}, \citenamefont {Bichler},\ and\ \citenamefont
  {Kotthaus}}]{weig04}%
  \BibitemOpen
  \bibfield  {author} {\bibinfo {author} {\bibfnamefont {E.~M.}\ \bibnamefont
  {Weig}}, \bibinfo {author} {\bibfnamefont {R.~H.}\ \bibnamefont {Blick}},
  \bibinfo {author} {\bibfnamefont {T.}~\bibnamefont {Brandes}}, \bibinfo
  {author} {\bibfnamefont {J.}~\bibnamefont {Kirschbaum}}, \bibinfo {author}
  {\bibfnamefont {W.}~\bibnamefont {Wegscheider}}, \bibinfo {author}
  {\bibfnamefont {M.}~\bibnamefont {Bichler}}, \ and\ \bibinfo {author}
  {\bibfnamefont {J.~P.}\ \bibnamefont {Kotthaus}},\ }\href {\doibase
  10.1103/PhysRevLett.92.046804} {\bibfield  {journal} {\bibinfo  {journal}
  {Phys. Rev. Lett.}\ }\textbf {\bibinfo {volume} {92}},\ \bibinfo {pages}
  {046804} (\bibinfo {year} {2004})}\BibitemShut {NoStop}%
\bibitem [{\citenamefont {LeRoy}\ \emph {et~al.}(2004)\citenamefont {LeRoy},
  \citenamefont {Lemay}, \citenamefont {Kong},\ and\ \citenamefont
  {Dekker}}]{leroy04}%
  \BibitemOpen
  \bibfield  {author} {\bibinfo {author} {\bibfnamefont {B.~J.}\ \bibnamefont
  {LeRoy}}, \bibinfo {author} {\bibfnamefont {S.~G.}\ \bibnamefont {Lemay}},
  \bibinfo {author} {\bibfnamefont {J.}~\bibnamefont {Kong}}, \ and\ \bibinfo
  {author} {\bibfnamefont {C.}~\bibnamefont {Dekker}},\ }\href {\doibase
  10.1038/nature03046} {\bibfield  {journal} {\bibinfo  {journal} {Nature}\
  }\textbf {\bibinfo {volume} {432}},\ \bibinfo {pages} {371} (\bibinfo {year}
  {2004})}\BibitemShut {NoStop}%
\bibitem [{\citenamefont {Leturcq}\ \emph {et~al.}(2009)\citenamefont
  {Leturcq}, \citenamefont {Stampfer}, \citenamefont {Inderbitzin},
  \citenamefont {Durrer}, \citenamefont {Hierold}, \citenamefont {Mariani},
  \citenamefont {Schultz}, \citenamefont {von Oppen},\ and\ \citenamefont
  {Ensslin}}]{leturcq09}%
  \BibitemOpen
  \bibfield  {author} {\bibinfo {author} {\bibfnamefont {R.}~\bibnamefont
  {Leturcq}}, \bibinfo {author} {\bibfnamefont {C.}~\bibnamefont {Stampfer}},
  \bibinfo {author} {\bibfnamefont {K.}~\bibnamefont {Inderbitzin}}, \bibinfo
  {author} {\bibfnamefont {L.}~\bibnamefont {Durrer}}, \bibinfo {author}
  {\bibfnamefont {C.}~\bibnamefont {Hierold}}, \bibinfo {author} {\bibfnamefont
  {E.}~\bibnamefont {Mariani}}, \bibinfo {author} {\bibfnamefont {M.~G.}\
  \bibnamefont {Schultz}}, \bibinfo {author} {\bibfnamefont {F.}~\bibnamefont
  {von Oppen}}, \ and\ \bibinfo {author} {\bibfnamefont {K.}~\bibnamefont
  {Ensslin}},\ }\href {\doibase 10.1038/nphys1234} {\bibfield  {journal}
  {\bibinfo  {journal} {Nature Physics}\ }\textbf {\bibinfo {volume} {5}},\
  \bibinfo {pages} {327} (\bibinfo {year} {2009})}\BibitemShut {NoStop}%
\bibitem [{\citenamefont {Mariani}\ and\ \citenamefont {von
  Oppen}(2009)}]{mariani}%
  \BibitemOpen
  \bibfield  {author} {\bibinfo {author} {\bibfnamefont {E.}~\bibnamefont
  {Mariani}}\ and\ \bibinfo {author} {\bibfnamefont {F.}~\bibnamefont {von
  Oppen}},\ }\href {\doibase 10.1103/PhysRevB.80.155411} {\bibfield  {journal}
  {\bibinfo  {journal} {Phys. Rev. B}\ }\textbf {\bibinfo {volume} {80}},\
  \bibinfo {pages} {155411} (\bibinfo {year} {2009})}\BibitemShut {NoStop}%
\bibitem [{\citenamefont {Steele}\ \emph {et~al.}(2009)\citenamefont {Steele},
  \citenamefont {H\"{u}ttel}, \citenamefont {Witkamp}, \citenamefont {Poot},
  \citenamefont {Meerwaldt}, \citenamefont {Kouwenhoven},\ and\ \citenamefont
  {van~der Zant}}]{Steele09}%
  \BibitemOpen
  \bibfield  {author} {\bibinfo {author} {\bibfnamefont {G.~A.}\ \bibnamefont
  {Steele}}, \bibinfo {author} {\bibfnamefont {A.~K.}\ \bibnamefont
  {H\"{u}ttel}}, \bibinfo {author} {\bibfnamefont {B.}~\bibnamefont {Witkamp}},
  \bibinfo {author} {\bibfnamefont {M.}~\bibnamefont {Poot}}, \bibinfo {author}
  {\bibfnamefont {H.~B.}\ \bibnamefont {Meerwaldt}}, \bibinfo {author}
  {\bibfnamefont {L.~P.}\ \bibnamefont {Kouwenhoven}}, \ and\ \bibinfo {author}
  {\bibfnamefont {H.~S.~J.}\ \bibnamefont {van~der Zant}},\ }\href {\doibase
  10.1126/science.1176076} {\bibfield  {journal} {\bibinfo  {journal}
  {Science}\ }\textbf {\bibinfo {volume} {325}},\ \bibinfo {pages} {1103}
  (\bibinfo {year} {2009})}\BibitemShut {NoStop}%
\bibitem [{\citenamefont {H\"{u}ttel}\ \emph {et~al.}(2009)\citenamefont
  {H\"{u}ttel}, \citenamefont {Steele}, \citenamefont {Witkamp}, \citenamefont
  {Poot}, \citenamefont {Kouwenhoven},\ and\ \citenamefont {van~der
  Zant}}]{huttel09}%
  \BibitemOpen
  \bibfield  {author} {\bibinfo {author} {\bibfnamefont {A.~K.}\ \bibnamefont
  {H\"{u}ttel}}, \bibinfo {author} {\bibfnamefont {G.~A.}\ \bibnamefont
  {Steele}}, \bibinfo {author} {\bibfnamefont {B.}~\bibnamefont {Witkamp}},
  \bibinfo {author} {\bibfnamefont {M.}~\bibnamefont {Poot}}, \bibinfo {author}
  {\bibfnamefont {L.~P.}\ \bibnamefont {Kouwenhoven}}, \ and\ \bibinfo {author}
  {\bibfnamefont {H.~S.~J.}\ \bibnamefont {van~der Zant}},\ }\href {\doibase
  10.1021/nl900612h} {\bibfield  {journal} {\bibinfo  {journal} {Nano letters}\
  }\textbf {\bibinfo {volume} {9}},\ \bibinfo {pages} {2547} (\bibinfo {year}
  {2009})}\BibitemShut {NoStop}%
\bibitem [{\citenamefont {Laird}\ \emph {et~al.}(2012)\citenamefont {Laird},
  \citenamefont {Pei}, \citenamefont {Tang}, \citenamefont {Steele},\ and\
  \citenamefont {Kouwenhoven}}]{Laird12}%
  \BibitemOpen
  \bibfield  {author} {\bibinfo {author} {\bibfnamefont {E.~A.}\ \bibnamefont
  {Laird}}, \bibinfo {author} {\bibfnamefont {F.}~\bibnamefont {Pei}}, \bibinfo
  {author} {\bibfnamefont {W.}~\bibnamefont {Tang}}, \bibinfo {author}
  {\bibfnamefont {G.~A.}\ \bibnamefont {Steele}}, \ and\ \bibinfo {author}
  {\bibfnamefont {L.~P.}\ \bibnamefont {Kouwenhoven}},\ }\href {\doibase
  10.1021/nl203279v} {\bibfield  {journal} {\bibinfo  {journal} {Nano Letters}\
  }\textbf {\bibinfo {volume} {12}},\ \bibinfo {pages} {193} (\bibinfo {year}
  {2012})}\BibitemShut {NoStop}%
\bibitem [{\citenamefont {Park}\ \emph {et~al.}(2000)\citenamefont {Park},
  \citenamefont {Park}, \citenamefont {Lim}, \citenamefont {Anderson},
  \citenamefont {Alivisatos}, \citenamefont {McEuen} \emph {et~al.}}]{park}%
  \BibitemOpen
  \bibfield  {author} {\bibinfo {author} {\bibfnamefont {H.}~\bibnamefont
  {Park}}, \bibinfo {author} {\bibfnamefont {J.}~\bibnamefont {Park}}, \bibinfo
  {author} {\bibfnamefont {A.}~\bibnamefont {Lim}}, \bibinfo {author}
  {\bibfnamefont {E.}~\bibnamefont {Anderson}}, \bibinfo {author}
  {\bibfnamefont {A.}~\bibnamefont {Alivisatos}}, \bibinfo {author}
  {\bibfnamefont {P.}~\bibnamefont {McEuen}},  \emph {et~al.},\ }\href
  {\doibase 10.1038/35024031} {\bibfield  {journal} {\bibinfo  {journal}
  {Nature}\ }\textbf {\bibinfo {volume} {407}},\ \bibinfo {pages} {57}
  (\bibinfo {year} {2000})}\BibitemShut {NoStop}%
\bibitem [{\citenamefont {Paulsson}\ \emph {et~al.}(2005)\citenamefont
  {Paulsson}, \citenamefont {Frederiksen},\ and\ \citenamefont
  {Brandbyge}}]{paulsson}%
  \BibitemOpen
  \bibfield  {author} {\bibinfo {author} {\bibfnamefont {M.}~\bibnamefont
  {Paulsson}}, \bibinfo {author} {\bibfnamefont {T.}~\bibnamefont
  {Frederiksen}}, \ and\ \bibinfo {author} {\bibfnamefont {M.}~\bibnamefont
  {Brandbyge}},\ }\href {\doibase 10.1103/PhysRevB.72.201101} {\bibfield
  {journal} {\bibinfo  {journal} {Phys. Rev. B}\ }\textbf {\bibinfo {volume}
  {72}},\ \bibinfo {pages} {201101} (\bibinfo {year} {2005})}\BibitemShut
  {NoStop}%
\bibitem [{\citenamefont {Viljas}\ \emph {et~al.}(2005)\citenamefont {Viljas},
  \citenamefont {Cuevas}, \citenamefont {Pauly},\ and\ \citenamefont
  {H\"afner}}]{viljas}%
  \BibitemOpen
  \bibfield  {author} {\bibinfo {author} {\bibfnamefont {J.~K.}\ \bibnamefont
  {Viljas}}, \bibinfo {author} {\bibfnamefont {J.~C.}\ \bibnamefont {Cuevas}},
  \bibinfo {author} {\bibfnamefont {F.}~\bibnamefont {Pauly}}, \ and\ \bibinfo
  {author} {\bibfnamefont {M.}~\bibnamefont {H\"afner}},\ }\href {\doibase
  10.1103/PhysRevB.72.245415} {\bibfield  {journal} {\bibinfo  {journal} {Phys.
  Rev. B}\ }\textbf {\bibinfo {volume} {72}},\ \bibinfo {pages} {245415}
  (\bibinfo {year} {2005})}\BibitemShut {NoStop}%
\bibitem [{\citenamefont {de~la Vega}\ \emph {et~al.}(2006)\citenamefont {de~la
  Vega}, \citenamefont {Martin-Rodero}, \citenamefont {Agra\"{i}t},\ and\
  \citenamefont {Yeyati}}]{vega}%
  \BibitemOpen
  \bibfield  {author} {\bibinfo {author} {\bibfnamefont {L.}~\bibnamefont
  {de~la Vega}}, \bibinfo {author} {\bibfnamefont {A.}~\bibnamefont
  {Martin-Rodero}}, \bibinfo {author} {\bibfnamefont {N.}~\bibnamefont
  {Agra\"{i}t}}, \ and\ \bibinfo {author} {\bibfnamefont {A.~L.}\ \bibnamefont
  {Yeyati}},\ }\href {\doibase 10.1103/PhysRevB.73.075428} {\bibfield
  {journal} {\bibinfo  {journal} {Phys. Rev. B}\ }\textbf {\bibinfo {volume}
  {73}},\ \bibinfo {pages} {075428} (\bibinfo {year} {2006})}\BibitemShut
  {NoStop}%
\bibitem [{\citenamefont {Pecchia}\ \emph {et~al.}(2004)\citenamefont
  {Pecchia}, \citenamefont {Carlo}, \citenamefont {Gagliardi}, \citenamefont
  {Sanna}, \citenamefont {Frauenheim},\ and\ \citenamefont
  {Guti{\'e}rrez}}]{pecchia}%
  \BibitemOpen
  \bibfield  {author} {\bibinfo {author} {\bibfnamefont {A.}~\bibnamefont
  {Pecchia}}, \bibinfo {author} {\bibfnamefont {A.~D.}\ \bibnamefont {Carlo}},
  \bibinfo {author} {\bibfnamefont {A.}~\bibnamefont {Gagliardi}}, \bibinfo
  {author} {\bibfnamefont {S.}~\bibnamefont {Sanna}}, \bibinfo {author}
  {\bibfnamefont {T.}~\bibnamefont {Frauenheim}}, \ and\ \bibinfo {author}
  {\bibfnamefont {R.}~\bibnamefont {Guti{\'e}rrez}},\ }\href {\doibase
  10.1021/nl048841h} {\bibfield  {journal} {\bibinfo  {journal} {Nano Lett.}\
  }\textbf {\bibinfo {volume} {4}},\ \bibinfo {pages} {2109} (\bibinfo {year}
  {2004})}\BibitemShut {NoStop}%
\bibitem [{\citenamefont {Kushmerick}\ \emph {et~al.}(2004)\citenamefont
  {Kushmerick}, \citenamefont {Lazorcik}, \citenamefont {Patterson},
  \citenamefont {Shashidhar}, \citenamefont {Seferos},\ and\ \citenamefont
  {Bazan}}]{kushmerick}%
  \BibitemOpen
  \bibfield  {author} {\bibinfo {author} {\bibfnamefont {J.~G.}\ \bibnamefont
  {Kushmerick}}, \bibinfo {author} {\bibfnamefont {J.}~\bibnamefont
  {Lazorcik}}, \bibinfo {author} {\bibfnamefont {C.~H.}\ \bibnamefont
  {Patterson}}, \bibinfo {author} {\bibfnamefont {R.}~\bibnamefont
  {Shashidhar}}, \bibinfo {author} {\bibfnamefont {D.~S.}\ \bibnamefont
  {Seferos}}, \ and\ \bibinfo {author} {\bibfnamefont {G.~C.}\ \bibnamefont
  {Bazan}},\ }\href {\doibase 10.1021/nl049871n} {\bibfield  {journal}
  {\bibinfo  {journal} {Nano Lett.}\ }\textbf {\bibinfo {volume} {4}},\
  \bibinfo {pages} {639} (\bibinfo {year} {2004})}\BibitemShut {NoStop}%
\bibitem [{\citenamefont {Wang}\ \emph {et~al.}(2004)\citenamefont {Wang},
  \citenamefont {Lee}, \citenamefont {Kretzschmar},\ and\ \citenamefont
  {Reed}}]{wang}%
  \BibitemOpen
  \bibfield  {author} {\bibinfo {author} {\bibfnamefont {W.}~\bibnamefont
  {Wang}}, \bibinfo {author} {\bibfnamefont {T.}~\bibnamefont {Lee}}, \bibinfo
  {author} {\bibfnamefont {I.}~\bibnamefont {Kretzschmar}}, \ and\ \bibinfo
  {author} {\bibfnamefont {M.~A.}\ \bibnamefont {Reed}},\ }\href {\doibase
  10.1021/nl049870v} {\bibfield  {journal} {\bibinfo  {journal} {Nano Lett.}\
  }\textbf {\bibinfo {volume} {4}},\ \bibinfo {pages} {643} (\bibinfo {year}
  {2004})}\BibitemShut {NoStop}%
\bibitem [{\citenamefont {Galperin}\ \emph {et~al.}(2007)\citenamefont
  {Galperin}, \citenamefont {Ratner},\ and\ \citenamefont {Nitzan}}]{galperin}%
  \BibitemOpen
  \bibfield  {author} {\bibinfo {author} {\bibfnamefont {M.}~\bibnamefont
  {Galperin}}, \bibinfo {author} {\bibfnamefont {M.~A.}\ \bibnamefont
  {Ratner}}, \ and\ \bibinfo {author} {\bibfnamefont {A.}~\bibnamefont
  {Nitzan}},\ }\href {\doibase 10.1088/0953-8984/19/10/103201} {\bibfield
  {journal} {\bibinfo  {journal} {J.\ Phys.: Condens.\ Matter}\ }\textbf
  {\bibinfo {volume} {19}},\ \bibinfo {pages} {103201} (\bibinfo {year}
  {2007})}\BibitemShut {NoStop}%
\bibitem [{\citenamefont {Tal}\ \emph {et~al.}(2008)\citenamefont {Tal},
  \citenamefont {Krieger}, \citenamefont {Leerink},\ and\ \citenamefont {van
  Ruitenbeek}}]{Tal08}%
  \BibitemOpen
  \bibfield  {author} {\bibinfo {author} {\bibfnamefont {O.}~\bibnamefont
  {Tal}}, \bibinfo {author} {\bibfnamefont {M.}~\bibnamefont {Krieger}},
  \bibinfo {author} {\bibfnamefont {B.}~\bibnamefont {Leerink}}, \ and\
  \bibinfo {author} {\bibfnamefont {J.~M.}\ \bibnamefont {van Ruitenbeek}},\
  }\href {\doibase 10.1103/PhysRevLett.100.196804} {\bibfield  {journal}
  {\bibinfo  {journal} {Phys. Rev. Lett.}\ }\textbf {\bibinfo {volume} {100}},\
  \bibinfo {pages} {196804} (\bibinfo {year} {2008})}\BibitemShut {NoStop}%
\bibitem [{\citenamefont {Fon}\ \emph {et~al.}(2002)\citenamefont {Fon},
  \citenamefont {Schwab}, \citenamefont {Worlock},\ and\ \citenamefont
  {Roukes}}]{fon}%
  \BibitemOpen
  \bibfield  {author} {\bibinfo {author} {\bibfnamefont {W.}~\bibnamefont
  {Fon}}, \bibinfo {author} {\bibfnamefont {K.~C.}\ \bibnamefont {Schwab}},
  \bibinfo {author} {\bibfnamefont {J.~M.}\ \bibnamefont {Worlock}}, \ and\
  \bibinfo {author} {\bibfnamefont {M.~L.}\ \bibnamefont {Roukes}},\ }\href
  {\doibase 10.1103/PhysRevB.66.045302} {\bibfield  {journal} {\bibinfo
  {journal} {Phys. Rev. B}\ }\textbf {\bibinfo {volume} {66}},\ \bibinfo
  {pages} {045302} (\bibinfo {year} {2002})}\BibitemShut {NoStop}%
\bibitem [{\citenamefont {Barman}\ and\ \citenamefont
  {Srivastava}(2006)}]{barman}%
  \BibitemOpen
  \bibfield  {author} {\bibinfo {author} {\bibfnamefont {S.}~\bibnamefont
  {Barman}}\ and\ \bibinfo {author} {\bibfnamefont {G.~P.}\ \bibnamefont
  {Srivastava}},\ }\href {\doibase 10.1103/PhysRevB.73.205308} {\bibfield
  {journal} {\bibinfo  {journal} {Phys. Rev. B}\ }\textbf {\bibinfo {volume}
  {73}},\ \bibinfo {eid} {205308} (\bibinfo {year} {2006})}\BibitemShut
  {NoStop}%
\bibitem [{\citenamefont {Khan}\ and\ \citenamefont {Allen}(1987)}]{Khan87}%
  \BibitemOpen
  \bibfield  {author} {\bibinfo {author} {\bibfnamefont {F.~S.}\ \bibnamefont
  {Khan}}\ and\ \bibinfo {author} {\bibfnamefont {P.~B.}\ \bibnamefont
  {Allen}},\ }\href {\doibase 10.1103/PhysRevB.35.1002} {\bibfield  {journal}
  {\bibinfo  {journal} {Phys. Rev. B}\ }\textbf {\bibinfo {volume} {35}},\
  \bibinfo {pages} {1002} (\bibinfo {year} {1987})}\BibitemShut {NoStop}%
\bibitem [{\citenamefont {Kahn}\ \emph {et~al.}(2001)\citenamefont {Kahn},
  \citenamefont {Kim},\ and\ \citenamefont {Stroscio}}]{kahn}%
  \BibitemOpen
  \bibfield  {author} {\bibinfo {author} {\bibfnamefont {D.}~\bibnamefont
  {Kahn}}, \bibinfo {author} {\bibfnamefont {K.~W.}\ \bibnamefont {Kim}}, \
  and\ \bibinfo {author} {\bibfnamefont {A.~M.}\ \bibnamefont {Stroscio}},\
  }\href {\doibase 10.1063/1.1356429} {\bibfield  {journal} {\bibinfo
  {journal} {J.\ Appl.\ Phys.}\ }\textbf {\bibinfo {volume} {89}},\ \bibinfo
  {pages} {5107} (\bibinfo {year} {2001})}\BibitemShut {NoStop}%
\bibitem [{\citenamefont {Suzuura}\ and\ \citenamefont {Ando}(2002)}]{suzuura}%
  \BibitemOpen
  \bibfield  {author} {\bibinfo {author} {\bibfnamefont {H.}~\bibnamefont
  {Suzuura}}\ and\ \bibinfo {author} {\bibfnamefont {T.}~\bibnamefont {Ando}},\
  }\href {\doibase 10.1103/PhysRevB.65.235412} {\bibfield  {journal} {\bibinfo
  {journal} {Phys.\ Rev.\ B}\ }\textbf {\bibinfo {volume} {65}},\ \bibinfo
  {pages} {235412} (\bibinfo {year} {2002})}\BibitemShut {NoStop}%
\bibitem [{\citenamefont {De~Martino}\ \emph {et~al.}(2009)\citenamefont
  {De~Martino}, \citenamefont {Egger},\ and\ \citenamefont
  {Gogolin}}]{martino}%
  \BibitemOpen
  \bibfield  {author} {\bibinfo {author} {\bibfnamefont {A.}~\bibnamefont
  {De~Martino}}, \bibinfo {author} {\bibfnamefont {R.}~\bibnamefont {Egger}}, \
  and\ \bibinfo {author} {\bibfnamefont {A.~O.}\ \bibnamefont {Gogolin}},\
  }\href {\doibase 10.1103/PhysRevB.79.205408} {\bibfield  {journal} {\bibinfo
  {journal} {Phys. Rev. B}\ }\textbf {\bibinfo {volume} {79}},\ \bibinfo
  {pages} {205408} (\bibinfo {year} {2009})}\BibitemShut {NoStop}%
\bibitem [{\citenamefont {Blencowe}(1999)}]{blencowe99}%
  \BibitemOpen
  \bibfield  {author} {\bibinfo {author} {\bibfnamefont {M.~P.}\ \bibnamefont
  {Blencowe}},\ }\href {\doibase 10.1103/PhysRevB.59.4992} {\bibfield
  {journal} {\bibinfo  {journal} {Phys.\ Rev.\ B}\ }\textbf {\bibinfo {volume}
  {59}},\ \bibinfo {pages} {4992} (\bibinfo {year} {1999})}\BibitemShut
  {NoStop}%
\bibitem [{\citenamefont {Santamore}\ and\ \citenamefont
  {Cross}(2001)}]{santamore01}%
  \BibitemOpen
  \bibfield  {author} {\bibinfo {author} {\bibfnamefont {D.~H.}\ \bibnamefont
  {Santamore}}\ and\ \bibinfo {author} {\bibfnamefont {M.~C.}\ \bibnamefont
  {Cross}},\ }\href {\doibase 10.1103/PhysRevLett.87.115502} {\bibfield
  {journal} {\bibinfo  {journal} {Phys.\ Rev.\ Lett.}\ }\textbf {\bibinfo
  {volume} {87}},\ \bibinfo {pages} {115502} (\bibinfo {year}
  {2001})}\BibitemShut {NoStop}%
\bibitem [{\citenamefont {Santamore}\ and\ \citenamefont
  {Cross}(2002)}]{santamore02}%
  \BibitemOpen
  \bibfield  {author} {\bibinfo {author} {\bibfnamefont {D.~H.}\ \bibnamefont
  {Santamore}}\ and\ \bibinfo {author} {\bibfnamefont {M.~C.}\ \bibnamefont
  {Cross}},\ }\href {\doibase 10.1103/PhysRevB.66.144302} {\bibfield  {journal}
  {\bibinfo  {journal} {Phys. Rev. B}\ }\textbf {\bibinfo {volume} {66}},\
  \bibinfo {pages} {144302} (\bibinfo {year} {2002})}\BibitemShut {NoStop}%
\bibitem [{\citenamefont {Lindenfeld}\ \emph {et~al.}(2011)\citenamefont
  {Lindenfeld}, \citenamefont {Eisenberg},\ and\ \citenamefont
  {Lifshitz}}]{lindenfeld11}%
  \BibitemOpen
  \bibfield  {author} {\bibinfo {author} {\bibfnamefont {Z.}~\bibnamefont
  {Lindenfeld}}, \bibinfo {author} {\bibfnamefont {E.}~\bibnamefont
  {Eisenberg}}, \ and\ \bibinfo {author} {\bibfnamefont {R.}~\bibnamefont
  {Lifshitz}},\ }\href {\doibase 10.1103/PhysRevB.84.064532} {\bibfield
  {journal} {\bibinfo  {journal} {Phys.\ Rev.\ B}\ }\textbf {\bibinfo {volume}
  {84}},\ \bibinfo {pages} {064532} (\bibinfo {year} {2011})}\BibitemShut
  {NoStop}%
\bibitem [{\citenamefont {Broughton}\ \emph {et~al.}(1997)\citenamefont
  {Broughton}, \citenamefont {Meli}, \citenamefont {Vashishta},\ and\
  \citenamefont {Kalia}}]{Broughton97}%
  \BibitemOpen
  \bibfield  {author} {\bibinfo {author} {\bibfnamefont {J.~Q.}\ \bibnamefont
  {Broughton}}, \bibinfo {author} {\bibfnamefont {C.~A.}\ \bibnamefont {Meli}},
  \bibinfo {author} {\bibfnamefont {P.}~\bibnamefont {Vashishta}}, \ and\
  \bibinfo {author} {\bibfnamefont {R.~K.}\ \bibnamefont {Kalia}},\ }\href
  {\doibase 10.1103/PhysRevB.56.611} {\bibfield  {journal} {\bibinfo  {journal}
  {Phys. Rev. B}\ }\textbf {\bibinfo {volume} {56}},\ \bibinfo {pages} {611}
  (\bibinfo {year} {1997})}\BibitemShut {NoStop}%
\bibitem [{\citenamefont {Saviot}\ \emph {et~al.}(2004)\citenamefont {Saviot},
  \citenamefont {Murray},\ and\ \citenamefont {de~Lucas}}]{murray2}%
  \BibitemOpen
  \bibfield  {author} {\bibinfo {author} {\bibfnamefont {L.}~\bibnamefont
  {Saviot}}, \bibinfo {author} {\bibfnamefont {D.~B.}\ \bibnamefont {Murray}},
  \ and\ \bibinfo {author} {\bibfnamefont {M.~d. C.~M.}\ \bibnamefont
  {de~Lucas}},\ }\href {\doibase 10.1103/PhysRevB.69.113402} {\bibfield
  {journal} {\bibinfo  {journal} {Phys. Rev. B}\ }\textbf {\bibinfo {volume}
  {69}},\ \bibinfo {pages} {113402} (\bibinfo {year} {2004})}\BibitemShut
  {NoStop}%
\bibitem [{\citenamefont {Combe}\ \emph {et~al.}(2007)\citenamefont {Combe},
  \citenamefont {Huntzinger},\ and\ \citenamefont {Mlayah}}]{combe}%
  \BibitemOpen
  \bibfield  {author} {\bibinfo {author} {\bibfnamefont {N.}~\bibnamefont
  {Combe}}, \bibinfo {author} {\bibfnamefont {J.~R.}\ \bibnamefont
  {Huntzinger}}, \ and\ \bibinfo {author} {\bibfnamefont {A.}~\bibnamefont
  {Mlayah}},\ }\href {\doibase 10.1103/PhysRevB.76.205425} {\bibfield
  {journal} {\bibinfo  {journal} {Phys. Rev. B}\ }\textbf {\bibinfo {volume}
  {76}},\ \bibinfo {eid} {205425} (\bibinfo {year} {2007})}\BibitemShut
  {NoStop}%
\bibitem [{\citenamefont {Ramirez}\ \emph {et~al.}(2008)\citenamefont
  {Ramirez}, \citenamefont {Heyliger}, \citenamefont {Rapp\'{e}},\ and\
  \citenamefont {Leisure}}]{ramirez}%
  \BibitemOpen
  \bibfield  {author} {\bibinfo {author} {\bibfnamefont {F.}~\bibnamefont
  {Ramirez}}, \bibinfo {author} {\bibfnamefont {P.~R.}\ \bibnamefont
  {Heyliger}}, \bibinfo {author} {\bibfnamefont {A.~K.}\ \bibnamefont
  {Rapp\'{e}}}, \ and\ \bibinfo {author} {\bibfnamefont {R.~G.}\ \bibnamefont
  {Leisure}},\ }\href {\doibase 10.1121/1.2823065} {\bibfield  {journal}
  {\bibinfo  {journal} {J.\ Acoust.\ Soc.\ Am.}\ }\textbf {\bibinfo {volume}
  {123}},\ \bibinfo {pages} {709} (\bibinfo {year} {2008})}\BibitemShut
  {NoStop}%
\bibitem [{\citenamefont {Yoon}\ \emph {et~al.}(2005)\citenamefont {Yoon},
  \citenamefont {Ru},\ and\ \citenamefont {Mioduchowski}}]{yoon}%
  \BibitemOpen
  \bibfield  {author} {\bibinfo {author} {\bibfnamefont {J.}~\bibnamefont
  {Yoon}}, \bibinfo {author} {\bibfnamefont {C.~Q.}\ \bibnamefont {Ru}}, \ and\
  \bibinfo {author} {\bibfnamefont {A.}~\bibnamefont {Mioduchowski}},\ }\href
  {\doibase 10.1115/1.1795814} {\bibfield  {journal} {\bibinfo  {journal} {J.\
  App.\ Mech.}\ }\textbf {\bibinfo {volume} {72}},\ \bibinfo {pages} {10}
  (\bibinfo {year} {2005})}\BibitemShut {NoStop}%
\bibitem [{\citenamefont {Chico}\ \emph {et~al.}(2006)\citenamefont {Chico},
  \citenamefont {P\'erez-\'Alvarez},\ and\ \citenamefont {Cabrillo}}]{chico}%
  \BibitemOpen
  \bibfield  {author} {\bibinfo {author} {\bibfnamefont {L.}~\bibnamefont
  {Chico}}, \bibinfo {author} {\bibfnamefont {R.}~\bibnamefont
  {P\'erez-\'Alvarez}}, \ and\ \bibinfo {author} {\bibfnamefont
  {C.}~\bibnamefont {Cabrillo}},\ }\href {\doibase 10.1103/PhysRevB.73.075425}
  {\bibfield  {journal} {\bibinfo  {journal} {Phys. Rev. B}\ }\textbf {\bibinfo
  {volume} {73}},\ \bibinfo {pages} {075425} (\bibinfo {year}
  {2006})}\BibitemShut {NoStop}%
\bibitem [{\citenamefont {Graff}(1975)}]{graff}%
  \BibitemOpen
  \bibfield  {author} {\bibinfo {author} {\bibfnamefont {K.~F.}\ \bibnamefont
  {Graff}},\ }\href@noop {} {\emph {\bibinfo {title} {Wave Motion in Elastic
  Solids}}}\ (\bibinfo  {publisher} {Dover Publications},\ \bibinfo {address}
  {New York},\ \bibinfo {year} {1975})\BibitemShut {NoStop}%
\bibitem [{\citenamefont {Landau}\ and\ \citenamefont
  {Lifshitz}(1986)}]{landl}%
  \BibitemOpen
  \bibfield  {author} {\bibinfo {author} {\bibfnamefont {L.~D.}\ \bibnamefont
  {Landau}}\ and\ \bibinfo {author} {\bibfnamefont {E.~M.}\ \bibnamefont
  {Lifshitz}},\ }\href@noop {} {\emph {\bibinfo {title} {Theory of
  Elasticity}}}\ (\bibinfo  {publisher} {Pergamon, Oxford},\ \bibinfo {year}
  {1986})\BibitemShut {NoStop}%
\bibitem [{\citenamefont {Fetter}\ and\ \citenamefont
  {Walecka}(1971)}]{fetter}%
  \BibitemOpen
  \bibfield  {author} {\bibinfo {author} {\bibfnamefont {A.~L.}\ \bibnamefont
  {Fetter}}\ and\ \bibinfo {author} {\bibfnamefont {J.~D.}\ \bibnamefont
  {Walecka}},\ }\href@noop {} {\emph {\bibinfo {title} {Quantum theory of many
  body systems}}}\ (\bibinfo  {publisher} {McGraw-Hill},\ \bibinfo {address}
  {New York},\ \bibinfo {year} {1971})\BibitemShut {NoStop}%
\bibitem [{\citenamefont {Simmons}\ and\ \citenamefont {Wang}(1971)}]{simmons}%
  \BibitemOpen
  \bibfield  {author} {\bibinfo {author} {\bibfnamefont {G.}~\bibnamefont
  {Simmons}}\ and\ \bibinfo {author} {\bibfnamefont {H.}~\bibnamefont {Wang}},\
  }\href@noop {} {\emph {\bibinfo {title} {Single Crystal Elastic Constants and
  Calculated Aggregate Properies: a Handbook}}}\ (\bibinfo  {publisher} {The
  M.I.T. Press},\ \bibinfo {address} {Cambridge},\ \bibinfo {year}
  {1971})\BibitemShut {NoStop}%
\bibitem [{\citenamefont {Ashcroft}\ and\ \citenamefont
  {Mermin}(1976)}]{ashcroft}%
  \BibitemOpen
  \bibfield  {author} {\bibinfo {author} {\bibfnamefont {N.~W.}\ \bibnamefont
  {Ashcroft}}\ and\ \bibinfo {author} {\bibfnamefont {N.~D.}\ \bibnamefont
  {Mermin}},\ }\href@noop {} {\emph {\bibinfo {title} {Solid state physics}}}\
  (\bibinfo  {publisher} {Saunders College},\ \bibinfo {address}
  {Philadelphia},\ \bibinfo {year} {1976})\BibitemShut {NoStop}%
\bibitem [{\citenamefont {Chang}\ \emph {et~al.}(1983)\citenamefont {Chang},
  \citenamefont {Ting}, \citenamefont {Tang},\ and\ \citenamefont
  {Hess}}]{chang}%
  \BibitemOpen
  \bibfield  {author} {\bibinfo {author} {\bibfnamefont {Y.-C.}\ \bibnamefont
  {Chang}}, \bibinfo {author} {\bibfnamefont {D.~Z.~Y.}\ \bibnamefont {Ting}},
  \bibinfo {author} {\bibfnamefont {J.~Y.}\ \bibnamefont {Tang}}, \ and\
  \bibinfo {author} {\bibfnamefont {K.}~\bibnamefont {Hess}},\ }\href {\doibase
  10.1063/1.93732} {\bibfield  {journal} {\bibinfo  {journal} {Appl.\ Phys.\
  Lett.}\ }\textbf {\bibinfo {volume} {42}},\ \bibinfo {pages} {76} (\bibinfo
  {year} {1983})}\BibitemShut {NoStop}%
\bibitem [{\citenamefont {Reggiani}\ \emph {et~al.}(1987)\citenamefont
  {Reggiani}, \citenamefont {Lugli},\ and\ \citenamefont {Jauho}}]{reggiani}%
  \BibitemOpen
  \bibfield  {author} {\bibinfo {author} {\bibfnamefont {L.}~\bibnamefont
  {Reggiani}}, \bibinfo {author} {\bibfnamefont {P.}~\bibnamefont {Lugli}}, \
  and\ \bibinfo {author} {\bibfnamefont {A.~P.}\ \bibnamefont {Jauho}},\ }\href
  {\doibase 10.1103/PhysRevB.36.6602} {\bibfield  {journal} {\bibinfo
  {journal} {Phys. Rev. B}\ }\textbf {\bibinfo {volume} {36}},\ \bibinfo
  {pages} {6602} (\bibinfo {year} {1987})}\BibitemShut {NoStop}%
\bibitem [{\citenamefont {Kim}\ \emph {et~al.}(1987)\citenamefont {Kim},
  \citenamefont {Mason},\ and\ \citenamefont {Hess}}]{kim}%
  \BibitemOpen
  \bibfield  {author} {\bibinfo {author} {\bibfnamefont {K.}~\bibnamefont
  {Kim}}, \bibinfo {author} {\bibfnamefont {B.~A.}\ \bibnamefont {Mason}}, \
  and\ \bibinfo {author} {\bibfnamefont {K.}~\bibnamefont {Hess}},\ }\href
  {\doibase 10.1103/PhysRevB.36.6547} {\bibfield  {journal} {\bibinfo
  {journal} {Phys. Rev. B}\ }\textbf {\bibinfo {volume} {36}},\ \bibinfo
  {pages} {6547} (\bibinfo {year} {1987})}\BibitemShut {NoStop}%
\bibitem [{\citenamefont {Ferrari}\ \emph {et~al.}(2006)\citenamefont
  {Ferrari}, \citenamefont {Asenov}, \citenamefont {Nedjalkov},\ and\
  \citenamefont {Jacoboni}}]{giulio}%
  \BibitemOpen
  \bibfield  {author} {\bibinfo {author} {\bibfnamefont {G.}~\bibnamefont
  {Ferrari}}, \bibinfo {author} {\bibfnamefont {A.}~\bibnamefont {Asenov}},
  \bibinfo {author} {\bibfnamefont {M.}~\bibnamefont {Nedjalkov}}, \ and\
  \bibinfo {author} {\bibfnamefont {C.}~\bibnamefont {Jacoboni}},\ }\href
  {\doibase 10.1007/s10825-006-0029-2} {\bibfield  {journal} {\bibinfo
  {journal} {J.\ Comput.\ Electron.}\ }\textbf {\bibinfo {volume} {5}},\
  \bibinfo {pages} {419} (\bibinfo {year} {2006})}\BibitemShut {NoStop}%
\end{thebibliography}%

\end{document}